\documentstyle[seceq,epsf]{ptptex}

\def\agt{\mathrel{\raise.3ex\hbox{$>$}\mkern-14mu\lower0.6ex\hbox{$\sim$}}}
\def\alt{\mathrel{\raise.3ex\hbox{$<$}\mkern-14mu\lower0.6ex\hbox{$\sim$}}}

\begin{document}

\newcommand{\beq}{\begin{equation}}
\newcommand{\eeq}{\end{equation}}
\newcommand{\beqn}{\begin{eqnarray}}
\newcommand{\eeqn}{\end{eqnarray}}
\newcommand{\pa}{\partial}
\newcommand{\vp}{\varphi}
\newcommand{\varep}{\varepsilon}
\newcommand{\ep}{\epsilon}


\title{Axisymmetric Simulations of 
Rotating Stellar Collapse in Full General Relativity \\
--- {\it Criteria for Prompt Collapse to Black Holes} ---}

\author{Masaru SHIBATA}

\inst{Department of Physics, 
University of Illinois at Urbana-Champaign, Urbana, IL 61801, USA \\
Department of Earth and Space Science,~Graduate School of
Science,~Osaka University, Toyonaka 560-0043, Japan}

\maketitle

\begin{abstract}

Motivated by a recent paper by the 
Potsdam numerical relativity group, 
we have constructed a new 
numerical code for hydrodynamic simulation of axisymmetric systems 
in full general relativity. In this code,  we solve 
the Einstein field equation using Cartesian coordinates with 
appropriate boundary conditions. On the other hand, 
the hydrodynamic equations are solved in cylindrical 
coordinates. Using this code, we perform simulations 
to study axisymmetric collapse of rotating stars, which thereby 
become black holes 
or new compact stars, in full general relativity. To investigate the 
effects of rotation on the criterion for prompt collapse to black holes, 
we first adopt a polytropic equation of state, $P=K\rho^{\Gamma}$, 
where $P$, $\rho$, and $K$ are the pressure, rest mass density, 
and polytropic constant, with $\Gamma=2$. In this case, 
the collapse is adiabatic ({i.e.}, no change in entropy), and 
we can focus on the bare effect of rotation. 
As the initial conditions, we prepare rigidly and 
differentially rotating stars 
in equilibrium and then decrease the pressure to induce collapse. 
In this paper, we consider cases in which $q \equiv J/M_g^2 < 1$, 
where $J$ and $M_g$ are the angular momentum and the gravitational mass. 
It is found that the criterion of black hole formation 
is strongly dependent on the angular momentum parameter $q$. 
For $q < 0.5$, the criterion is not strongly sensitive to $q$; 
more precisely, if 
the rest mass is slightly larger than the maximum allowed value 
of spherical stars, a black hole is formed. 
However, for $q \alt 1$, it changes significantly: For $q \simeq 0.9$, the 
maximum allowed rest mass becomes $\sim 70$--$80\%$ larger than that for 
spherical stars. These findings depend only weakly on 
the rotational profiles given initially. 
We then report the results for simulations employing 
a $\Gamma$-law equation of state $P=(\Gamma-1)\rho\varep$, where 
$\varepsilon$ is the specific internal energy, 
to study effects of shock heating. We find that the effects of 
shock heating are particularly important for preventing 
prompt collapse to black holes in the case of large $q$ 
[i.e., $q = O(1)$]. 
\end{abstract}



\section{Introduction}

In the 1980s, one of the most important issues in numerical relativity 
was to perform simulations of rotating stellar collapse 
with the assumption of axial symmetry. It seems that 
there were two motivations for such studies.\cite{NOK} 
One was to explore the final fate after 
gravitational collapse of rotating stars. 
If the cosmic censorship conjecture suggested by 
Penrose \cite{penrose} is correct, any collapsed star 
should be surrounded by an event horizon, and consequently 
settle down to a rotating Kerr black hole, according to 
the uniqueness theorem of 
Israel,\cite{israel} Carter \cite{carter} and others. \cite{wald} 
However, if this conjecture is not correct, 
such a star might be reduced to a state with naked 
singularities. Numerical relativity was required to resolve 
this problem. The other motivation was to compute 
gravitational wave forms in the collapse, 
because the gravitational collapse of a rotating star 
to a black hole or a neutron star 
is one of promising sources of gravitational waves. 

Simulations of rotating stellar collapse in full general relativity 
were first performed by Nakamura.\cite{Nakamura} 
Using the (2+1)+1 formalism developed by Maeda et al.,\cite{Maeda} 
he performed simulations of 
rotating collapse for massive stars whose masses are much larger than the 
maximum allowed mass for neutron star formation. 
He used cylindrical coordinates $(\varpi, z)$ with 
a non-uniform grid spacing. At that time, he was able to 
use at most a (42,42) grid resolution for $(\varpi, z)$ 
because of restricted computational resources. 
The interesting finding in his simulations was 
that the rotational parameter $q\equiv J/M_g^2$, where $J$ and $M_g$ 
are the angular momentum and gravitational mass, can be an 
important parameter for determining black hole formation: 
The results of his simulations 
suggest that if $q>1$, no black hole is formed, indicating that 
cosmic censorship holds. 
Unfortunately, he was not able to compute gravitational waves, 
because his computational resources were severely restricted and 
the formalism he used was not well suited for computing 
gravitational wave forms in the wave zone. 

To compute gravitational waves emitted during gravitational collapse to 
black holes, Stark and Piran \cite{SP} 
performed simulations similar to those of Nakamura, 
adopting spherical polar coordinates with 
a typical grid size $(100,16)$ for $(r, \theta)$. 
The distinguishing 
feature of their work is that they adopted the Bardeen-Piran 
formalism,\cite{BP} 
which is well suited for computation of gravitational waves 
in the wave zone. As a result of this 
choice of formalism, they succeeded in computing 
gravitational wave forms, and clarified that the wave forms  
are characterized by the quasi-normal mode of rotating black holes 
formed after gravitational collapse and that the total radiated 
energy of gravitational waves is at most $0.1\%$ of the 
gravitational mass of the system.\cite{SP} 

Since the completion of their work, 
no new work in this field has been done for about 15 years.\cite{acst} 
Although several questions they originally wished to answer 
have been answered by their simulations, there are still 
many unsolved issues in astrophysics and 
general relativity that can be investigated using 
axisymmetric hydrodynamic simulations in full general relativity.  
For example, the following issues should be addressed as open 
questions: 
(1) Previous simulations were performed using restricted initial 
conditions and equations of state.\cite{Nakamura,SP} 
It is not clear whether 
their results are independent of the initial velocity field, 
density profile, internal energy distribution, and equations of state. 
A systematic study considering different initial settings 
should be carried out for detailed exploration of 
the validity of the cosmic censorship conjecture. 
(2) Previous works have indicated 
that if $q > 1$, a black hole is not formed. 
However, these works did not carefully investigate 
how the criterion of black hole 
formation changes due to the effect of rotation for $q < 1$. 
(3) Stark and Piran computed gravitational wave forms in the 
case of black hole formation. There have been also many Newtonian works 
for computing gravitational wave forms 
in the formation of neutron stars that are not very massive and 
compact.\cite{Newton} 
However, there has been no fully general relativistic work 
for computing gravitational wave forms in neutron star formation. 
For the case in which the formed neutron star is so massive and 
compact that general relativistic effects play an important role 
even at the formation, fully general relativistic 
studies on gravitational waves 
could contribute to the understandings of gravitational-wave-astronomy. 
(4) Realistic simulations of the rotating collapse of massive stars, 
which thereby become black holes (or proto-neutron stars) 
in full general relativity, 
have not yet been performed. This should be done 
to understand mechanisms of black hole formation in nature. 
Actually, a study of the formation of rapidly rotating black holes with 
surrounding accretion disks in stellar core collapse 
is currently one of the hot topics in connection with 
a hypothetical scenario for the central 
engine of $\gamma$-ray bursts.\cite{hyper} 
(5) The formation mechanism of supermassive black holes 
of mass $10^6$--$10^9M_{\odot}$, where $M_{\odot}$ 
denotes the solar mass, is also an open issue. 
Recently, Baumgarte and Shapiro have proposed 
an interesting scenario in which rigidly rotating supermassive stars at 
mass shedding limits ({i.e.}, rigidly rotating stars 
of maximum rotational velocity) are possible progenitors of 
supermassive black holes.\cite{BS}  According to their scenario, 
after dissipating the internal energy due to thermal radiation, 
the supermassive stars become unstable 
at the point at which $R/M_g \sim 450$ and $q \simeq 0.97$, where $R$ 
denotes the circumferential radius at the equator. 
Since $q \sim 1$, it is not clear whether such a progenitor 
really collapses to a black hole. To confirm that the scenario is 
promising, simulation in full general relativity is necessary. 

All these arguments obviously remind us that 
there are still many astrophysical and 
general relativistic motivations 
for performing hydrodynamic simulations of the 
rotating collapse of massive 
and supermassive stars in full general relativity.

Before proceeding, we roughly summarize the present situation regarding 
our computational resources. 
Recently, we have been performing 3D numerical simulations in 
full general relativity with Cartesian coordinates $(x,y,z)$ for 
investigating merging processes of binary neutron stars.\cite{gr3d,SU} 
Our 3D simulations are performed 
on the FACOM VPP/300R at the National Astronomical Observatory of Japan 
(NAOJ), in which we can use about 30 GBytes memory. 
We assume a plane symmetry with respect to 
the equatorial plane ($z=0$) and use a grid of size 
$(2N+1,2N+1,N+1)$ in $(x,y,z)$ where $N$ is at most 160. 
Let us here estimate a desirable grid number $N$ for simulation of 
the merging of binary neutron stars. 
In the following, we assume to use a uniform grid. 

First, we wish to locate the outer boundary 
in the wave zone to compute the gravitational wave forms accurately. 
The typical wavelength of gravitational waves just before 
the merging of two neutron stars is about $50M_g$. 
Thus, the outer boundaries 
should be located for $> 50M_g$. Second, such a binary could result in  
a black hole, and to resolve the black hole formation accurately, 
we need at least 10 grid 
points in the gravitational radius $\sim M_g$. 
These considerations imply that 
if we use a uniform grid, 
we need $N > 500$ for an accurate computation. 
In the next generation machines, which will be available in a year, 
we will be able to use several hundred GBytes. In this case, 
$N = 400({\rm Memory}/400{\rm GBytes})^{1/3}$ would be feasible. This 
is close to the value needed for accurate computation. 
However, with this maximum number, 
it would take $\sim 100$ CPU hours for 
one simulation on such machines, in which the 
computational speed is several hundred GFlops. This 
implies that a {\it systematic survey}, 
performing simulations for $\alt 100$ models of different equations 
of state and compactness of neutron stars, 
would still be a difficult task to complete 
on a time scale of approximately a year. 
(Note that under normal circumstances, 
we can use at most $\sim 1000$ CPU hours 
per year.) 

In contrast, the memory and computational speed 
of present computational resources are large enough for 
axisymmetric simulations with $\sim (500, 500)$. 
In the following, we assume 
the use of cylindrical coordinates $(\varpi,z)$, with reflection 
symmetry with respect to $z=0$; {i.e.}, the computational domain 
is assumed to be in the region satisfying $0\leq \varpi \leq L_x$ 
and $0\leq z \leq L_{z}$, where $L_x$ and $L_z$ are constants.  
In this case, a simulation 
with grid size $\sim (500,500)$ can be performed on 
a machine with $\alt 1$ GBytes and $\sim 1$ GFlops 
using only $\sim 10$ CPU hours or less (see below). This implies that 
a systematic survey performing simulations for $\sim 100$ 
models with $N \sim 500$ can be completed for 
axisymmetric systems in several months. 
Using more powerful machines, such as FACOM VPP/300R, 
we could use grids of size about $(3000,3000)$. In this case, 
completion of highly resolved simulations would also be 
feasible.

One longstanding problem for axisymmetric 
simulations in full general relativity has been to develop 
methods in which the accuracy and stability 
for a long-duration simulation can be preserved. 
In axisymmetric simulations, we have in general used 
cylindrical and spherical coordinate 
systems, which have coordinate singularities at the origin and 
along the symmetric axis $\varpi=0$. 
At such singularities, we have often faced 
several troubles. For example, there appear singular terms 
in the (3+1) form of the Einstein equation, such as \cite{NOK}
\beq
{\gamma_{\varpi\varpi,\varpi} \over \varpi}~~{\rm and}~~
{1 \over \varpi^2}
\biggl(\gamma_{\varpi\varpi}-{\gamma_{\varphi\varphi} \over \varpi^2}
\biggr),
\eeq
where $\gamma_{ij}$ denotes the three metric. 
Because of the regularity condition along the symmetric axis $\varpi=0$, 
these terms are finite in reality, but it is necessary to introduce 
appropriate prescriptions to guarantee them to be finite. 
Moreover, at the coordinate singularities, we have to change 
the scheme for finite differencing, because there are no 
negative value of $\varpi$ and $r$. 
Under such changes, the numerical system 
can have a mismatching between the coordinate singularities and 
their neighborhoods, often resulting in the 
accumulation of numerical error and, thus, numerical instability. 
For a long-duration simulation, 
we have often been required to add artificial viscosities around the 
coordinate singularities to stabilize the numerical system.\cite{foot00} 

Recently, the Potsdam numerical relativity group has proposed a method 
to make a robust code for solving the Einstein field equation 
for axisymmetric systems.\cite{alcu} 
The essence of their idea is that Cartesian coordinates 
could be used even for axisymmetric systems if 
the Einstein field equation is solved only for 
$y=0$ (or $x=0$), using the boundary condition 
at $y=\pm \Delta y$ (or $x=\pm \Delta x$) provided by the 
axial symmetry. 
Since the Einstein field equation is written in the Cartesian 
coordinate system, we neither have singular terms nor have to change 
the finite differencing scheme anywhere, except at the outer boundaries. 
Thus, it is possible to perform a stable and accurate, 
long-duration simulation without any prescription or 
artificial viscosities. 
Another merit for us is that it is easy to construct such a code, 
because we have already possessed a 3D code and 
only need to modify its boundary conditions.

Using this method, we 
have constructed a new numerical code for axisymmetric spacetimes 
and performed simulations of 
rotating stellar collapse. As a first step, in this paper, 
we focus on simulations adopting simple initial conditions and 
simple equations of state 
to investigate the effects of rotation 
on the criteria for prompt collapse to black holes. 
As initial conditions, we prepare rigidly and differentially 
rotating stars 
in equilibrium states which are not too compact. We prepare 
a large number of equilibrium states 
that have different values of $q$ but 
identical values of the rest mass. 
Then, we reduce the pressure to induce gravitational collapse. 
We pay particular attention to collapse for $q < 1$ in this paper 
[{i.e.}, we focus on the issue (2) mentioned above], 
because black holes are not likely to be 
formed for $q > 1$.\cite{Nakamura,SP} 
For most computations, we adopt a polytropic equation 
of state [see Eq.~(\ref{PEOS})]; 
namely, we do not take into account the effect of 
shock heating. In this treatment, 
it is easy to investigate the bare effect of rotation for black hole 
formation. For some computations, 
we adopt a $\Gamma$-law equation of state [see Eq.~(\ref{GEOS})] 
to investigate the effect of 
shock heating. To roughly mimic a moderately stiff 
equation of state for neutron stars, 
the adiabatic constant $\Gamma$ is set to $2$. 
Giving the numerical results for this experiment, we demonstrate 
that our present strategy and current computational 
resources are sufficient to systematically 
perform stable and well-resolved hydrodynamic simulations 
in axisymmetric numerical relativity.

The paper is organized as follows. 
In \S 2 we describe the formulation and numerical methods. 
In \S 3 we give the initial conditions which 
we adopt in this paper. In \S 4 we present the numerical results. 
Section 5 is devoted to summary and discussion. 
Throughout this paper, we adopt units in which $G=1=c$ where $G$ and $c$ 
denote the gravitational constant and speed of light.  
We use Cartesian coordinates, $x^k=(x, y, z)$ 
as the spatial coordinates, with $r=\sqrt{x^2+y^2+z^2}$, 
$\varpi=\sqrt{x^2+y^2}$ and $\varphi=\tan^{-1}(y/x)$. $t$ denotes
the coordinate time. In the following, we use a uniform grid with 
grid spacing $\Delta x = \Delta y = \Delta z$ or 
$\Delta x = \Delta y = 2 \Delta z$. 
Greek indices $\mu, \nu, \cdots$ 
denote $x, y, z$ and $t$, 
small Latin indices $i,j,k, \cdots$ denote $x, y$ and $z$, and 
capital Latin indices $A$ and $B$ denote $x$ and $z$. 

\section{Methods}

\subsection{Summary of formulation}

For solving the Einstein field equation, 
we use the same formulation and gauge condition as
in Refs.~\citen{gw3p2,gr3d} and \citen{gr3drot}, 
to which the reader may refer for details and basic
equations.  The fundamental variables for the geometry 
used in this paper are as follows:
\beqn
\alpha~ &&: {\rm lapse~ function}, \nonumber \\
\beta^k~ &&: {\rm shift~ vector}, \nonumber \\
\gamma_{ij}~ &&:{\rm metric~ in~ 3D~ spatial~ hypersurface},\nonumber \\ 
\gamma~ &&=e^{12\phi}={\rm det}(\gamma_{ij}), \nonumber \\
\tilde \gamma_{ij}~ &&=e^{-4\phi}\gamma_{ij}, \nonumber \\
K_{ij}~ &&:{\rm extrinsic~curvature}.\nonumber 
\eeqn
The variables, $\tilde \gamma_{ij}$, $\phi$, 
$\tilde A_{ij}\equiv e^{-4\phi}(K_{ij}-\gamma_{ij}K_k^{~k}/3)$, 
together with the three auxiliary functions 
$F_i\equiv \delta^{jk} \pa_k \tilde \gamma_{ij}$ and the
trace of the extrinsic curvature $K_k^{~k}$, 
evolve according to a free evolution code.  

As the source of the Einstein equation, we use an ideal fluid. 
The fundamental variables are defined as follows: 
\beqn
\rho~ &&:{\rm rest~ mass~ density},\nonumber \\
\varep~ &&: {\rm specific~ internal~ energy}, \nonumber \\
P~ &&:{\rm pressure}, \nonumber \\
h~ &&=1+\varepsilon + P/\rho,\nonumber \\
u^{\mu}~ &&:~{\rm four~ velocity};~ \hat u_k =h u_k,\nonumber \\
v^{k}~ && ={u^k \over u^0};~\Omega=v^{\varphi}, \nonumber \\
w~  && =\alpha u^0=\sqrt{1+\gamma^{ij}u_i u_j}. 
\eeqn

In this paper, we adopt two kinds of equations of state 
for the dynamical simulations. One is a polytropic equation of state, 
\beq
P=K\rho^{\Gamma}, \label{PEOS}
\eeq
where $K$ is fixed to a constant value throughout the evolution. 
The other is a $\Gamma$-law equation of state of the form
\beq
P=(\Gamma-1)\rho \varep. \label{GEOS} 
\eeq 
In both cases, we set the adiabatic constant as $\Gamma=2$ 
({i.e.}, the polytropic index $n$ is given by 
$n=1/(\Gamma-1)=1$) as a qualitative 
approximation of moderately stiff equations of state for neutron stars.
We note that if we prepare a polytropic star as the initial conditions 
and the system evolves in an adiabatic manner without shocks, 
the equation of state is preserved in the form of Eq.~(\ref{PEOS}), 
even using Eq.~(\ref{GEOS}); {i.e.}, 
the value $P/\rho^{\Gamma} \equiv K'(x^{\mu})$ for any fluid 
element remains a constant ($=K$).

In the simulation with the equation of state (\ref{PEOS}), 
the entropy of each fluid element is conserved during the time evolution. 
For this reason, 
we refer to the collapse with Eq.~(\ref{PEOS}) as ``adiabatic 
collapse'' in the following.

The slicing and spatial gauge conditions we use in this paper are 
basically the same as those adopted in our previous series of 
papers;\cite{gw3p2,gr3d,gr3drot} {i.e.}, we 
impose an ``approximate'' maximal slice condition ($K_k^{~k} \simeq 0$)
and an ``approximate'' minimum distortion (AMD) gauge condition 
($\tilde D_i (\pa_t \tilde \gamma^{ij}) \simeq 0$, 
where $\tilde D_i$ is the
covariant derivative with respect to $\tilde \gamma_{ij}$). 
For the case in which a rotating star significantly contracts and 
$\phi$ increases by a large factor, we 
slightly modify the spatial gauge condition in order to improve the 
spatial resolution around the highly relativistic region, 
as in Refs.~\citen{gw3p2} and \citen{gr3drot}. 
In this paper, we modify the gauge according to 
\beqn
&&\beta^x=\beta^x_{\rm AMD}-f(t){x \over \varpi+\epsilon}
\beta^{\varpi'}_{\rm AMD},\\
&&\beta^y=\beta^y_{\rm AMD}-f(t){y \over \varpi+\epsilon}
\beta^{\varpi'}_{\rm AMD},\\
&&\beta^z=\beta^z_{\rm AMD}[1-f(t)],
\eeqn
where $\epsilon$ is 
a constant much less than $\Delta x$ and $\Delta z$, 
$\beta^{k}_{\rm AMD}$ is 
$\beta^k$ in the AMD gauge condition, 
$\beta^{\varpi'}_{\rm AMD}=
(\beta^x_{\rm AMD} x+\beta^y_{\rm AMD} y)/(\varpi+\epsilon)$, and 
$f(t)$ is chosen as
\beq
f(t)=\left\{
\begin{array}{ll}
\displaystyle 
1 & {\rm for}~\phi_0 \geq 1.0,\\
2\phi_0 -1& {\rm for}~0.5 \leq \phi_0 \leq 1.0,  \\
0 & {\rm for}~\phi_0 < 0.5, \\
\end{array}
\right. 
\eeq
where $\phi_0$ is $\phi$ at $r=0$. 
Namely, for large $\phi_0 \geq 1$, $\beta^z=0$ and 
$\beta^{\varpi} \simeq 0$, but $\beta^{\varphi} \not=0$. 

We impose axial symmetry, following Ref.~\citen{alcu}. 
First, we define the computational domain as $0 \leq x,z \leq L$ and 
$-\Delta y \leq y \leq \Delta y$, 
where $L$ denotes the location of the outer boundaries, and 
reflection symmetry with respect to the $z=0$ plane is assumed. 
In this computational domain, we need only 3 points 
in the $y$ direction, $0$ and $\pm \Delta y$. 
We here determine that the Einstein equation 
is solved only in the $y=0$ plane.  
Then, the boundary conditions at $y=\pm \Delta y$ that 
are necessary in evaluating $y$-derivatives 
are supplied from the assumption of axial symmetry as 
\beqn
Q_{xx} && =Q_{xx}^{(0)}\cos^2\varphi-Q_{xy}^{(0)}\sin 2\varphi 
+Q_{yy}^{(0)}\sin^2\varphi,\nonumber \\
Q_{xy} && ={1 \over 2}Q_{xx}^{(0)}\sin 2\varphi
+Q_{xy}^{(0)}\cos 2\varphi -{1 \over 2} Q_{yy}^{(0)}\sin 2\varphi,
\nonumber\\
Q_{yy} && =Q_{xx}^{(0)}\sin^2\varphi+Q_{xy}^{(0)}\sin 2\varphi 
+Q_{yy}^{(0)}\cos^2\varphi,\nonumber\\
Q_{xz} && =Q_{xz}^{(0)}\cos\varphi-Q_{yz}^{(0)}\sin\varphi ,\hskip 1cm
Q_{yz}  =Q_{xz}^{(0)}\sin\varphi+Q_{yz}^{(0)}\cos\varphi ,\nonumber \\
Q_{x} && =Q_{x}^{(0)}\cos\varphi-Q_{y}^{(0)}\sin\varphi ,\hskip 1cm
Q_{y}  =Q_{x}^{(0)}\sin\varphi+Q_{y}^{(0)}\cos\varphi ,\nonumber \\
Q_{zz} && =Q_{zz}^{(0)},\hskip 1cm
Q_{z} =Q_{z}^{(0)},\hskip 1cm   Q  = Q^{(0)},\nonumber 
\eeqn
where $\varphi=\varphi(x)=\tan^{-1}[\pm\Delta y/\sqrt{x^2+(\Delta y)^2}]$, 
$Q_{ij}$, $Q_i$ and $Q$ are an arbitrary tensor, vector and 
scalar in the 3D spatial hypersurface, 
and $Q_{ij}^{(0)}, Q_i^{(0)}~{\rm and}~Q^{(0)}$ 
denote the values of $Q_{ij}, Q_i~{\rm and}~Q$ at 
$(\sqrt{x^2+(\Delta y)^2},0,z)$, which are interpolated using 
Lagrange's formula \cite{recipe} 
with three nearby points along the $x$ direction 
({i.e.}, $x \pm \Delta x$ and $x$). At $x=L$, we use only two points, 
$x-\Delta x$ and $x$, for the extrapolation. 
We note that in using the interpolation and extrapolation, we 
assume that the geometric variables are smooth functions. 
This assumption is justified as long as black holes are not
formed. 
In the case of black hole formation, however, some of 
the geometric variables around the black hole horizons become 
very steep. In such a situation, the interpolation (and extrapolation) 
becomes less accurate. 

To impose the gauge conditions, as well as to solve constraint 
equations in preparing the initial conditions, we 
solve scalar and vector elliptic-type equations of 
the form \cite{gr3d}
\beqn
&& \Delta Q = S, \label{Poissons}\\
&& \Delta Q_i=S_i,\label{Poissonv}
\eeqn
where $\Delta$ denotes the Laplacian in the flat 3D space, and 
$S$ and $S_i$ denote the source terms. Using the interpolation 
mentioned above, $\pa_{yy} Q$ and $\pa_{yy} Q_i$ 
are evaluated in the finite differencing as 
\beqn
&&\pa_{yy} Q = 2{ Q^{(0)}-Q(x,0,z) \over (\Delta y)^2}, \hskip 1.8cm
\pa_{yy} Q_z = 2{ Q_z^{(0)}-Q_z(x,0,z) \over (\Delta y)^2},\nonumber \\
&&\pa_{yy} Q_x = 2{ Q_x^{(0)}|\cos\varphi|-Q_x(x,0,z) \over (\Delta y)^2},~~
\pa_{yy} Q_y = 2{ Q_y^{(0)}|\cos\varphi|-Q_y(x,0,z) \over (\Delta y)^2}.
\nonumber
\eeqn
On the other hand, the finite differencing in the $x$ and $z$ 
directions, $\pa_{xx} Q_i$ and $\pa_{zz} Q_i$, are 
written in the standard form as 
\beqn
&&{Q_i(x+\Delta x, 0, z)-2 Q_i(x,0,z)+Q_i(x-\Delta x,0,z) 
\over (\Delta x)^2}, \nonumber \\
&&~ \nonumber \\
&&{Q_i(x, 0, z+\Delta z)-2 Q_i(x,0,z)+Q_i(x,0,z-\Delta z) 
\over (\Delta z)^2}.\nonumber 
\eeqn
Thus, in the finite differencing form for each component of 
Eq.~(\ref{Poissonv}), 
only one component of $Q_i$ is included, 
implying that each component of the vector elliptic-type 
equation is solved independently, 
as in the case of the scalar elliptic equation.

For computation of hydrodynamic equations, on the other hand, we 
do not use the above method. Instead, we solve the 
equations in cylindrical coordinates. 
These equations are written in the $y=0$ plane as
\beqn
&&\pa_t \rho_* + {1 \over x}\pa_x(\rho_*x v^x)+\pa_z(\rho_* v^z)=0,
\label{continuity}\\
&&\pa_t (\rho_* \hat u_A) + {1 \over x}\pa_x(\rho_* x \hat u_A v^x)
+\pa_z(\rho_* \hat u_A v^z) \nonumber \\
&&~~={\rho_* \hat u_y v^y \over x}\delta_{Ax}
-\alpha e^{6\phi}\pa_A (P+P_{\rm art}) \nonumber \\
&&~~~~-\rho_* \biggl[w h \pa_A \alpha - \hat u_j\pa_A \beta^j
+{\alpha e^{-4\phi} \hat u_i \hat u_j \over 2 w h} 
\pa_A \tilde \gamma^{ij}
-{2\alpha h (w^2-1) \over w} \pa_A \phi \biggr],\label{eulerA}\\
&&\pa_t (\rho_* \hat u_y) 
+ {1 \over x^2}\pa_x(\rho_* x^2 \hat u_y v^x)+\pa_z(\rho_* \hat u_y v^z)
=0,\label{eulery}\\
&& \pa_t e_* + {1 \over x}\pa_x(e_* x v^x)+\pa_z(e_* v^z)=
\dot e_{\rm art},\label{energy}
\eeqn
where $\rho_* \equiv \rho w e^{6\phi}$ and 
$e_* \equiv (\rho\varepsilon)^{1/\Gamma}w e^{6\phi}$. 
$P_{\rm art}$ and $\dot e_{\rm art}$ denote artificial viscous 
terms, and are added in the same manner as those in Ref.~\citen{gr3d} 
only for the $\Gamma$-law equations of state 
to consider effects of shock heating. 
Equations (\ref{continuity})--(\ref{energy}) represent 
the continuity equation, Euler equation for the 
$x$ and $z$ directions ($A=x$ or $z$), Euler equation for the $y$ 
direction (or angular momentum equation), 
and energy (or entropy) equation, respectively. 
The hydrodynamic equations are solved using van Leer's scheme 
for the advection terms \cite{Leer} in this paper. 
We note that for the polytropic equations of state (\ref{PEOS}), 
we do not have to solve Eq.~(\ref{energy})\cite{foot} 
nor add the artificial viscous terms. 

There are two  reasons 
why we adopt these equations instead of the equations in 
Cartesian coordinates. One is that in the simulation, 
conservation of the rest mass and 
angular momentum of the system during the evolution 
is easily guaranteed in these forms with help of 
standard numerical methods. 
The other is that the hydrodynamic variables often become 
non-smooth in the presence of shocks in contrast to the geometric 
variables. In such cases, the axisymmetric conditions imposed by 
the interpolation (and extrapolation) are less accurate.

With these axisymmetric equations, 
however, there exist coordinate singularities along the 
symmetric axis ($x=0$) which might 
cause a numerical instability for a long-duration simulation. 
In order to avoid such instability, we here add 
small artificial viscous terms 
\beq
\nu_{\rm vis}\rho_* \Delta \hat u_A 
\eeq
to Eq.~(\ref{eulerA}) in order to stabilize the system. 
Here, $\Delta$ denotes the flat Laplacian again, and 
$\nu_{\rm vis}$ is a small constant 
such that the typical dissipation time scale due to the 
viscosity, $(\Delta x)^2/\nu_{\rm vis}$, is much longer than the 
dynamical time scale of the system, $\sim \rho_c^{-1/2}$, 
where $\rho_c$ denotes the central density at $t=0$. 

We note that there are no $y$-derivatives in Eqs.~(\ref{continuity})
--(\ref{energy}), so that we do not have to assign 
grid points at $\pm \Delta y$ for the hydrodynamic variables 
({i.e.}, we can save memory).

To check the accuracy of the numerical results, we monitor the 
conservation of the total rest mass $M_*$, 
gravitational mass $M_g$, and angular momentum $J$, 
which are computed in the $y=0$ plane  as 
\beqn
&&M_*=4\pi \int_0^{L_x} x dx \int_0^{L_z} d z \rho_*, \\ 
&&M_g=-2\int_0^{L_x} x dx \int_0^{L_z} dz \biggl[ -2\pi E e^{5\phi}
+{e^{\phi} \over 8} \tilde R 
-{e^{5\phi}\over 8}\Bigl\{K_{ij}K^{ij}-(K_k^{~k})^2\Bigr\}
\biggr],~~~~~\\
&&J=4\pi \int_0^{L_x} x^2 dx \int_0^{L_z} dz \rho_* \tilde u_y,
\eeqn
where $E=\rho h w^2-P$ and $\tilde R$ is the Ricci scalar 
with respect to $\tilde \gamma_{ij}$. 
$M_*$ should be conserved in all systems. 
Because of the axial symmetry, $J$ should also be conserved. 
On the other hand, $M_g$ should not be conserved 
because of gravitational radiation. 
However, the total radiated energy 
of gravitational waves is likely to be quite 
small in the axisymmetric spacetime, so that we can consider 
$M_g$ as an approximately conserved quantity. 
We also monitor the conservation of the 
specific angular momentum spectrum \cite{SP2} 
\beq
M_*(j_0)=4\pi \int_{j \leq j_0} xdx dz \rho_*,
\eeq
where $j$ is the specific angular momentum computed as 
$x\hat u_y(=h u_{\varphi})$ and $j_0$ denotes a particular value for $j$. 
We stop simulations before violation of the 
conservation of these quantities is large.

\subsection{Test simulations}

Following Ref.~\citen{gr3d}, 
we have performed several test simulations, such as 
the spherical collapse of a pressure-free dust to a black hole, 
radial oscillation of stable 
spherical stars, and quasi-radial oscillation of stable rotating stars.
As in Ref.~\citen{gr3d}, we have confirmed that 
the collapse of dust can be accurately computed 
up to the formation of the apparent horizon and that 
the frequency of the fundamental mode for radial oscillation of 
spherical stars is obtained. We have 
also checked that the frequency of the fundamental mode 
for quasi-radial oscillation of rapidly rotating 
stars agrees with that computed in Refs.~\citen{gr3d} and \citen{gr3drot}. 
From these results, we are convinced 
that our axisymmetric code is sufficiently robust to perform 
long-duration simulations stably and to provide reliable results.

\section{Initial conditions for rotating stars}

Initial conditions are prepared in the following manner. 
First, we adopt rigidly as well as  differentially rotating
stars in (approximately) equilibrium states using the 
polytropic equation of state $P=K_{\rm i}\rho^{\Gamma}$, 
with $\Gamma=2$ and 
$K_{\rm i}=1$. Approximate equilibrium states are 
obtained by choosing a conformally
flat spatial metric, {i.e.}, assuming $\gamma_{ij}=e^{4\phi}
\delta_{ij}$. (See, {e.g.},  Ref.~\citen{gr3d} for the 
equations to be solved.)  
This approach is computationally convenient, 
and, as illustrated in Ref.~\citen{CST0}, it 
provides an excellent approximation
to exact equilibrium configurations. 
In particular, we here 
prepare equilibrium states which are not very compact, so that 
effects of the neglected spatial components of the metric 
are not important. Hence, the approximate initial conditions 
can be almost exact equilibrium states. 
We also note that this approach provides valid initial data for 
full general relativistic simulations in the sense that 
the solution satisfies the constraint equations exactly. 

Following previous studies,\cite{rotstar} 
we fix a differentially rotational profile according to 
\beq
F(\Omega)\equiv u^0u_{\varphi}=A^2(\Omega_0-\Omega), 
\eeq
where $A$ is an arbitrary constant 
which describes the length scale over which $\Omega$ changes, 
and $\Omega_0$ is the angular velocity on the rotational axis. 
In this paper, $A$ is chosen to be $\varpi_e/3$ 
and $\varpi_e$, where $\varpi_e$ denotes the 
coordinate radius of rotating stars at the equator. 
In the Newtonian limit $u^0 \rightarrow 1$ and 
$u_{\varphi} \rightarrow \varpi^2 \Omega$, the rotational 
profile reduces to 
\beq
\Omega = \Omega_0 {A^2 \over \varpi^2 + A^2}. \label{diffomega} 
\eeq
Thus, for $A=\varpi_e/3$ and $\varpi_e$, 
$\Omega$ at $\varpi=\varpi_e$ (hereafter 
referred to as $\Omega_e$) is about one tenth and half of $\Omega_0$, 
respectively ({cf.}, Tables II and III). 
Note that for $A \rightarrow \infty$, the rotational profile 
approaches the rigid rotation. 

We prepare a large 
number of equilibrium states, changing the angular momentum 
parameter of the system, $q$, 
but fixing the rest mass $M_*$ to 0.05. In Tables I--III, 
several quantities 
which characterize the properties of rotating stars are listed 
for $A=\infty$ (rigid rotation), $A=\varpi_e/3$, and $A=\varpi_e$. 
Here, the rotational kinetic energy $T$ and the 
gravitational binding energy $W (>0)$ are defined as
\beqn
T&&={1 \over 2}\int \rho_* j \Omega d^3x,\\
W&&=\int \rho_* \varep d^3x+T+M_*-M_g.
\eeqn
We note that for unit in which 
$K_{\rm i}=1$, the maximum allowed rest masses 
of spherical stars (hereafter $\bar M_{* {\rm smax}}$) 
and rigidly rotating stars are 
about $0.180$ and $0.207$, respectively, with 
$R/M_g \sim 5$.\cite{CST}  Hence, the 
rotating stars prepared here as initial conditions 
are not very compact and massive. We also note that 
for a rigidly rotating star at the mass shedding limit 
(i.e., a rigidly rotating star 
of maximum rotational velocity), the 
ratio of the polar coordinate length to the equatorial coordinate length  
is $R_{\rm ratio} \simeq 0.57$ for $M_*=0.05$ with $q \simeq 1.22$. 
For differentially rotating stars with $A = \varpi_e/3$ and $\varpi_e$, 
even a toroidal star of $R_{\rm ratio}=0$ and $q \gg 1$ can be obtained. 
Since $R_{\rm ratio} \geq 0.65$ and $q < 1$, 
it is found that the rotation of stars here is not very rapid. 

We note that $T/W$ for differentially rotating stars is larger than 
that for rigidly rotating stars for an identical value of $q$. 
During collapse, $T/W$ is expected to roughly increase in 
proportion to $R^{-1}$ when $J$ and $M_g$ are approximately 
conserved. 
A star for which $T/W$ is larger than the critical value ($\sim 0.25$) 
will be unstable against the non-axisymmetric dynamical 
instability.\cite{diff}   Thus, even if $q$ and $M_*$ are 
identical, a collapsing star of differentially rotating initial 
conditions is likely to be more susceptible 
to the non-axisymmetric instability than a rigidly rotating one.

To induce gravitational collapse, 
we initially reduce the pressure by changing $K$ from $K_{\rm i}=1$ 
to a value less than unity 
without changing the profile of $\rho_*$ and $u_i$. 
(Note that we recompute the constraint equations 
whenever we modify the initial equilibrium configurations.) 
In the case of the polytropic equations of state with units for which 
$G=1=c$, 
$\bar M_* \equiv M_* K^{-n/2}$ is a non-dimensional quantity, 
and the maximum allowed value is 0.180 ($=\bar M_{*\rm smax}$) 
for spherical stars and 0.207 for rigidly rotating stars, as 
mentioned above. Thus, if 
$K^{-1/2}$ is larger than about $0.18/0.05$, a collapsing star 
of small $q$ (i.e., $q \ll 1$) is likely to collapse to a black hole 
in the absence of shocks.\cite{foot2}   
For rapidly rotating stars with $q=O(1)$, the criterion for $K^{-1/2}$ 
is expected to be larger by $O(1)$, due to effects of rotation, 
because for $q > 1$, no black hole would be formed.\cite{Nakamura,SP}  
In the presence of shocks, $K'(=P/\rho^{\Gamma})$ 
increases at shocks, so that the threshold value of $K^{-1/2}$ 
for black hole formation is increased. 

We should mention that we do not consider 
this numerical experiment to be realistic for  
description of stellar collapses in nature. In realistic stellar 
collapses, the equations of state are much more complicated, effects 
of cooling and heating due to microscopic processes are 
important, the initial stellar radius is much larger, and 
the initial density configuration and rotational profile 
may be much more complicated. In this paper, we 
focus on the dynamics of rotating collapse in general relativity, 
and pay particular 
attention to qualitative effects of rotation on the criterion for 
prompt collapse to black holes. We expect that the present study 
with its simple equations of states and initial conditions
will be helpful for understanding qualitative features of 
realistic rotating stellar collapses.

\section{Numerical results}

In numerical simulations, we 
basically adopt a fixed uniform grid in which 
$\Delta x=\Delta y=\Delta z=$constant with grid size 
$(x, z)=(N+1, N+1)$. For collapse of rapidly rotating stars, 
the products after the collapses are often accompanied by 
very thin disks. To improve resolution in the direction 
perpendicular to the disk, we choose 
$\Delta x=\Delta y=2\Delta z=$constant using grid size $(N+1,2N+1)$ 
in such cases. To check the convergence 
of our numerical results, we adopt different grid resolutions as 
$N=180,~270$, 360 and 500. Typically, we choose $N=360$ as 
the maximum grid number. Irrespective of the grid size, 
the grid covers the same computational domain, $0 \leq x,~z \leq L$; 
namely, the resolution is improved by increasing $N$. 
We always prepare rotating stars whose surface at the equator 
($\varpi=\varpi_e$) is located at $2L/3$ as initial conditions. 

In numerical simulations, we have used the FACOM VX/4R machine in 
the data processing center of NAOJ. This is a vector-parallel 
machine composed of four vector processors, but we have used only one 
of four processors in the present numerical experiment. 
The typical memory and CPU time in one simulation with typical grid size 
$(361, 361)$ are about 400 MBytes and $25$ hours for 15000 time steps. 
(We note that with help of parallelization, 
the CPU hours could be reduced, although we have not yet carried this out 
for the present numerical code.)

\subsection{Adiabatic collapse for rigidly rotating initial data}

Simulations of adiabatic collapses 
using polytropic equations of state 
and rigidly rotating initial data sets 
have been performed for models (B)--(G) described in Table I. 
For each set of initial conditions, 
the simulations were carried out changing the 
decrease factor of the pressure, $K$, over a wide range. 
In Fig.~1, we summarize the products after collapse: 
The horizontal and vertical axes represent 
$q$ and $K^{-1/2}$, respectively. 
The filled circles and crosses indicate that 
the products are a new star and 
a black hole, respectively. The filled triangles imply that 
we were not able to judge the nature of 
the products because a very thin disk 
is formed in the equatorial plane, which cannot be well 
resolved even with the maximum grid number.\cite{foot5}   
(As for the process of the disk formation, 
see, e.g., the first four panels of Figs.~4 and 5.) 
The dotted line denotes $K^{-1/2}=0.18/0.05$. The dashed curve  
denotes the approximate threshold which distinguishes between 
new star and black hole formation. 
In this numerical experiment, the threshold curve 
can be approximated by a quadratic function of $q$ 
as $K^{-1/2}=1.9q^2+3.4$ for $q < 0.5$, and 
it is a steeper function for larger $q$. 

As we expected (see discussion above), the criterion for $K^{-1/2}$ 
for prompt collapse to black holes is $\sim 3.6$ 
for small $q$ (i.e., $q < 0.5$). 
This implies that for $q < 0.5$, the effect of rotation is not 
very important for supporting the self-gravity of formed rotating stars. 
However, with increasing $q$, the critical value of 
$K^{-1/2}$ increases, and for $q \simeq 0.9$, 
it becomes $\sim 6$--$6.5$, which implies that a black hole is not formed 
after collapse even if $\bar M_*~(=K^{-1/2}M_*)$ 
is $\sim 75\%$ larger than $\bar M_{* {\rm smax}}$. 
For model (B), the critical value of $K^{-1/2}$ can be $\sim 7$ or 
even larger; {i.e.}, $\bar M_*$ of the resulting star can be 
twice as large as $\bar M_{* {\rm smax}}$. 
For this model, the maximum allowed value of 
$K^{-1/2}$ might be much larger, but it is difficult to 
perform accurate simulations for such a large decrease factor, 
because the thickness of the formed disk is too small to 
be well resolved.\cite{foot5}

In Figs.~2 and 3, we display the 
time evolution of the central density $\rho(r=0)$ \cite{foot4}
and $\alpha$ at the origin for model (C) with $K^{-1/2}=
5.92$ and 6.71 and for 
model (E) with $K^{-1/2} = 3.74$ and $3.87$. 
The results shown here are obtained with $(361,361)$ (solid curves), 
$(271,271)$ (dotted curves), $(181, 181)$ (dashed curves), 
and $(361,721)$ (dotted-dashed curve). 
For model (C) with $K^{-1/2} = 5.92$ and 
model (E) with $K^{-1/2}=3.87$, 
the collapse is halted during the evolution, 
resulting in an oscillating, compact star. On the other hand, 
for model (C) with $K^{-1/2} = 6.71$ and for model (E) with 
$K^{-1/2} = 3.87$, the stars collapse to 
black holes in the dynamical time scale. 
As Figs.~2 and 3 show, these results depend only weakly on 
the resolution, implying that convergence is achieved fairly well. 

In Figs.~4--7, we show snapshots of the contour lines 
and velocity fields in the $x$-$z$ planes at selected time 
steps for model (C) with $K^{-1/2} = 5.92$ (Fig.~4) 
and 6.71 (Fig.~5),  and for 
model (E) with $K^{-1/2} = 3.74$ (Fig.~6) and $3.87$ (Fig.~7). 
All the results presented here were 
obtained with $(361, 361)$ grid resolution. 
For model (C) with larger initial decrease of the pressure, 
the rotating star collapses 
to a highly flattened object in the early stage. When the density 
increases sufficiently, the central region stops contracting, 
to form a core. On the other hand, the outer region, which 
forms a thin disk, is still collapsing. 
If the decrease factor of the pressure is not very large, 
the core subsequently expands, sweeping the surrounding disk, and 
finally an oscillating, oblate star is formed. 
On the other hand, for a sufficiently large decrease factor 
({i.e.}, $K^{-1/2} \agt 6.5 $), 
the core cannot expand, due to the 
insufficient pressure, resulting in a black hole after a fraction of 
matter accretion. 
In this case, the disk of a small fraction of mass 
appears to be formed around the black hole 
at $r \simeq 3$--$10M_g$ (see also Fig.~9 (a) and discussion below). 

We note that even for model (C), the product is not a flattened 
object, but spheroidal one, if $K^{-1/2} \alt 4$. In such cases, 
the core radius is larger, and the evolution process is similar to that 
described by Fig.~6. The formation of a highly flattened object is a 
feature in the collapse for large $q$ [i.e., $q=O(1)$] 
and for near critical 
values of $K^{-1/2}$ (i.e., $K^{-1/2} \agt 6$), in other words, 
for the highly supramassive case 
in the sense $\bar M_* \gg \bar M_{* {\rm smax}}$.

For model (E), in which $q$ is not very large, on the other hand, 
the collapse proceeds in a nearly spherical manner with 
tiny disks. When the density sufficiently increases, a nearly 
spherical core is formed. Since the decrease factor 
is smaller than that for model (C), the radius of the core is 
larger. For $K^{-1/2} = 3.74$, the core stops 
contracting, mainly due to the pressure force, subsequently 
forming an oscillating, nearly spherical star. 
For a larger decrease factor, the collapse cannot be halted, 
and a black hole is finally formed, swallowing almost all the fluid 
elements. 

The noteworthing difference between the evolution for 
models (C) and (E) for near critical values of $K$ is with regard to 
the rest mass of the core in the early stage 
of its formation. 
For (C), the rest mass is not very large, because the fraction of 
the rest mass in the surrounding disk is fairly large (see 
the third panel of Fig.~4). 
This implies that the core does not have to support all of the 
rest mass of the system, and consequently 
the internal energy to support the self-gravity can be small. 
On the other hand, for (E), almost all the fluid elements accrete 
to the core simultaneously (see the third panel of Fig.~6), 
and the fraction of the rest mass which the core 
has to support is larger. 
This appears to be part of reason why the critical value of $K^{-1/2}$ 
for model (C) can be larger than that for model (E).

We should note that the product shown in Fig.~4 
can have a large $T/W$, because 
according to a simple scaling law assuming the 
conservation of the angular momentum and the mass, 
we can expect it to roughly 
increase as $\sim (T/W)_i(R_i/R)$, where $(T/W)_i$ and $R_i$ are 
$T/W$ and $R$ at $t=0$. 
Since $T/W \sim 0.06$ at $t=0$ (see Table I), 
$T/W$ in a formed oscillating star would be $\sim 0.2$--$0.3$, 
and non-axisymmetric instabilities should be taken into account. 
A formed star is evidently {\it secularly}  
unstable with respect to gravitational wave emission to form a bar.\cite{SF} 
However, the secular time scale is 
much longer than the dynamical time scale, $\sim \rho_c^{-1/2}$, 
so that this would not be relevant even in a non-axisymmetric 
simulation performed only for $\sim 3\rho_c^{-1/2}$. 
A star also might be {\it dynamically} unstable 
to form a bar or spiral arms.\cite{diff}    Using the results 
found in a previous paper,\cite{diff}  we can roughly examine 
the dynamical stability. From Fig.~2, the normalized central 
density $\bar \rho(r=0) \equiv K\rho(r=0)$ is found to be 
between $0.05$ and $0.15$, while the 
normalized rest mass $\bar M_*$ is about $0.3$. 
In Ref.~\citen{diff}, we have found that  if $\bar \rho \alt 0.1$, 
a star with $\bar M_* \agt 0.3$ and $\Gamma=2$ is dynamically unstable 
({cf.}, Fig.~2 of Ref.~\citen{diff}). Thus, such a star is located 
near the threshold of the dynamical stability. 
Therefore, the process of the collapse might change if 
non-axisymmetric simulations were carried out.

In Fig.~8, we display the angular velocity $\Omega$ 
as a function of $\varpi$ in the equatorial plane 
at selected times for model (C) with $K^{-1/2}=5.92$. 
It is found that the star is differentially rotating, although 
it was rigidly rotating at $t=0$. 
Since the formed star is not stationary but in an oscillating 
state, $\Omega$ varies with time and becomes a steeper function 
when the central density is large. 
As shown in a previous paper,\cite{BSS}   
the effect of differential rotation can significantly increase 
the maximum allowed rest mass much beyond $\bar M_{* {\rm smax}}$. 
This seems to be the reason why this highly supramassive object 
of rest mass much larger than $\bar M_{* \rm smax}$ did not 
collapse to a black hole in this simulation. 

We, however, note that the differential rotation will be changed to 
rigid rotation due to a magnetic field or viscosity,  
or that the angular momentum 
will be dissipated by the magnetic dipole radiation 
for a realistic neutron star formed after the collapse.\cite{BSS}   
Thus, even if they are formed, 
any real supramassive stars of differential rotation 
will eventually collapse to black holes 
due to dissipation processes in the secular time scale.

In Figs.~9 (a) and (b), we plot the fraction of the rest mass inside 
a coordinate radius $r$, $M_*(r)/M_*$, as a function of time 
(a) for model (C) with $K^{-1/2}= 6.71$ and  
(b) for model (E) with $K^{-1/2}= 3.87$. 
In both cases, black holes are formed on the dynamical time scale. 
Here, $M_*(r)$ is defined as
\beq
M_*(r)=4\pi \int_{r'<r} x' dx'  dz' \rho_*. 
\eeq
The dotted, solid and dashed curves denote the results 
for $(271, 271)$, $(361, 361)$ and $(361, 721)$ grid resolutions, 
respectively. 
As expected from Fig.~7, almost all the fluid elements are
swallowed inside the apparent horizon for model (E). 
On the other hand, a small fraction of mass could form a disk 
around a black hole for model (C).
Since the simulation crashed soon after the formation of 
the apparent horizon at $t\sim 1.15\rho^{-1/2}$, 
it is not possible to draw a definite 
conclusion from Fig.~9(a). However, Fig.~9(a) appears to suggest 
that $\sim 5\%$ of the total rest mass might form a disk around 
a black hole for $r \sim 3$--$10 M_g$.\cite{foot3}   
The reason for these results can be explained from 
the specific angular momentum distribution at $t=0$ as follows. 

In Fig.~10, we show $M_*(j)/M_*$ as a function of $j/M_g$ for 
models (C) and (E) (note that $M_g$ here denotes the initial value). 
For (C), we show it at selected time steps 
to demonstrate that the specific angular momentum 
spectrum $M_*(j)$ is preserved during the evolution fairly accurately. 
In both models, the specific angular momentum 
$j$ is a monotonically increasing function of $\varpi$, and 
at the equator, it is 
$\sim 3.6M_g$ for model (C) and $\sim 1.8 M_g$ for model (E). 
Taking into account that 
$J$ is a conserved quantity, $M_g$ is almost conserved, and 
the disk mass occupies only a small fraction of the total mass, we 
can expect that 
$q$ of the formed black hole should be approximately equal to the 
initial value. Thus, the specific angular momentum at 
the innermost stable circular orbit in the 
equatorial plane around the formed Kerr black holes, 
which is the minimum allowed value for a test particle orbiting 
a Kerr black hole, 
is $\sim 2.1 M_g$ for model (C) and $\sim 2.9M_g$ for (E).\cite{ST}   
It is immediately found that the specific angular momentum of no fluid 
elements is large enough to form a disk around the black hole 
for model (E). For model (C), a fraction of 
fluid elements of $j$ between 
$\sim 2.1M_g$ and $3.6M_g$ can form a disk, but Fig.~10 shows that 
the mass fraction is only $\sim 6\%$, which is 
in approximate agreement with the fraction of 
the disk mass found in the numerical simulation. 

\subsection{Adiabatic collapse for differentially rotating initial data}

Simulations of adiabatic collapses using polytropic equations of state 
and differentially rotating initial data sets 
have been performed for models (H)--(K) described in Table II and 
for models (L)--(O) in Table III. 
For each set of initial conditions, 
the simulations were carried out changing the 
decrease factor of the pressure for a wide range. 
In Figs.~11 and 12, we summarize the products formed after collapse 
for $A=\varpi_e/3$ and $A=\varpi_e$ 
in the same manner as for Fig.~1. Note that the dashed curves 
denote the approximate threshold for black hole formation 
({i.e.}, below the dashed curves, a stable star is formed) 
for rigidly rotating initial data sets. 
It is found that the threshold for the 
differentially rotating initial conditions 
almost coincides with that for the rigidly rotating ones. 
Therefore, for the moderate range of rotational profiles 
we have investigated ($\varpi_e/3 \leq A \leq \infty$), 
the criterion of black hole formation depends only weakly on 
the initial rotational profile.  

On the other hand, the products after the collapse 
depend strongly on the initial rotational profile, in particular, 
for $q\simeq 0.9$: 
For rigidly rotating initial data, the star remains 
a spheroid in the collapse even if $q \simeq 0.9$. On the other hand, 
it becomes a toroid for differentially rotating initial data 
with $A=\varpi_e/3$. 
To illustrate this fact, we display snapshots of the contour lines 
and velocity fields in the $x$-$z$ planes at selected time 
steps for model (H) with $K^{-1/2}= 6.32$ in Fig.~13.  
In this case,  the star does not form a black hole, but a toroid. 
As in the case of rigidly rotating initial data, 
fluid elements which are initially located far from $\varpi=0$ 
collapse toward the central region in the early phase. 
On the other hand,  those near the symmetric axis do not simply collapse 
toward the center, but slightly expand in the $\varpi$ direction. 
This seems to be due to the effect of strong centrifugal force 
near the $z$ axis (note that $\Omega$ at $\varpi=0$ is initially 
about 4 times 
larger than that for the rigidly rotating cases of identical $q$). 
As a result of these types of motion, 
the collapsed object forms a ring-shape 
core in the early stage (cf., the third panel). 
The ring subsequently collapses toward 
the center, to become a very compact toroid (cf., the fifth and 
sixth panels). However, the centrifugal force 
and/or pressure hang up the collapse, and subsequently the ring 
expands with the formation of a jet along the symmetric axis 
(cf., the seventh panel). 
Once it reaches a maximum expansion at $t \sim 1.5\rho_{\rm max}^{-1/2}$, 
the ring begins to contract again to form 
a compact toroid (cf., the eighth panel), 
and this oscillating motion is repeated. 

Here, we note that the maximum normalized density of the ring 
is very small $\bar \rho_{\rm max} \sim 0.03$ at 
$t \sim 1.5\rho_{\rm max}^{-1/2}$, 
while $\bar M_*$ is very large, $\sim 3.1$. 
This implies that the ring would be dynamically unstable 
with respect to formation of a bar and/or 
spiral arms, or with respect to 
fission in a non-axisymmetric simulation.\cite{diff} 
This result is different from that in the case of 
rigidly rotating initial data, in which a collapsed object 
seems to be marginally stable, and 
it suggests that a collapsed object of 
differentially rotating initial data could be more susceptible 
to a non-axisymmetric, dynamical instability than that of 
rigidly rotating initial data, even if $q$ is identical initially. 

For model (H) with $K^{-1/2}>7$, 
black holes are formed. 
For these cases, the evolution processes until 
$t \sim \rho_{\rm max}^{-1/2}$ is essentially the same as 
that shown in Fig.~13, but the star subsequently collapses, 
without bounce. 
In this black hole formation case, almost all the fluid elements 
are swallowed inside the black holes quickly. This can be understood 
from the sixth panel of Fig.~13. 
This result is in contrast to that in the black hole formation 
for model (C).  The reason is as follows. In the case of a 
steep differentially rotational profile at $t=0$, 
the specific angular momentum near the equator is much 
smaller than that for rigidly rotating cases. 
For model (H), the specific angular momentum at the equator is 
$0.94M_g$, which is smaller than the value at the innermost 
stable circular orbit of a Kerr black hole with $q=0.9$ and mass $M_g$. 
Consequently, it is unlikely for a disk to be formed. 
We argue this point in \S 5 in detail. 

In Fig.~14, we show snapshots of the contour lines 
and velocity fields in the $x$-$z$ plane at selected time 
steps for model (J) with $K^{-1/2}= 3.74$, 
in which a black hole is not formed.  
Since $q$ is not very large $(\sim 0.48)$ in this case, 
a ring is not formed throughout the entire evolution. However, 
the effect of the rotation appears to be significant.  
Compared with the core formed in the simulation of model (E), 
in which $q$ is nearly equal to that for model (J), 
the shape of the core is much more deformed, due to the 
effect of the differential rotation initially given 
(compare the second panel with the second panel in Fig.~6). 
Also, a violent non-spherical oscillation is found after 
formation of the new star. These results imply that 
even if $q$ and $M_*$ are identical initially, 
the shape and motion of the 
products after collapse depend strongly on 
the initial rotational profiles.

\subsection{Non-adiabatic case}

Simulations of non-adiabatic collapses using 
a $\Gamma$-law equation of state have been performed for 
models (C)--(E) and (H)--(J) described in Tables I and II.  
In Figs.~15 and 16, we summarize the products after collapses 
for (C)--(E) and (H)--(J) 
in the same manner as for Fig.~1. The dashed curve 
denotes the approximate threshold for black hole formation 
in the adiabatic collapses. 
Obviously, the shock heating raises the criterion for $K^{-1/2}$ 
for prompt formation of black holes. 
It is found that the effect is in particular significant 
for larger $q$. 
We note that for model (C), we cannot judge the nature of the product 
for $K^{-1/2} \geq 10$, because a very thin disk 
is formed in the equatorial plane 
and cannot be well resolved even with $(501, 501)$
grid resolution. 
For model (H), formation of black holes is not found 
for $K^{-1/2} \leq 10$. 
For larger $K^{-1/2}$ (i.e., $K^{-1/2} \gg 10$), 
black holes would be formed for both models, 
but to find this formation, simulations with finer resolution 
are necessary to resolve the thin disk and/or thin ring. 

Even in the presence of the shock heating, the 
formation process of black holes is not modified drastically. 
In Fig.~17, we show snapshots of the contour lines 
and velocity fields in the $x$-$z$ planes at selected time 
steps for model (E) with $K^{-1/2} = 5.00$ as an example. 
Although shocks are generated at the formation of the central core 
(cf., the second panel), 
they stall due to the subsequent accretion of matter, 
and hence the collapse is never halted, 
resulting in a black hole on the dynamical time scale.

On the other hand, the process in the case of no black hole formation 
is considerably affected by shocks, in particular, for $q \sim 0.9$. 
In Figs.~18--21, we show snapshots of the contour lines 
and velocity fields in the $x$-$z$ planes at selected time 
steps for model (C) with $K^{-1/2} = 8.37$,  
for model (E) with $K^{-1/2} = 4.47$, for model (H) with $K^{-1/2}=9.49$, 
and for model (J) with $K^{-1/2}=4.00$. 

In the case of small $q$, spheroidal shocks are formed 
around the surface of the central core (cf., the first panel 
in Figs.~19 and 21). 
This implies that the shock heating is effective only around the 
spheroidal shell. Indeed, $K'(x^{\mu})$ around the center of the core 
remains nearly constant throughout the evolution. 
Therefore, the shock heating is not very effective for increase of 
the internal energy in the high density region. Instead, 
the energy generated by shocks at the shell is effectively used to 
provide the energy to the surrounding matter, which subsequently 
expands outward. 

On the other hand, the formation mechanism of shocks 
for $q \simeq 0.9$ and $K^{-1/2} M_* \gg \bar M_{*\rm smax}$ 
is completely different from that for 
smaller $q$ (see Figs.~18 and 20): In the very early 
phase, the shocks around the small spheroidal core (for model (C)) or 
disk-plus-ring (for model (H)) are formed. 
Here, the shape of the shocks depends strongly on the rotational profile 
initially given. The reason why the mass fraction around 
the shock forming, high density region is small is that 
a large fraction of mass, which is initially located at large $\varpi$, 
has not yet accreted, due to the centrifugal force, and 
the collapse proceeds mainly in the $z$ direction. 
Since the mass of the high density region is small, 
prompt formation of a black hole does not occur in spite of 
the small internal energy. 

The shocks formed in the early phase 
subsequently expand in the $z$ direction for model (C) and 
toward the $z$ axis for model (H), 
because a large amount of matter still accretes near 
the equatorial plane and prevents 
the shocks from expanding outward. The mass fraction near the symmetric 
axis is not very large, so that this shock expansion is not efficiently 
used for sweeping the accreting matter around the cores. 
The evolution processes after a substantial fraction of 
matter residing in the outer part 
has accreted to the small cores are different for the two cases. 
For model (C), the shocks formed in the interface between the 
small core and the surrounding disk initiate expansion that sweeps 
the matter (cf., the fourth and fifth panels in Fig.~18). 
However, the expansion is not very powerful 
because a part of the internal energy has already been used in the first 
expansion in the $z$ direction and these second shocks are not very strong. 
For model (H), the evolution process is more complicated. 
During the accretion of the matter from the outer part to the 
core, the toroidal core starts contracting toward 
the $z$ axis without significant shock formation (cf., the 
third panel in Fig.~20);  
i.e., in this case, shocks which sweep accreting matter are 
not formed. 
Then, the inner surface of the toroid collides at the $z$ axis 
(cf., the fourth panel), and 
as a result, a shock is formed along the $z$ axis. 
This shock subsequently expands 
mainly in the $\varpi$ direction (cf., the fifth panel), 
contributing to an increase of the internal energy for 
a large fraction of mass elements in the toroid. 
The toroid oscillates nearly periodically, so that 
this process is repeated (cf., the sixth--ninth panels). 
For both models (C) and (H), several different types of shocks are formed 
in different stages incoherently. Consequently, 
the motion of the shocks is less coherent and less 
efficient in sweeping the accreting matter than 
for the small $q$ case. This result appears to represent 
an extreme example in which rapid rotation prevents the formation of 
coherent shocks.\cite{YS}   

The above described mechanism of shock formation also explains the 
reason that the threshold value of $K^{-1/2}$ is much 
larger than that in the absence of shocks only for the large $q$ case. 
For small $q$, 
shocks are generated near the surface of a central dense core, 
so that the entropy can be increased only 
in the surface region of low density. 
Consequently, the shock heating is not very effective 
for increasing the internal energy of the central core. 
On the other hand, for large $q$ with large pressure 
decrease ({i.e.}, $K^{-1/2}M_* \gg \bar M_{*\rm smax}$), 
shocks can be generated near high density regions, {i.e.}, 
around the small cores and the disks. 
Indeed, $K'$ is increased by a large factor for most fluid elements. 
This is due to the highly non-spherical motion 
induced by the large centrifugal force, as mentioned above. 
As a result, a large amount 
of the matter is affected by the shock heating, resulting 
in a large increase of the total internal energy. 
Thus, although rapid rotation prevents the formation of coherent 
shocks, it could contribute to a shock heating that is efficient 
for preventing prompt collapse to black holes.

\section{Summary}

We constructed a new numerical code for axisymmetric hydrodynamic 
simulations in full general relativity. In this code, 
the Einstein field equation is solved using Cartesian coordinates 
with appropriate boundary conditions, 
while the hydrodynamic equations are solved in cylindrical 
coordinates. It was shown from numerical experiments 
that this method is sufficiently robust to perform stable and well-resolved 
computations for rotating stellar collapse. 

Using this code, we have investigated the criterion for 
black hole formation after rotating stellar collapses. 
It is found that the effect of rotation could 
play an important role in preventing prompt formation of black holes. 
For the case in which the angular momentum parameter $q$ is smaller than 
$\sim 0.5$, a black hole is formed if the total rest mass is slightly 
larger than the maximum allowed rest mass of spherical stars 
($\bar M_{* {\rm smax}}$), 
implying that the effect of rotation is not very important. 
On the other hand, the threshold of the total rest mass 
can be significantly increased by the 
effect of rotation for $q \alt 1$: {e.g.}, for 
$q\sim 0.9$, the threshold value is about $70$--$80\%$ larger 
than $\bar M_{* {\rm smax}}$. The self-gravity 
of such large mass products 
is supported by rapid, differential rotation. 
These results 
are found to depend only weakly on the initial rotational profile 
within the range that we have investigated. 

We have found that the formation mechanism of shocks 
depends strongly on the rotational parameter $q$. 
For collapse of slowly rotating stars of $q < 0.5$, 
shocks are formed at the spheroidal shell around a 
high density core. As a result, shock heating is effective 
only around the shell and is not very helpful for increasing the 
internal energy of the central core. 
On the other hand, shocks can be formed in a 
highly non-spherical manner and near the high density region 
in the collapse of rapidly rotating, supramassive stars for which 
$q=O(1)$ and $K^{-1/2}M_* \gg \bar M_{*\rm smax}$. 
Since the centrifugal force is large in this case, 
a large amount of matter at large $\varpi$ cannot 
accrete immediately, and 
the high density regions formed in the early stage could 
have highly non-spherical shapes. 
As a result, shocks can be formed in a highly non-spherical manner 
around the high density region. 
(The shapes of the shocks depend strongly on 
the initial rotational profile.) 
Because of the matter accretion around the equatorial plane, 
shocks cannot expand coherently, and hence they cannot 
sweep the accreting matter effectively. 
However, shocks can contribute to the heating for a large number 
of the mass elements, because they are formed near the high density 
regions. Consequently, they prevent prompt 
collapse to black holes even for highly supramassive stars 
with $K^{-1/2}M_* \gg \bar M_{*\rm smax}$. 

The numerical results reported in this paper indicate that 
the disk mass around formed black holes is not very large 
even for $q \simeq 0.9$. 
In a different setting, with different equations of state, 
different initial radii, and different initial rotational profiles, 
results could be modified. 
It is an interesting issue to clarify what kind of initial 
conditions are preferable for obtaining a system 
of a black hole surrounded by massive disks, because 
the formation of disks around a black hole \cite{hyper} 
could play an important role in nature. 
Here, we carry out a simple analysis to 
address that equations of state and initial velocity 
profiles of a star before collapse 
play a crucial role in determining the disk mass.

In the following, we assume that (1) rotating stars at $t=0$ 
are not very compact, (2) black holes are formed only for $q <1$, 
and (3) the fraction of the disk mass is 
much smaller than the total mass and that the value of $q$ of 
a formed black hole is nearly equal to the initial value. 
Due to the assumption (1), the angular momentum can be approximately 
computed from initial data as 
\beq
J \simeq \int dV \rho_* \varpi^2 \Omega, 
\eeq
where we use $h \simeq 1$ and $j \simeq \varpi^2 \Omega$. 
For simplicity, here, we consider $\Omega$ to be of the same form as 
in Eq.~(\ref{diffomega}). 
For the case that $A \ll \varpi_e$, 
$J \simeq M_* A^2\Omega_0 \simeq M_* \varpi_e^2\Omega_e$, where 
$\Omega_e$ is the angular velocity at $\varpi=\varpi_e$. 
Then, $q$ can be approximately written as $\varpi_e^2 \Omega_e/M_g$, 
where we use $M_g \simeq M_*$, due to the assumption (1). 
For the case $A \gg \varpi_e$, on the other hand, 
the star is in a state of nearly rigid rotation ($\Omega_0=\Omega_e$), and 
$J = M_* \langle \varpi^2 \rangle \Omega_e$, where 
\beq
\langle \varpi^2 \rangle \equiv {1 \over M_*} \int \rho_* \varpi^2 dV. 
\eeq
Consequently, $q \simeq \langle \varpi^2 \rangle \Omega_e/M_g$. 
Then, the specific angular momentum $j$ can be 
approximately written as 
\beq
j(\varpi) \simeq \varpi^2 \Omega_e 
\simeq \left\{
\begin{array}{ll}
\displaystyle 
M_g q & {\rm for}~A \ll \varpi_e ,\\
M_g q \varpi^2/\langle \varpi^2 \rangle
& {\rm for}~A \gg \varpi_e. 
\end{array}
\right.\label{type2}
\eeq
Thus, $j(\varpi)/M_g=q$ for $A \ll \varpi_e$, and 
$j(\varpi)/M_g \leq q\varpi_e^2/\langle \varpi^2 \rangle$ for 
$A \gg \varpi_e$. 
To form a disk around a black hole, $j/M_g$ has to be larger than 
$\ell_{\rm ISCO}$, where $\ell_{\rm ISCO}$ denotes 
$j/M_g$ at the innermost 
circular orbit around a black hole which 
depends on $q$. ($\ell_{\rm ISCO}=1$ for $q=1$, $\simeq 2.1$ 
for $q=0.9$, and $\simeq 2\sqrt{3}$ for $q \ll 1$.) 
This implies that for $A \ll \varpi_e$, a disk cannot 
be formed, irrespective of the value of $q$ for 
$q < 1$. For a star which was 
initially in rigid rotation, 
a disk can be formed if $q \varpi_e^2/\langle \varpi^2 \rangle =O(1)$. 
For equilibrium stars with stiff equations of state, 
the density profile is not very centrally peaked, so that 
$\varpi_e^2/\langle \varpi^2 \rangle$ is of $O(1)$. This implies  
that $q$ has to be of $O(1)$ for disk formation. 
We presented such an example in \S 4.1. 
On the other hand, $q$ may be smaller for equilibrium stars 
of soft equations of state for which 
the density profile is centrally peaked, and 
$\varpi_e^2/\langle \varpi^2 \rangle$ can be much larger than unity. 
We note that $\varpi_e^2/\langle \varpi^2 \rangle$ 
depends weakly on $\varpi_e$ for non-relativistic stars of an 
identical equation of state. Thus, this conclusion is almost 
independent of the initial stellar radius. 
From these arguments, we can recognize that to obtain disks 
around a black hole, irrespective of the initial stellar radius,
(i) stars before collapse should not have a steep differential 
rotational profile, but rather a nearly rigidly rotational profile, 
(ii) the value of $q$ of stars before collapse should be of $O(1)$ 
if the equation of state is stiff, or stars before 
collapse should have a soft equation of state if $q \ll 1$. 
In realistic situations, a progenitor of a supernova just before 
collapse has a soft equation of state with $\Gamma \simeq 1.3$. 
Thus, a disk would be formed as long as the initial rotational 
profile is not too steep. 

Although these conclusions would be qualitative correct, 
nothing quantitative can be clarified at this time. In particular, we 
assumed that the disk mass is much smaller than the black hole mass 
in this analysis, but this might be incorrect for the case of 
soft equations of state. This is because the star is centrally 
condensed in this case, and it may be the case that 
only the central region collapses 
to form a black hole of a small mass fraction. Thus, 
to compute the disk mass in rotating stellar collapse correctly, 
numerical simulations are obviously necessary. 

In this paper, we have studied rotating stellar collapse 
using simple equations of state and simple initial conditions 
which are not very realistic. We believe that the significant 
effects of rotation found here are qualitatively correct, 
but they might not quantitatively realistic. 
To obtain a result applicable to 
realistic systems, we have to perform simulations 
using more realistic equations of state and initial conditions. 
We plan to perform such simulations in the future.

\vskip 5mm
\begin{center}
{\bf Acknowledgements}
\end{center}
\vskip 5mm

The author thanks Thomas Baumgarte and Stu Shapiro for discussion. 
Numerical computations were performed on the FACOM VX/4R
machines in the data processing center of NAOJ. 
The author gratefully acknowledges support by 
JSPS (Fellowships for Research Abroad) and the 
hospitality of the Department 
of Physics, University of Illinois at Urbana-Champaign.

\clearpage

\vskip 5mm
\noindent 
\begin{table}
\caption{Central density $\rho_c$, 
gravitational mass $M_g$, $q \equiv J/M_g^2$, 
$T/W$, $R/M_g$ (where $R$ is the circumferential radius at the 
equator), angular velocity $\Omega$, 
and ratio of polar coordinate radius to equatorial radius 
($R_{\rm ratio}$) for rigidly rotating stars. 
$M_*$ is fixed to 0.05, and the polytropic constant $K$ is fixed 
to $1$. }
\end{table}

\vskip 5mm
\noindent
\begin{center}
\begin{tabular}{|c|c|c|c|c|c|c|c|} \hline\hline
\hspace{3mm} $\rho_{c}$ \hspace{3mm} &
\hspace{2mm} $M_g$ \hspace{2mm} &
\hspace{1mm} $q$ \hspace{1mm} & 
\hspace{1mm} $T/W$  \hspace{1mm} &
\hspace{0mm} $R/M_g$    \hspace{0mm} & 
\hspace{1mm} $\Omega$   \hspace{1mm} &
\hspace{0mm} $R_{\rm ratio}$ \hspace{0mm} & 
\hspace{0mm} Model    \hspace{0mm} 
\\ \hline 
0.018750  & 0.04909 & 0.978 & 0.0715  & 30.04 & 0.0892 & 0.717 &(B) \\ \hline 
0.019316  & 0.04907 & 0.902 & 0.0628  & 29.07 & 0.0861 & 0.750 &(C) \\ \hline 
0.020757  & 0.04903 & 0.702 & 0.0411  & 27.02 & 0.0748 & 0.833 &(D) \\ \hline 
0.022179  & 0.04899 & 0.473 & 0.0200  & 25.40 & 0.0557 & 0.917 &(E) \\ \hline 
0.022868  & 0.04898 & 0.327 & 0.00986 & 24.70 & 0.0403 & 0.958 &(F) \\ \hline 
0.023408  & 0.04897 & 0.143 & 0.00194 & 24.19 & 0.0183 & 0.992 &(G) \\ \hline 
\end{tabular}
\end{center}

\vskip 1cm 

\noindent 
\begin{table} 
\caption{The same as Table I, but for differentially 
rotating stars with $A=\varpi_e/3$. The rest mass is also fixed to 
$0.05$. $\rho_{\rm max}$, instead of $\rho_c$, is given in the first 
column. In the sixth column, $\Omega_0$ and $\Omega_e$ are 
given.}
\end{table}

\vskip 5mm
\noindent
\begin{center}
\begin{tabular}{|c|c|c|c|c|c|c|c|} \hline\hline
\hspace{3mm} $\rho_{\rm max}$ \hspace{3mm} &
\hspace{2mm} $M_g$ \hspace{2mm} &
\hspace{1mm} $q$ \hspace{1mm} & 
\hspace{1mm} $T/W$  \hspace{1mm} &
\hspace{0mm} $R/M_g$    \hspace{0mm} & 
\hspace{1mm} $(\Omega_0,\Omega_e)$   \hspace{1mm} &
\hspace{0mm} $R_{\rm ratio}$ \hspace{0mm} & 
\hspace{0mm} Model    \hspace{0mm} 
\\ \hline 
0.015286  & 0.04910 & 0.900 & 0.0767 & 27.51 & (0.3668,0.0334) & 0.650 &(H) 
\\ \hline 
0.017432  & 0.04905 & 0.715 & 0.0528 & 26.31 & (0.3611,0.0329) & 0.750 &(I) 
\\ \hline 
0.020327  & 0.04901 & 0.487 & 0.0268 & 25.14 & (0.2470,0.0223) & 0.867 &(J) 
\\ \hline 
0.022776  & 0.04897 & 0.230 & 0.00640 & 24.32 & (0.1265,0.0114) & 0.967 &(K) 
\\ \hline 
\end{tabular}
\end{center}

\vskip 1cm 

\noindent 
\begin{table} 
\caption{The same as Table II, but for differentially 
rotating stars with $A=\varpi_e$. The rest mass is also fixed to 
$0.05$. }
\end{table}

\vskip 5mm
\noindent
\begin{center}
\begin{tabular}{|c|c|c|c|c|c|c|c|} \hline\hline
\hspace{3mm} $\rho_{\rm max}$ \hspace{3mm} &
\hspace{2mm} $M_g$ \hspace{2mm} &
\hspace{1mm} $q$ \hspace{1mm} & 
\hspace{1mm} $T/W$  \hspace{1mm} &
\hspace{0mm} $R/M_g$    \hspace{0mm} & 
\hspace{1mm} $(\Omega_0,\Omega_e)$   \hspace{1mm} &
\hspace{0mm} $R_{\rm ratio}$ \hspace{0mm} & 
\hspace{0mm} Model \hspace{0mm}   
\\ \hline 
0.018646  & 0.04907 & 0.893 & 0.0635 & 27.76 & (0.1239,0.0580) & 0.767 &(L) 
\\ \hline 
0.020153  & 0.04904 & 0.717 & 0.0438 & 26.50 & (0.1096,0.0512) & 0.833 &(M) 
\\ \hline 
0.021913  & 0.04900 & 0.478 & 0.0210 & 25.18 & (0.0813,0.0379) & 0.917 &(N) 
\\ \hline 
0.023066  & 0.04897 & 0.252 & 0.00610 & 24.38 & (0.0458,0.0213) & 0.975 &(O) 
\\ \hline 
\end{tabular}
\end{center}

\clearpage 
\begin{figure}[t]
\begin{center}
\epsfxsize=3.5in
\leavevmode
\epsffile{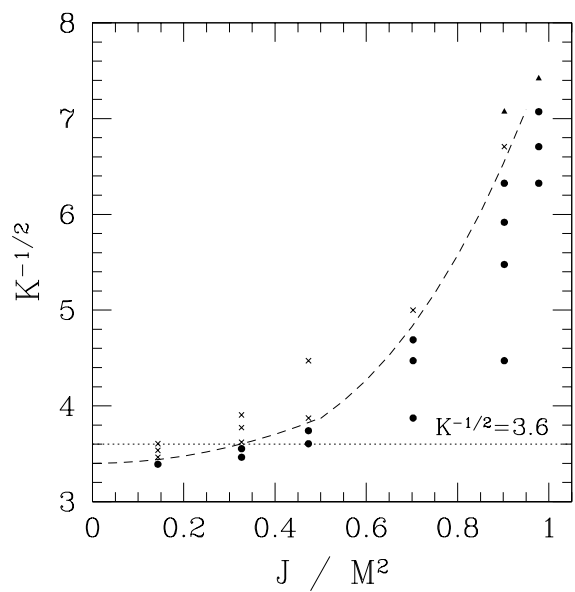}
\end{center}
\caption{Summary of the products in adiabatic collapse 
with rigidly rotating initial data. 
The horizontal and vertical axes denote $q=J/M_g^2$ and $K^{-1/2}$. 
The filled circles and crosses indicate that 
the products are a new compact star and 
a black hole, respectively. The filled triangles imply that 
we could not judge the nature of the products. 
The dotted line and dashed curves denote $K^{-1/2}=18/5$ and 
the approximate threshold for black hole formation, respectively. 
}
\end{figure}

\clearpage

\begin{figure}[t]
\begin{center}
\epsfxsize=3.5in
\leavevmode
\epsffile{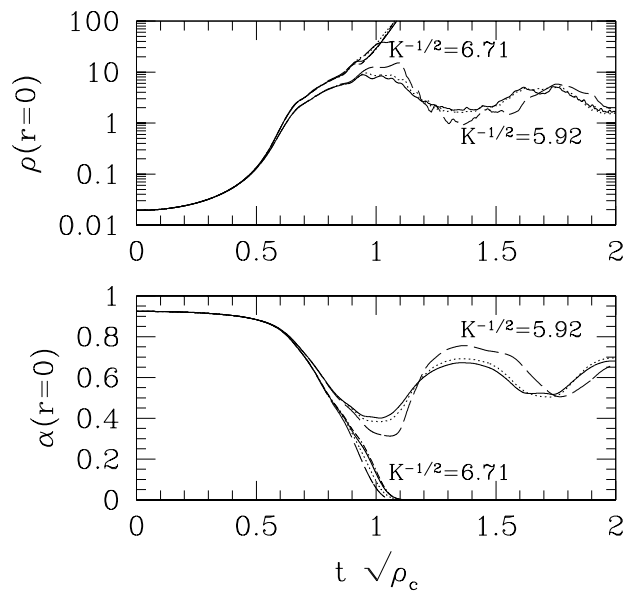}
\end{center}
\caption{Time evolution of $\rho$ and $\alpha$ at $r=0$ for 
model (C) with $K^{-1/2}=5.92$ and 6.71. 
The solid, dotted, dashed, and dotted-dashed curves represent the results 
for $(361, 361)$, $(271, 271)$, $(181, 181)$ and $(361, 721)$ 
grid resolutions, respectively. 
}
\end{figure}

\begin{figure}[t]
\begin{center}
\epsfxsize=3.5in
\leavevmode
\epsffile{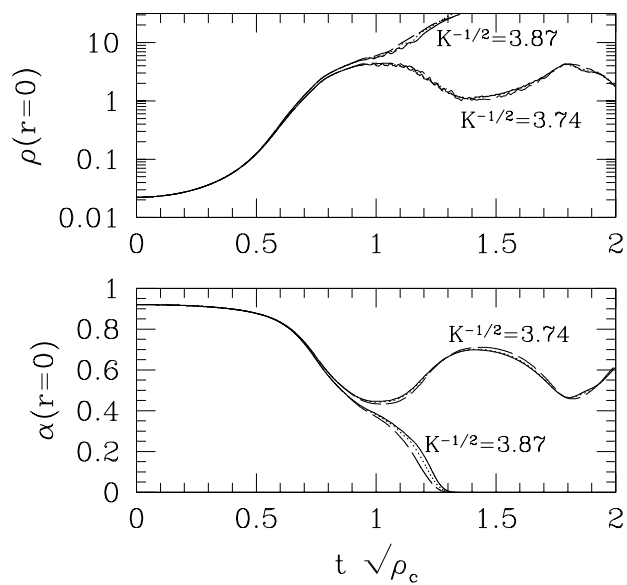}
\end{center}
\caption{The same as Fig.~2, but for 
model (E) with $K^{-1/2}=3.74$ and 3.87. 
}
\end{figure}

\clearpage
\begin{figure}[t]
\begin{center}
\epsfxsize=1.8in
\leavevmode
\epsffile{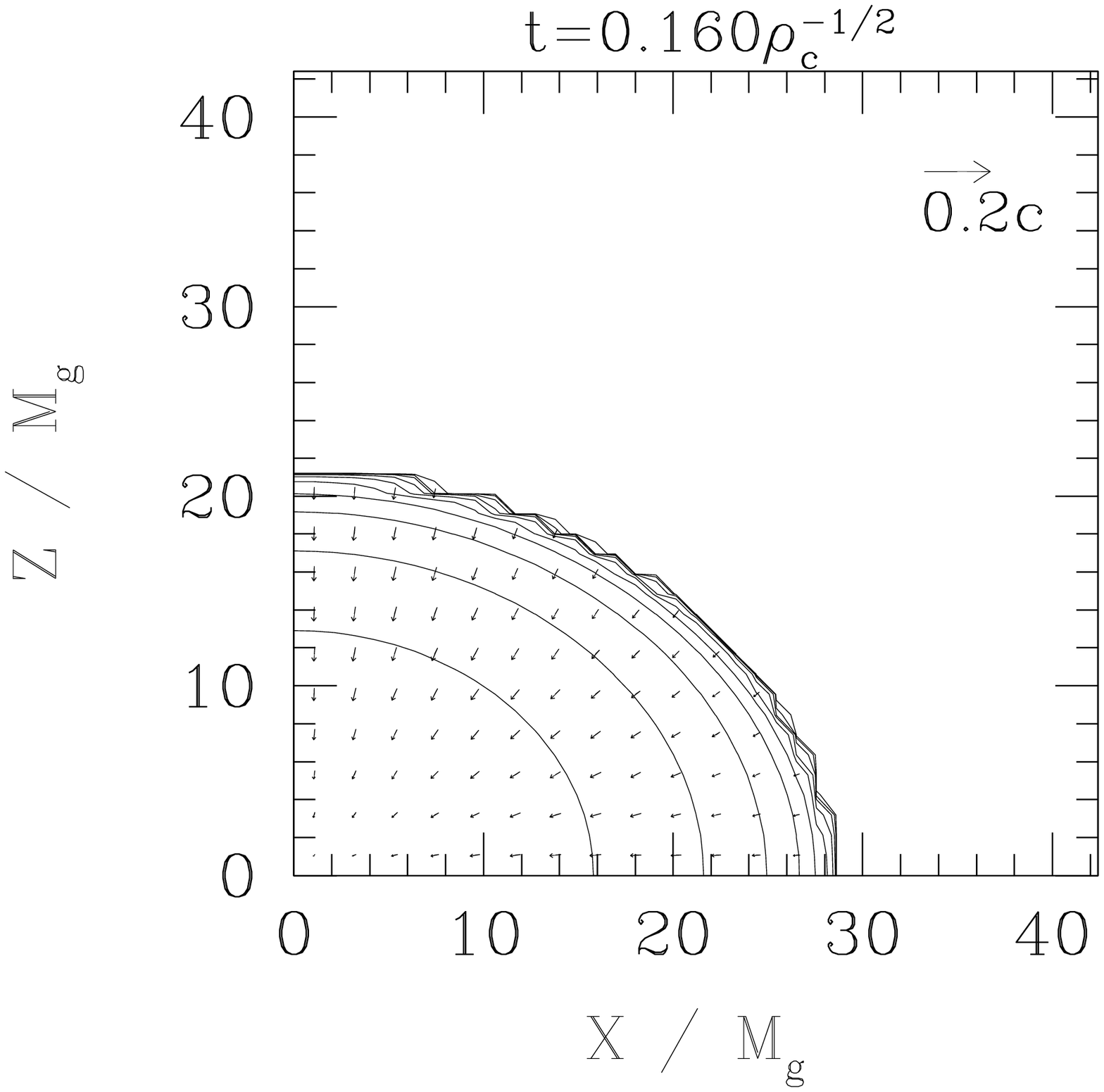}
\epsfxsize=1.8in
\leavevmode
\epsffile{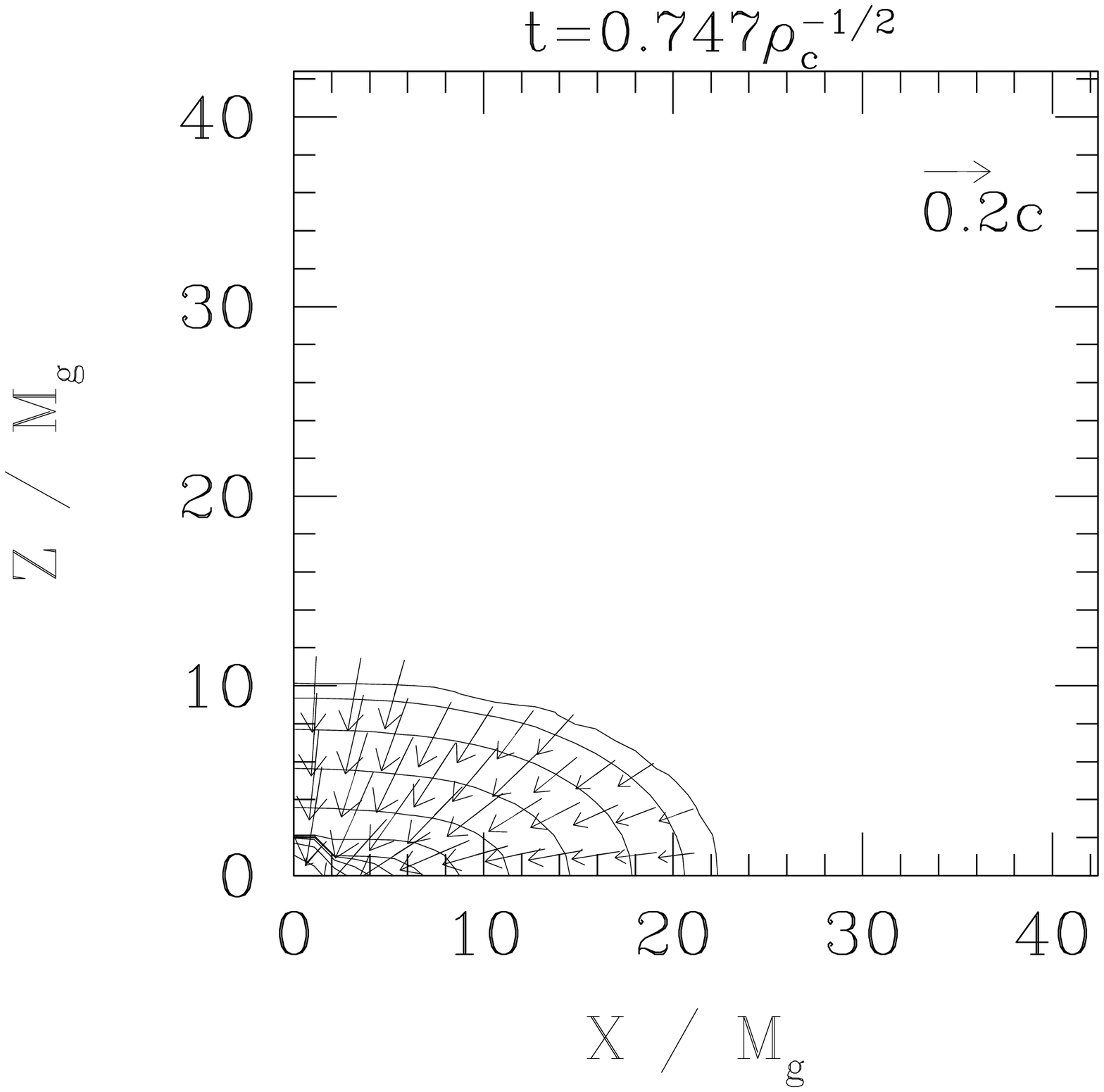}
\epsfxsize=1.8in
\leavevmode
\epsffile{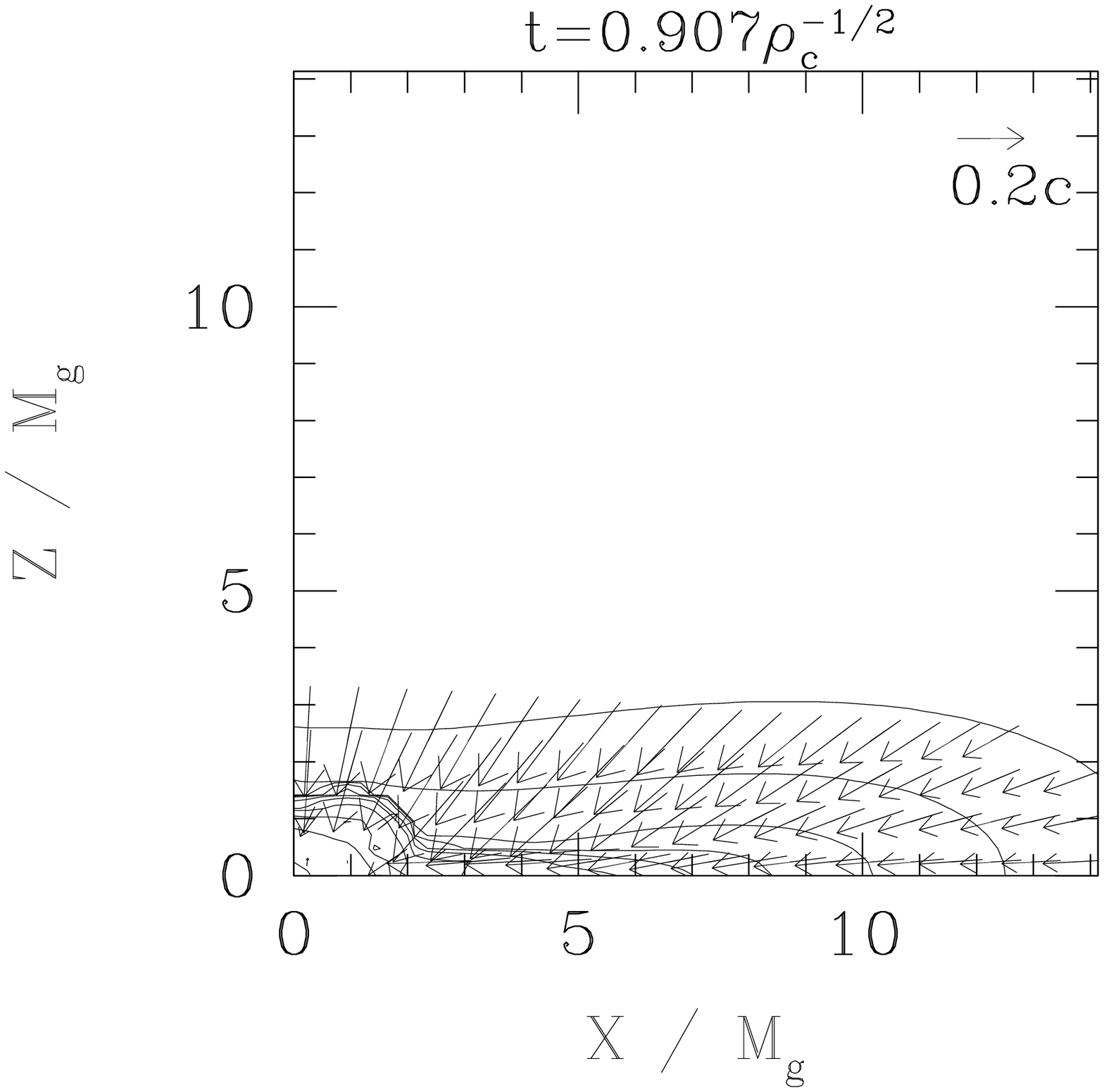}\\
\epsfxsize=1.8in
\leavevmode
\epsffile{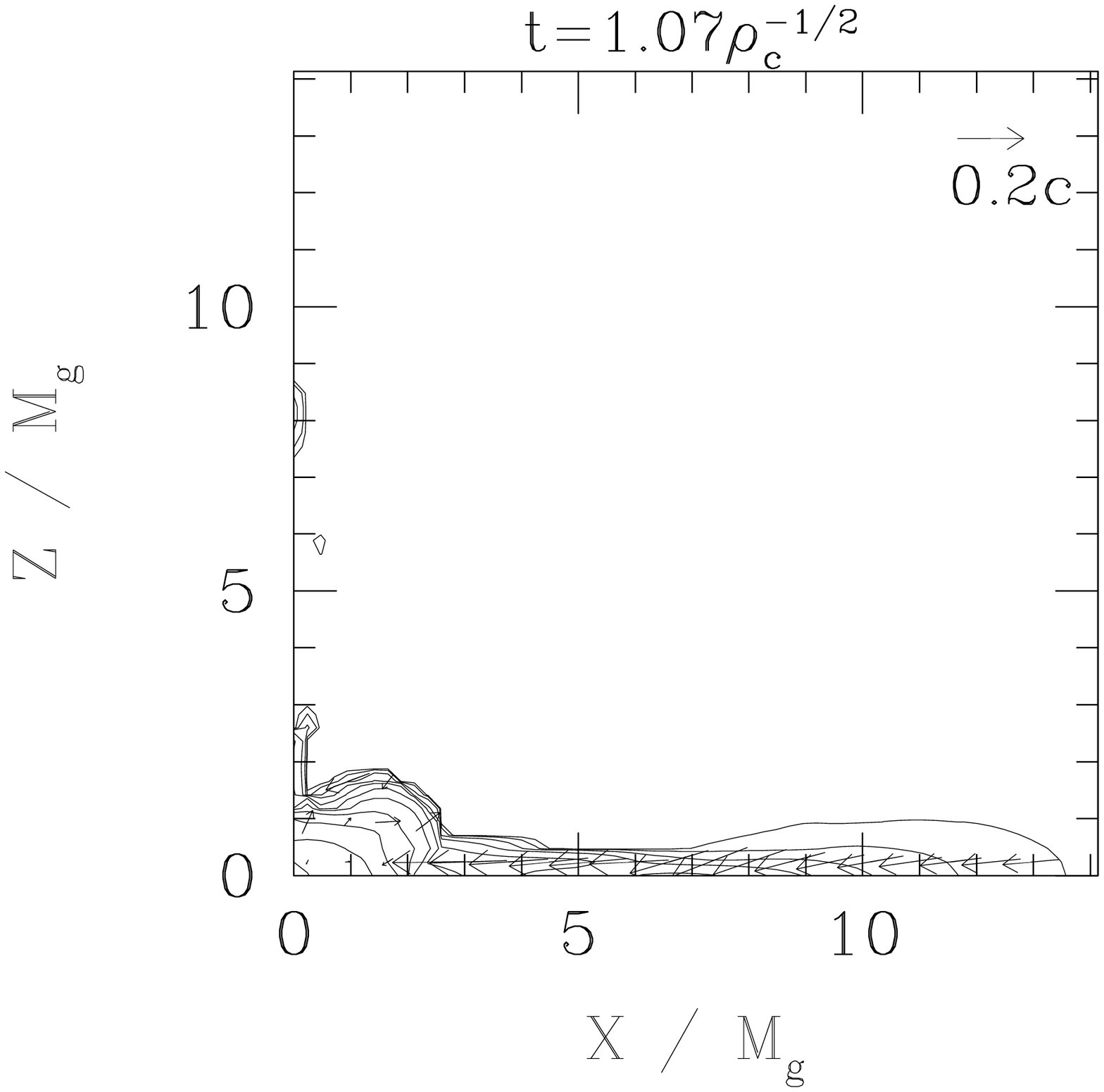}
\epsfxsize=1.8in
\leavevmode
\epsffile{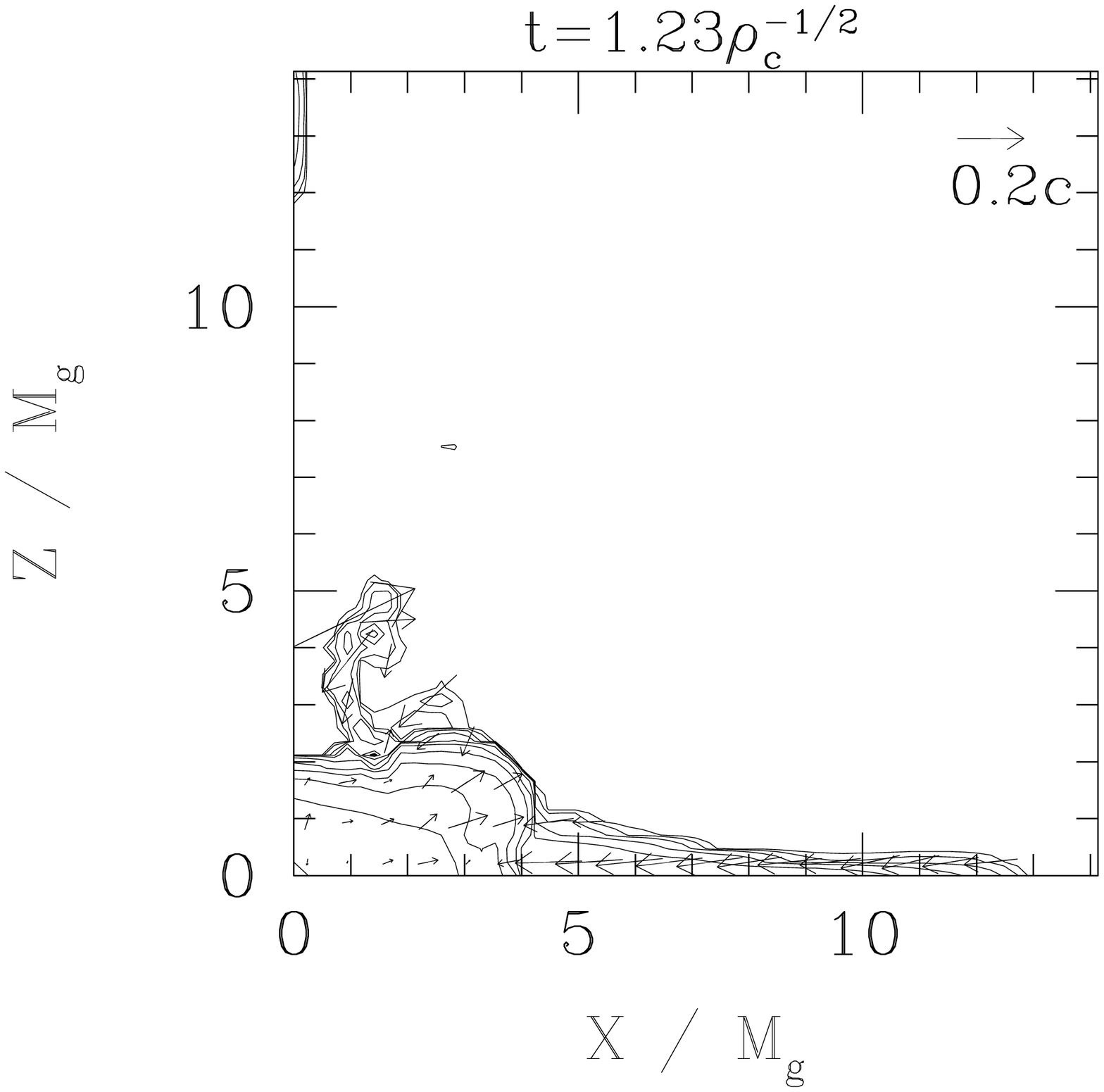}
\epsfxsize=1.8in
\leavevmode
\epsffile{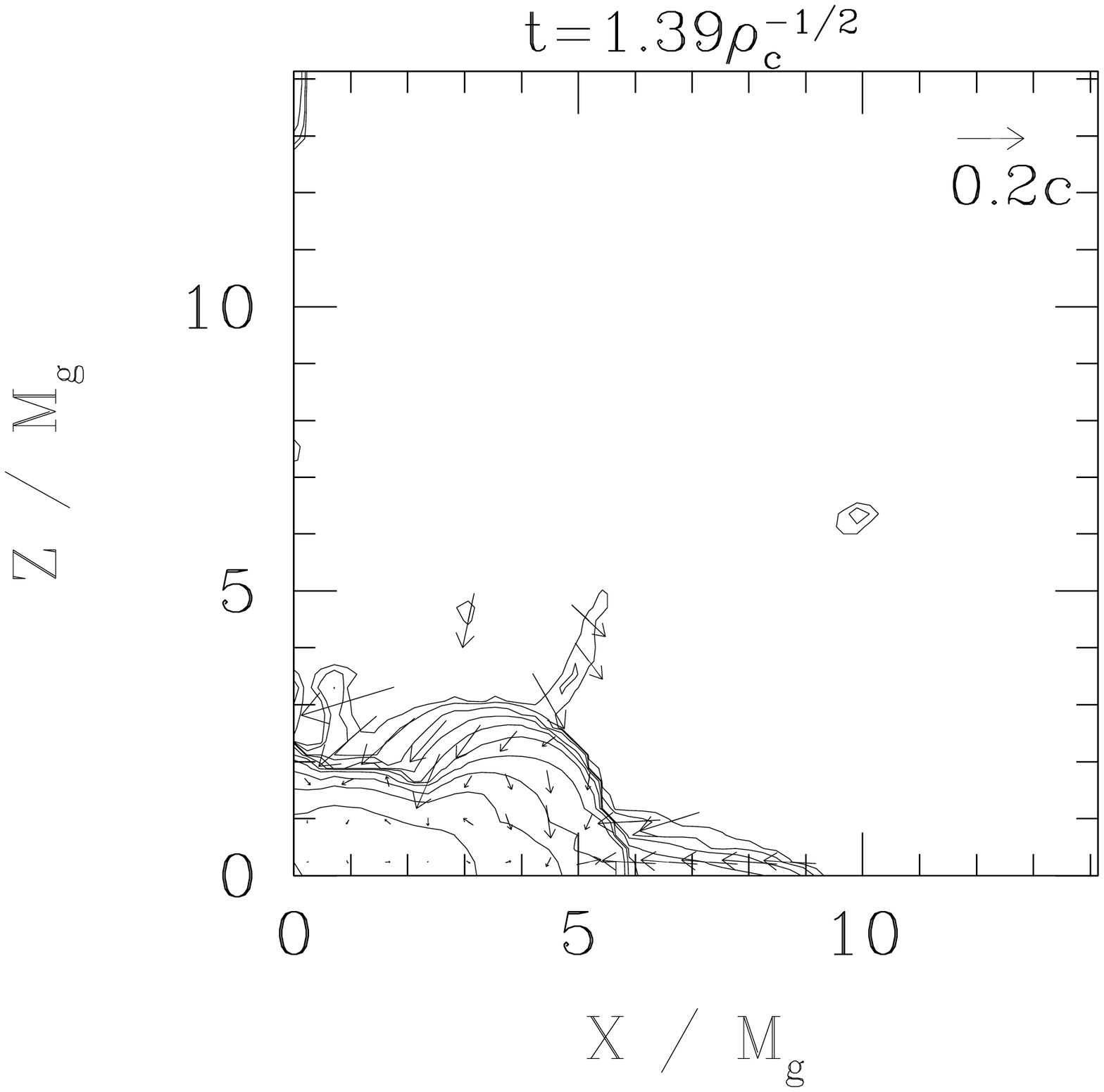}\\
\epsfxsize=1.8in
\leavevmode
\epsffile{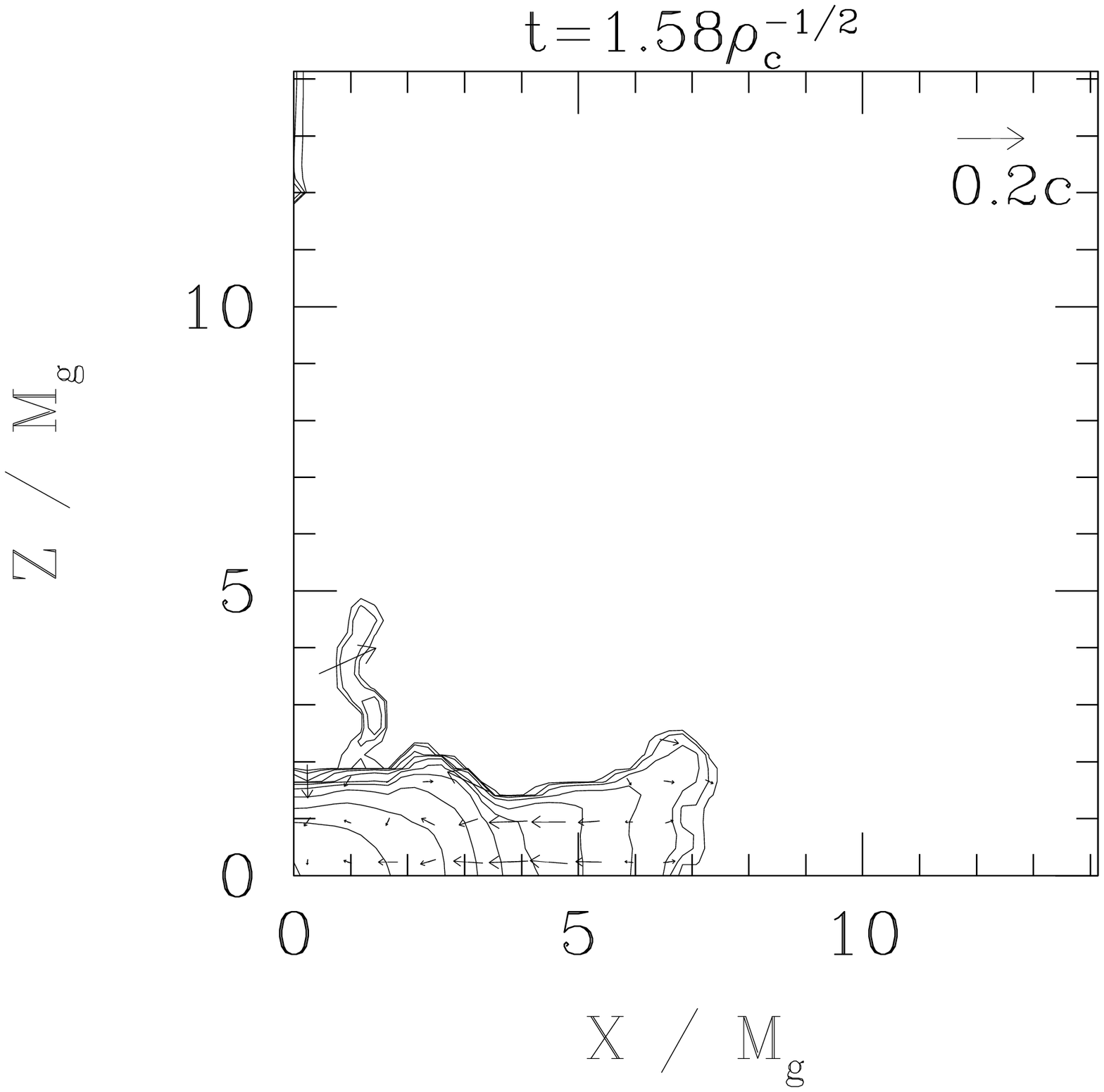}
\epsfxsize=1.8in
\leavevmode
\epsffile{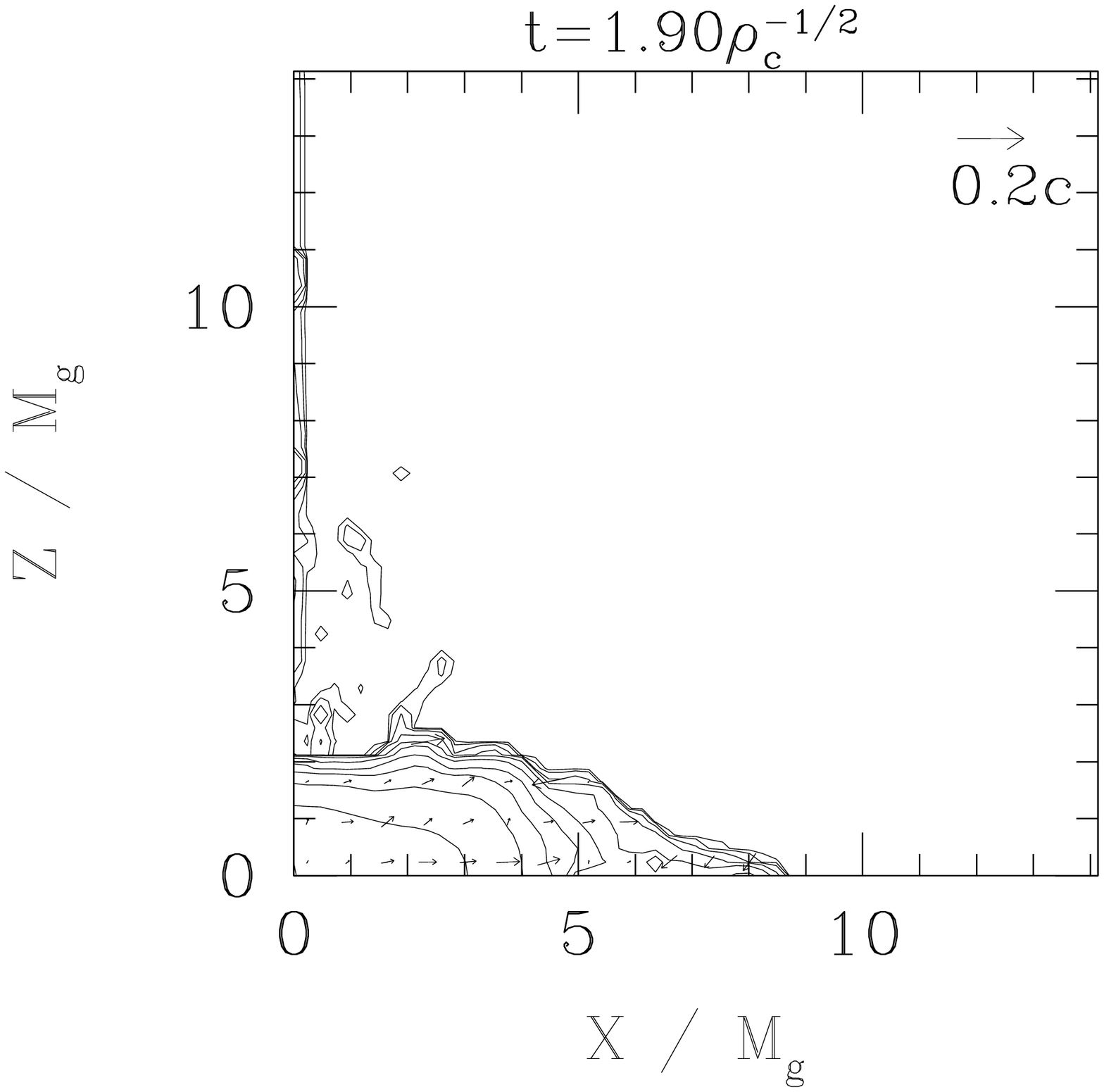}
\epsfxsize=1.8in
\leavevmode
\epsffile{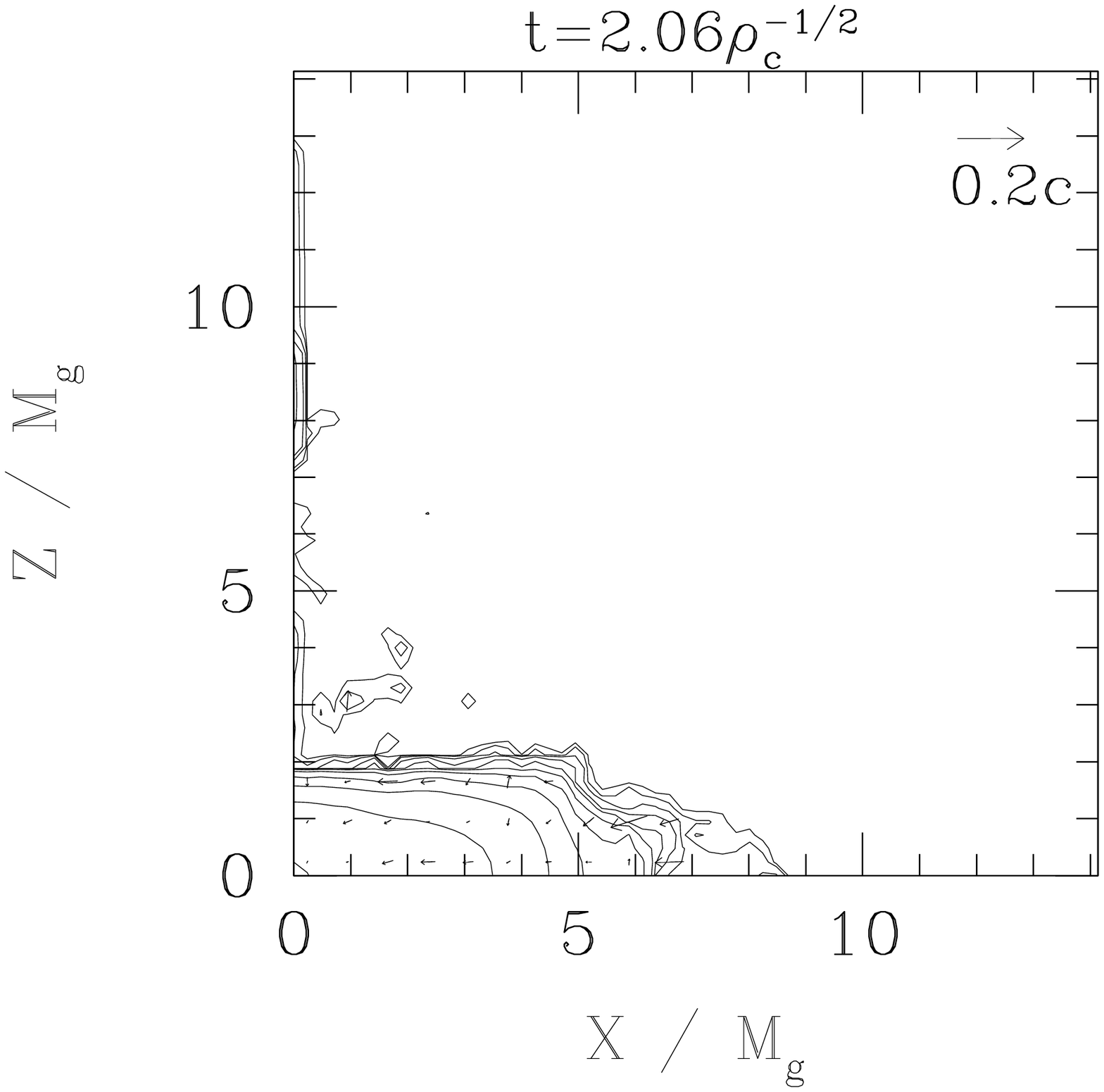}
\end{center}
\caption{Snapshots of density contours for $\rho_*$ and 
the velocity flow for $v^i$ in the $(x,z)$ plane 
for model (C) with $K^{-1/2}=5.92$. 
The contour lines 
are drawn for $\rho_*/\rho_{*~{\rm max}}=10^{-0.4j}$ 
for $j=0,1,2,\cdots,10$, 
where $\rho_{*~{\rm max}}$ 
is 1.15, 422, 2054, 2705, 360, 264, 876, 315 and 244 times larger than 
$\rho_{*~{\rm max}}$ at $t=0$. 
The length of the velocity vector (the arrows) is normalized to $0.2c$.
The time and length appear in units of $\rho_c^{-1/2}$ and $GM_g/c^2$ 
at $t=0$, respectively. }
\end{figure}

\clearpage
\begin{figure}[t]
\begin{center}
\epsfxsize=1.8in
\leavevmode
\epsffile{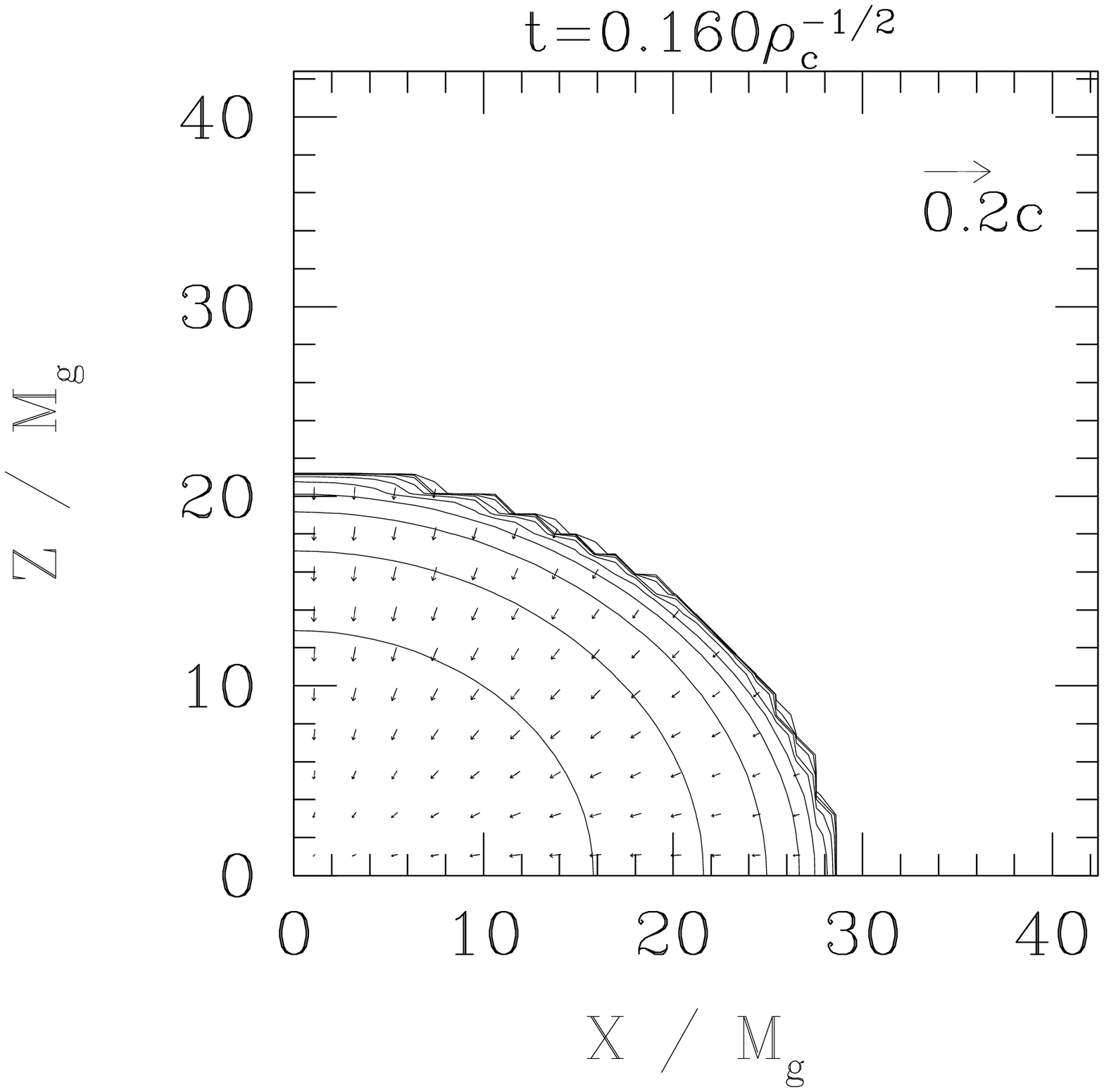}
\epsfxsize=1.8in
\leavevmode
\epsffile{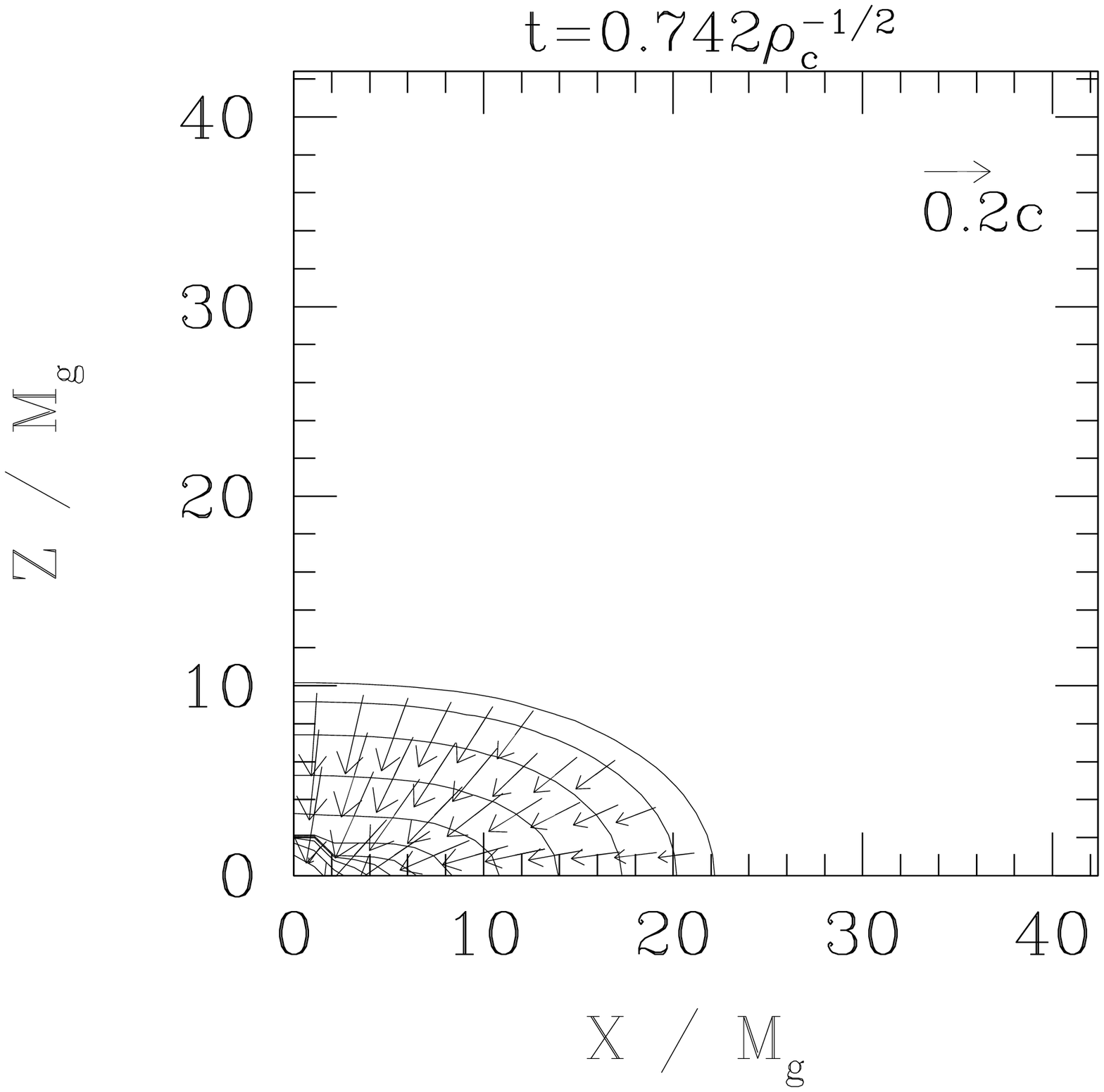}
\epsfxsize=1.8in
\leavevmode
\epsffile{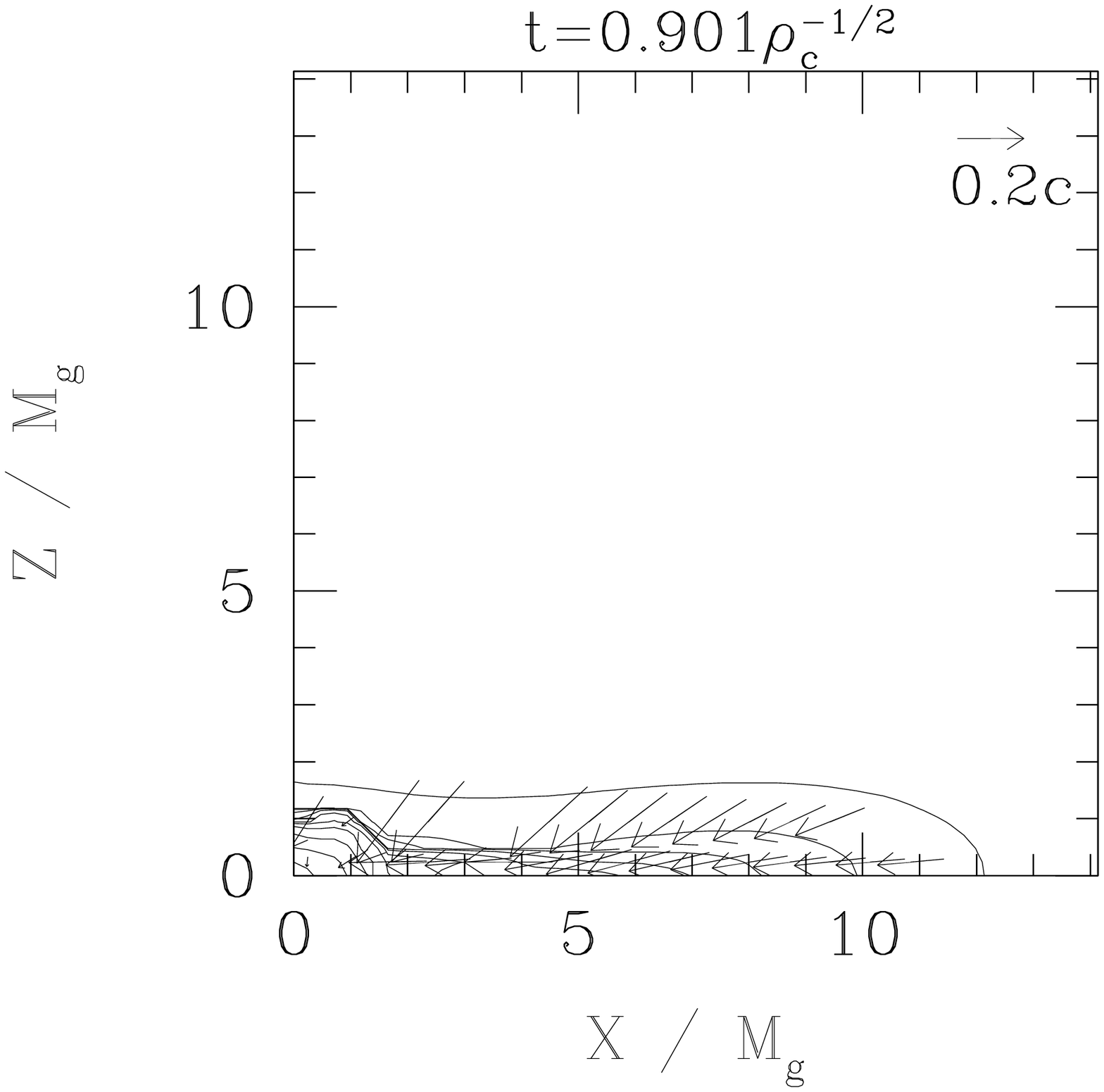}\\
\epsfxsize=1.8in
\leavevmode
\epsffile{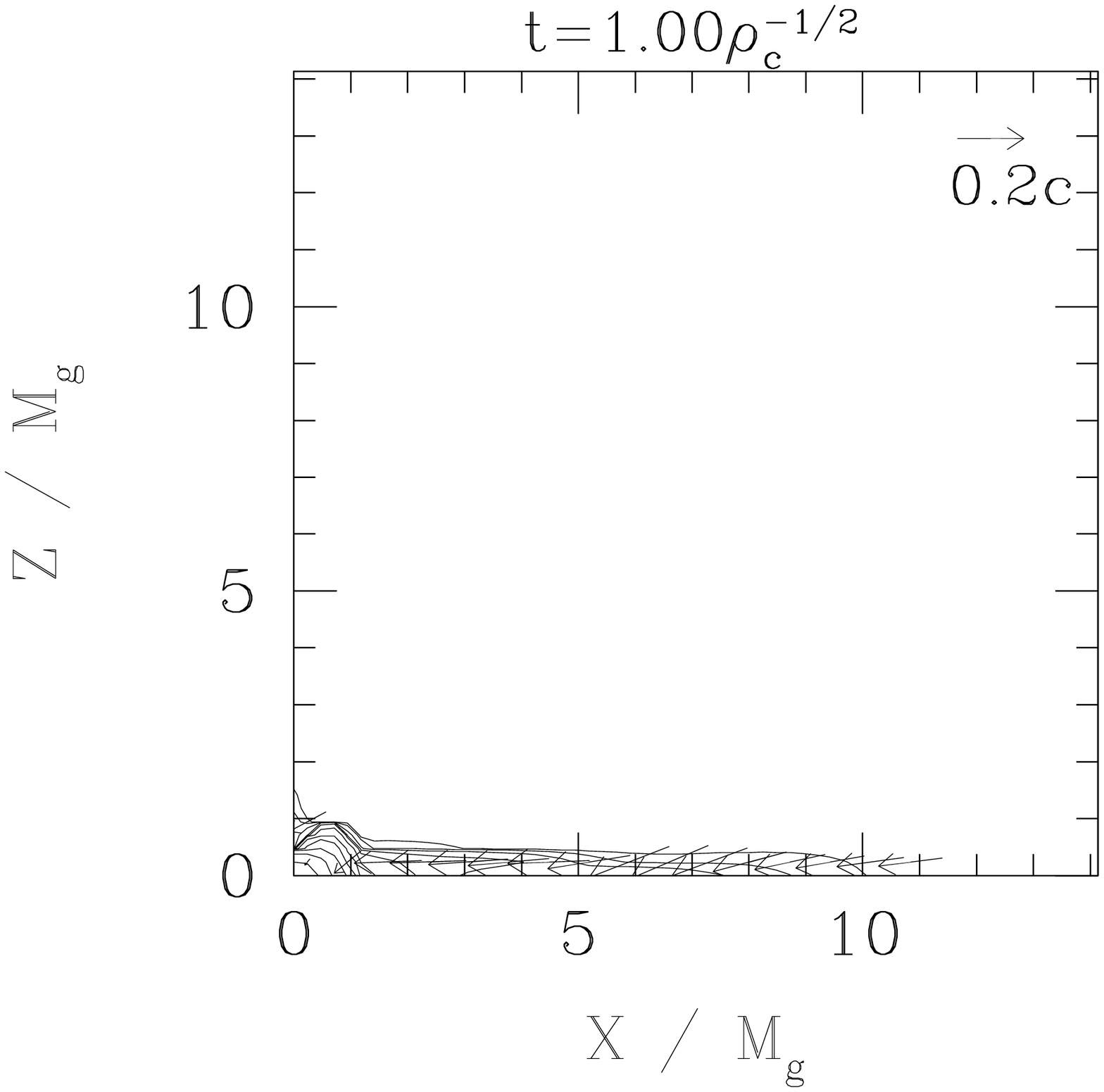}
\epsfxsize=1.8in
\leavevmode
\epsffile{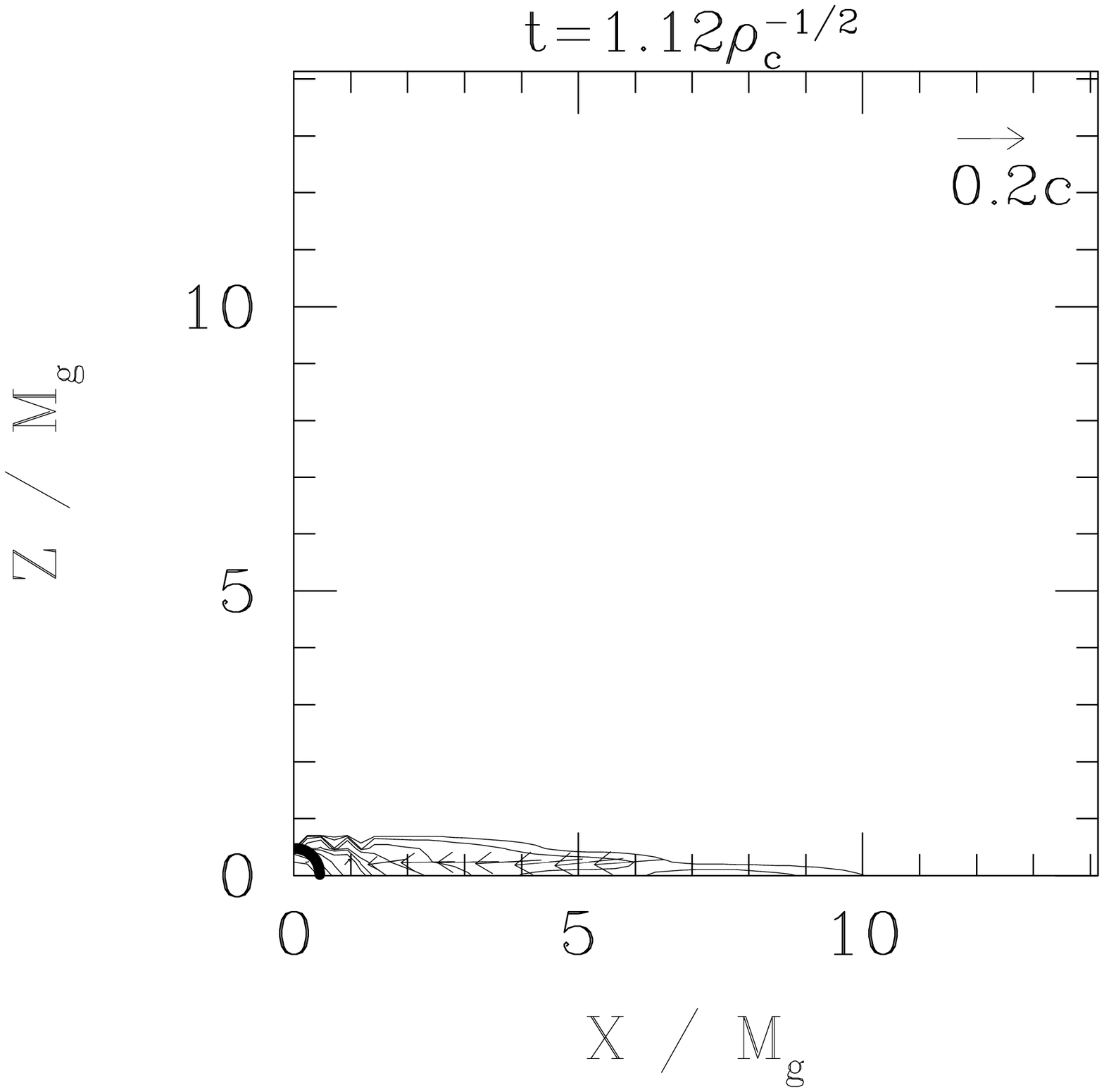}
\epsfxsize=1.8in
\leavevmode
\epsffile{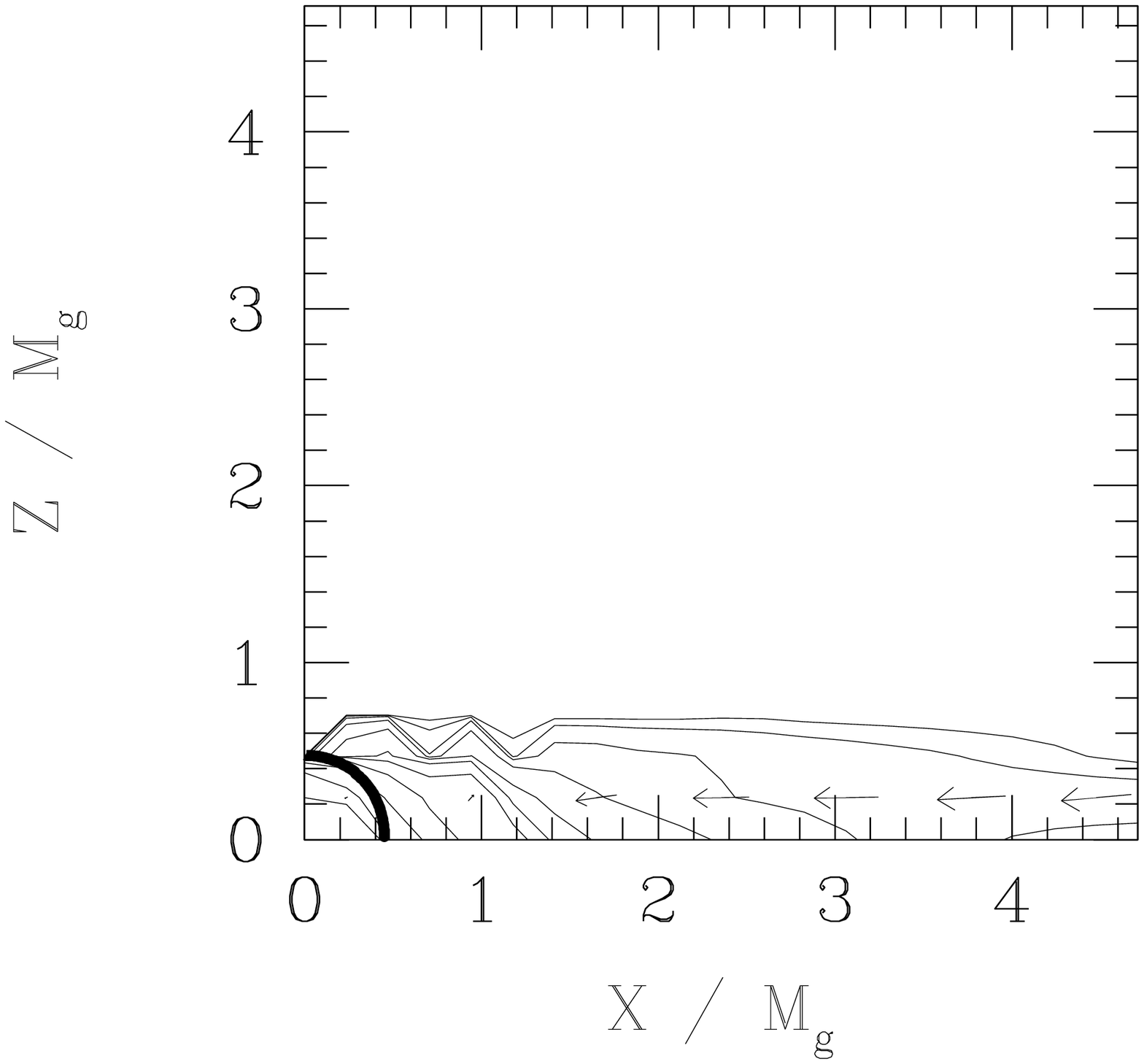}
\end{center}
\caption{The same as Fig.~4, but 
for model (C) with $K^{-1/2}=6.71$. 
The contour lines 
are drawn for $\rho_*/\rho_{*~{\rm max}}=10^{-0.4j}$ 
for $j=0,1,2,\cdots,10$, where $\rho_{*~{\rm max}}$ 
is 1.15, 598, 7830, $6.37\times 10^4$, and $3.89\times 10^5$ 
times larger than $\rho_{*~{\rm max}}$ at $t=0$. 
The last panel is the magnification of the 5th panel. 
The thick solid curve in the last two panels indicates 
the apparent horizon. }
\end{figure}

\clearpage
\begin{figure}[t]
\begin{center}
\epsfxsize=1.8in
\leavevmode
\epsffile{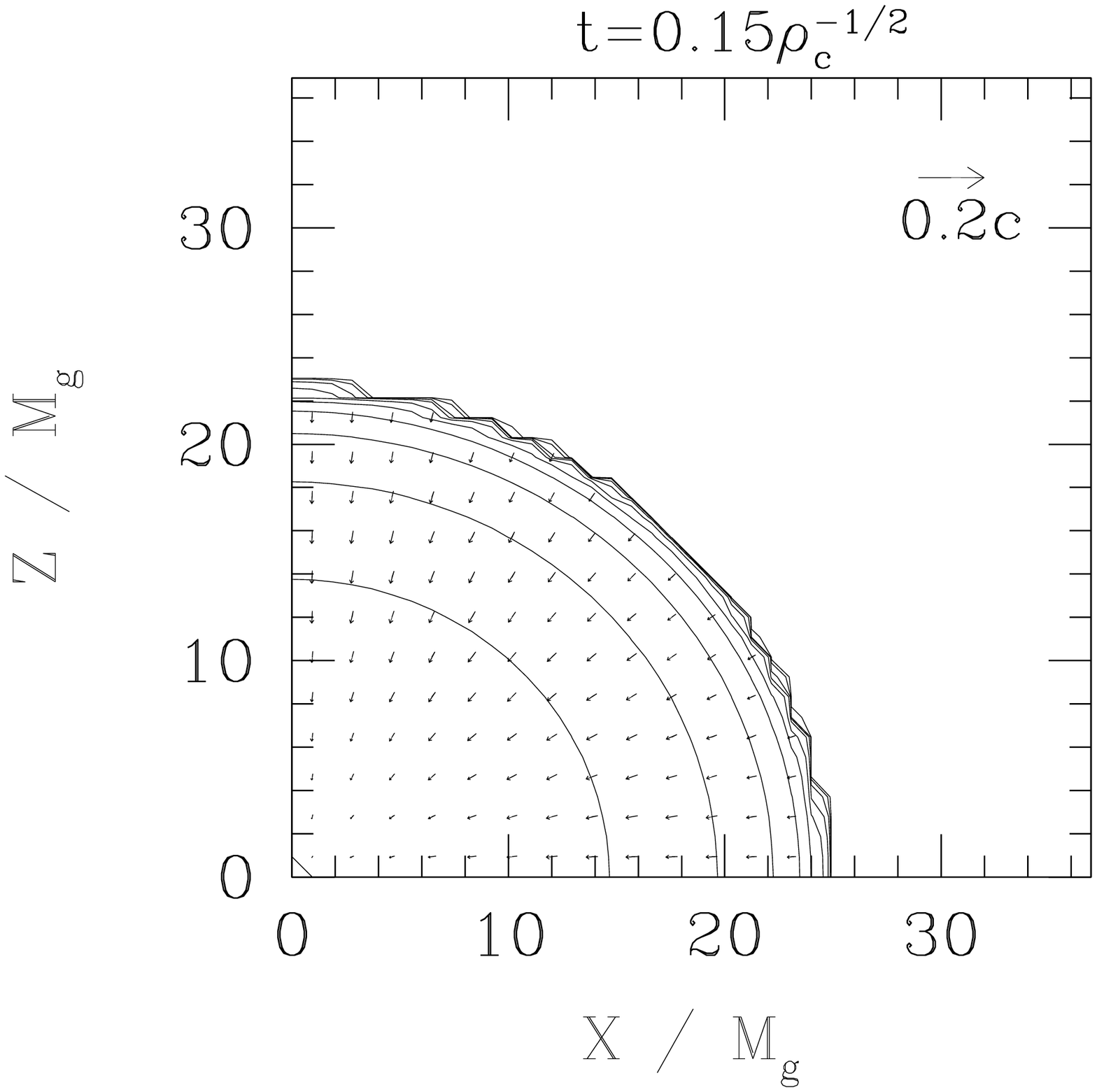}
\epsfxsize=1.8in
\leavevmode
\epsffile{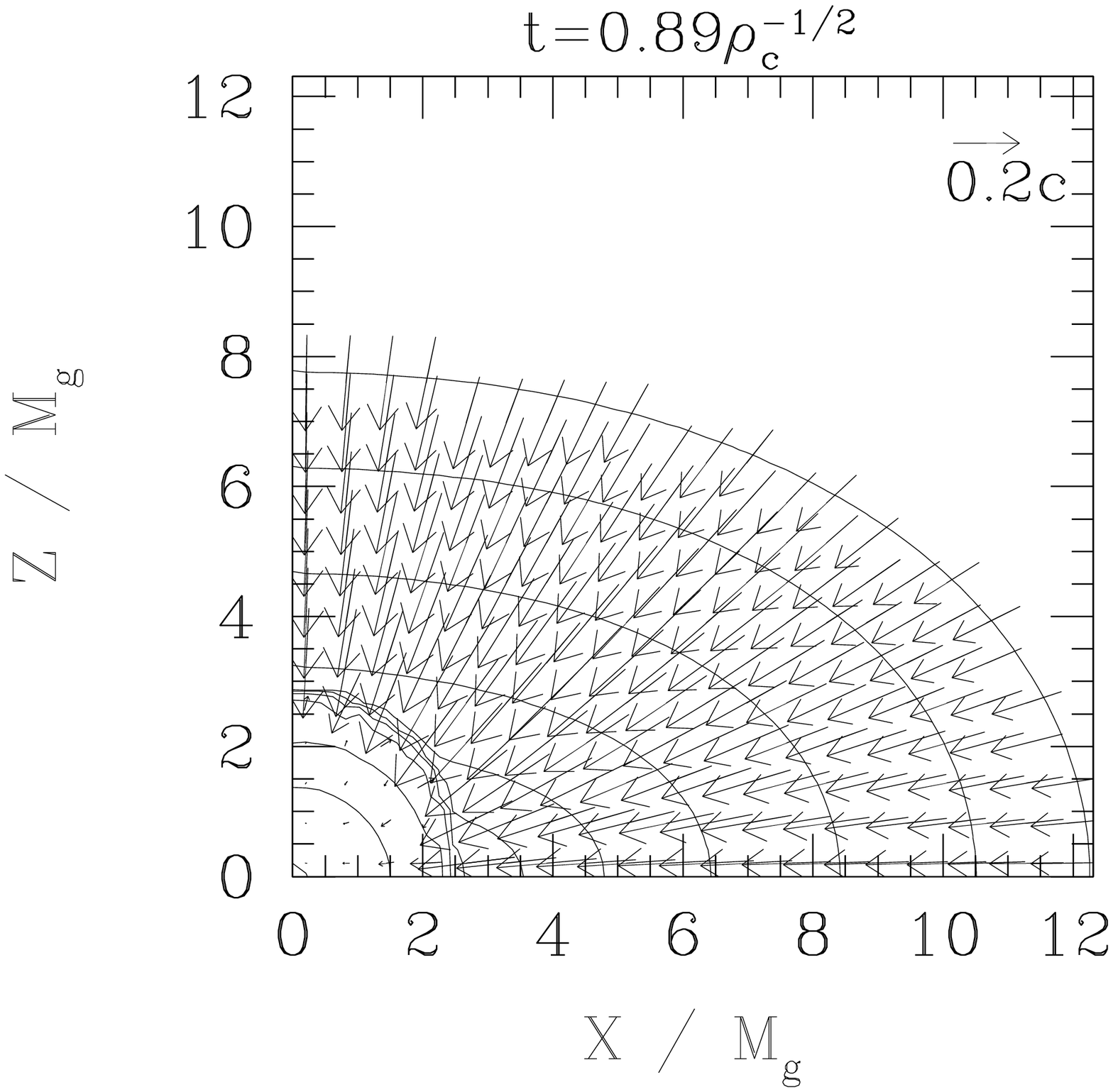}
\epsfxsize=1.8in
\leavevmode
\epsffile{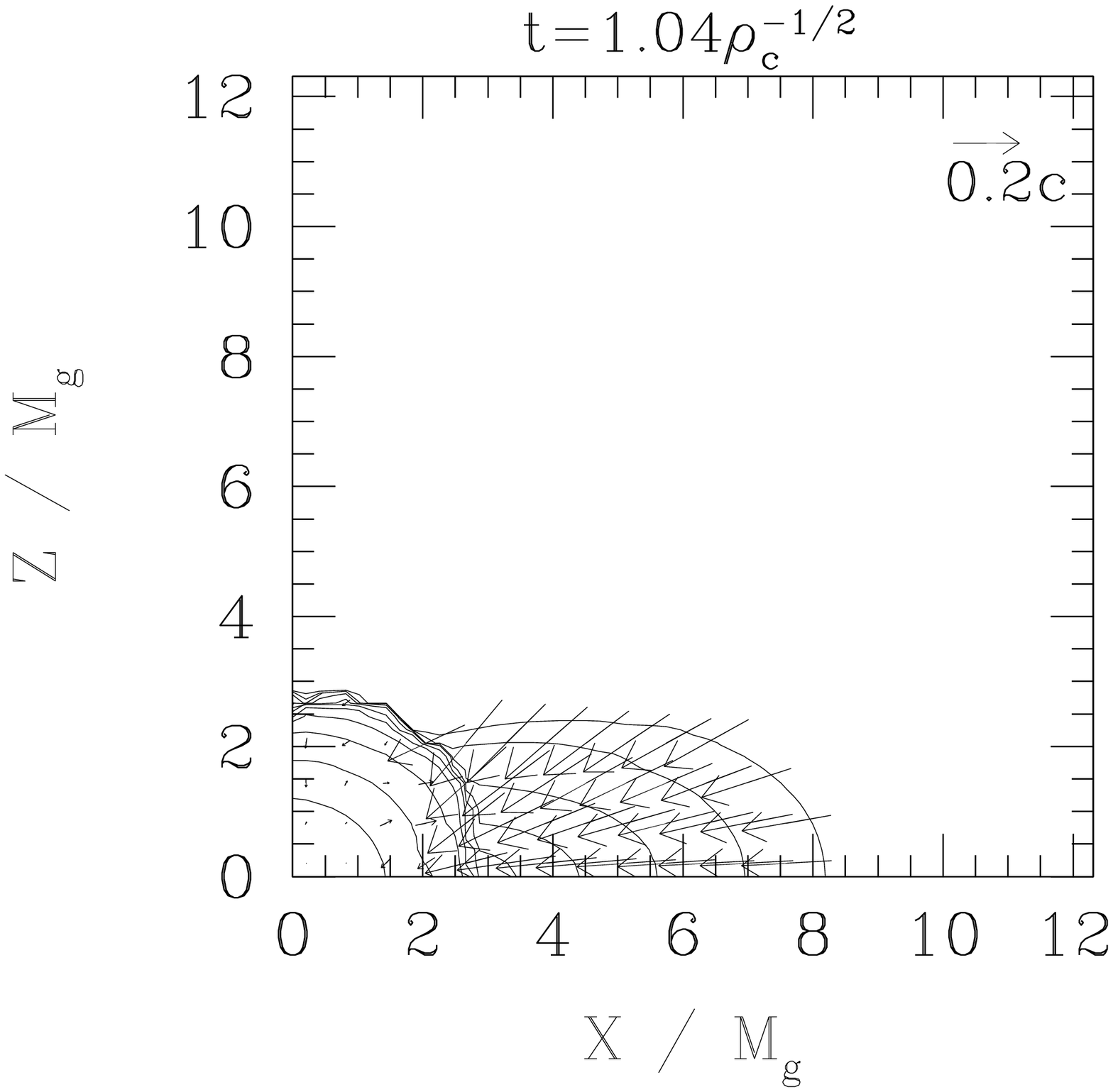}\\
\epsfxsize=1.8in
\leavevmode
\epsffile{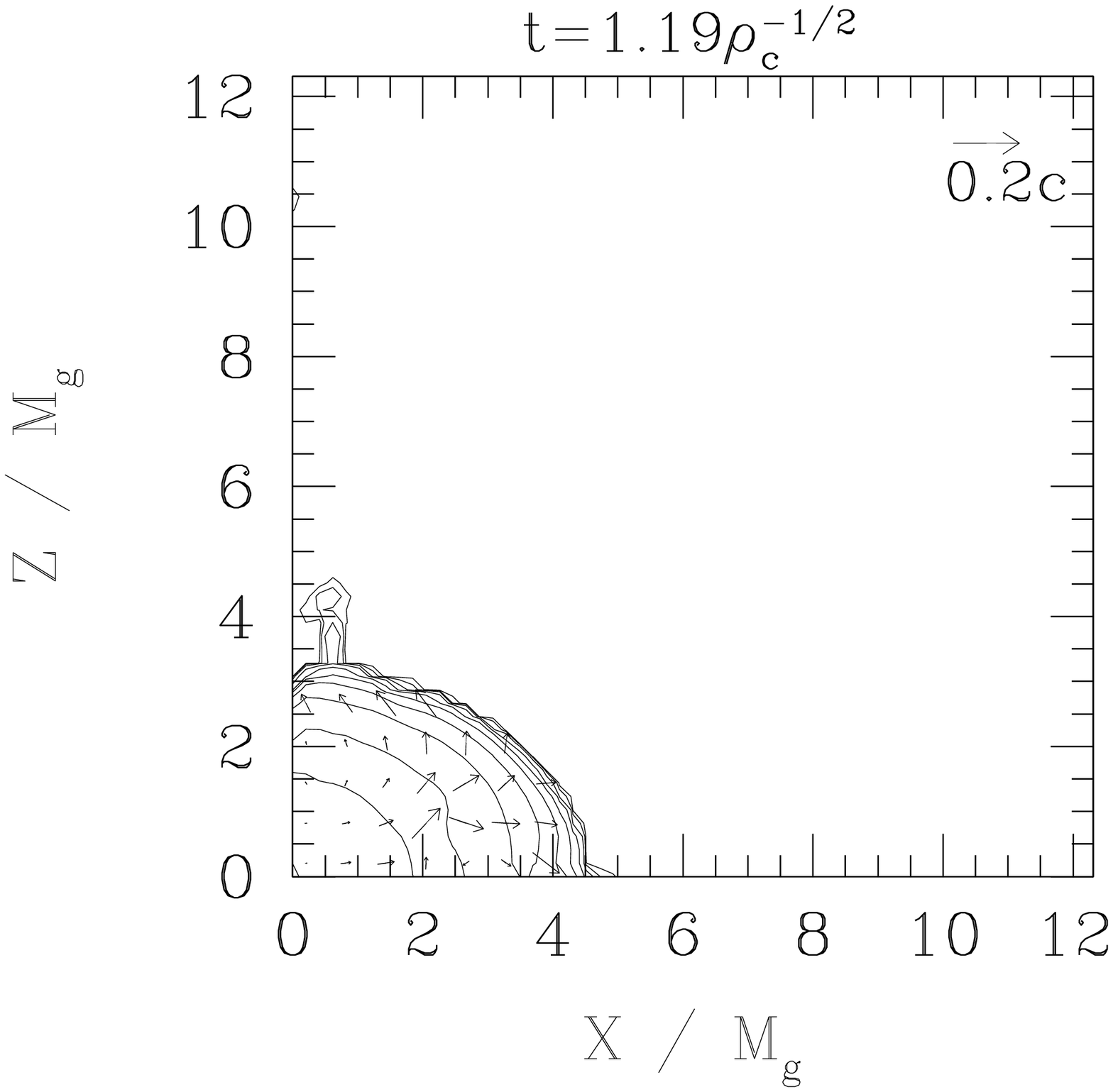}
\epsfxsize=1.8in
\leavevmode
\epsffile{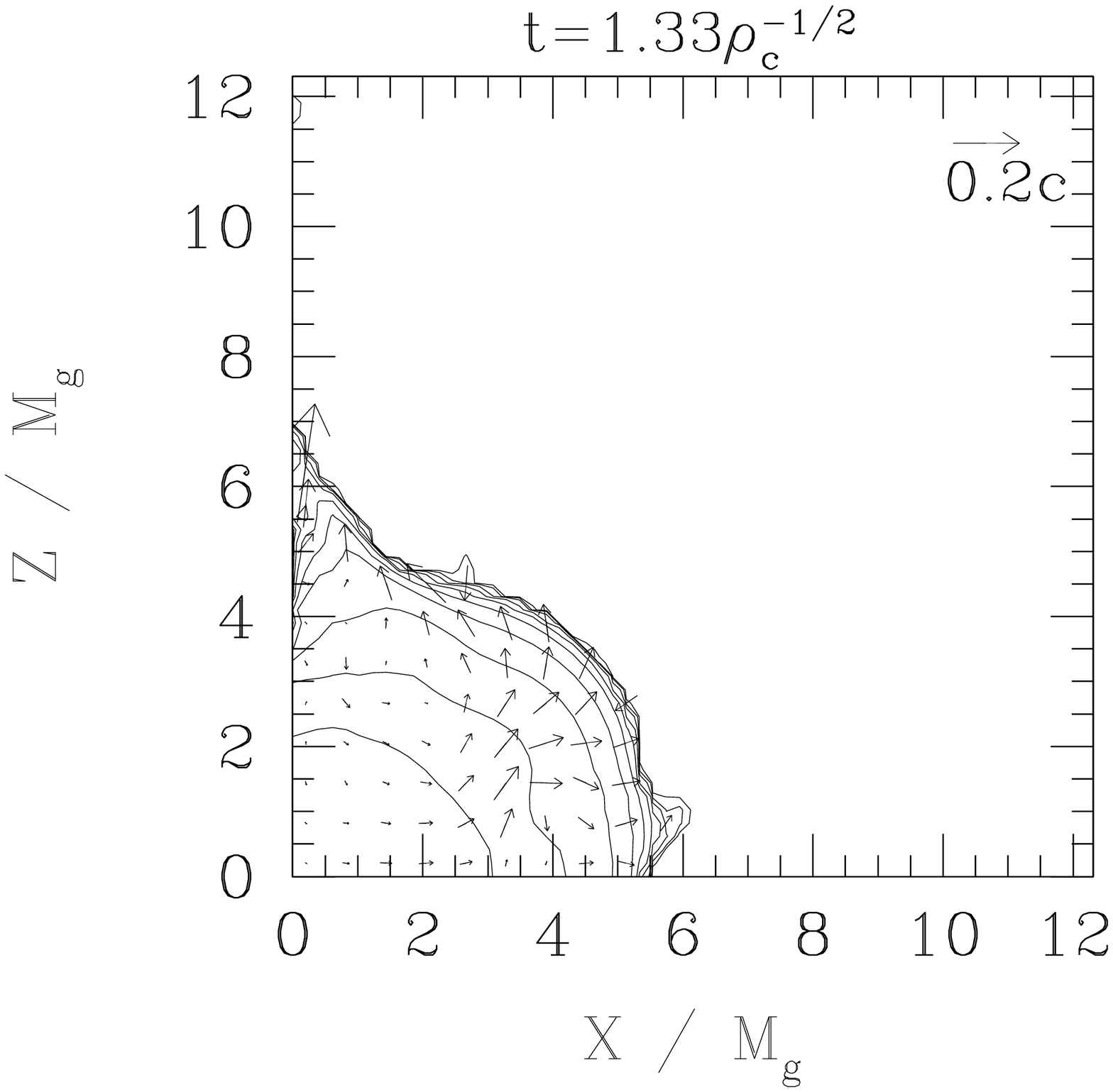}
\epsfxsize=1.8in
\leavevmode
\epsffile{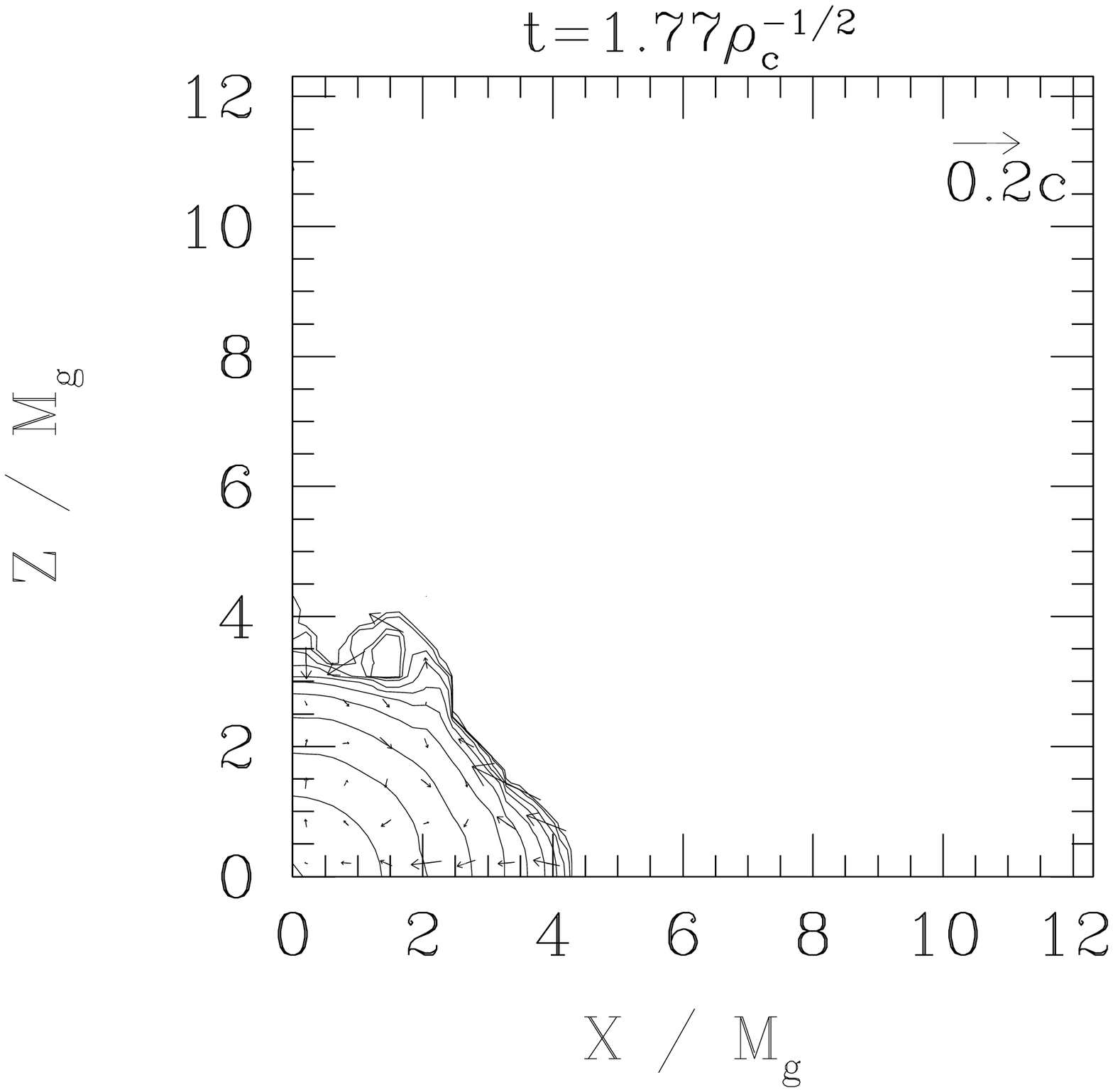}
\end{center}
\caption{The same as Fig.~4, but 
for model (E) with $K^{-1/2}=3.74$. The contour lines 
are drawn for $\rho_*/\rho_{*~{\rm max}}=10^{-0.4j}$ 
for $j=0,1,2,\cdots,10$, where $\rho_{*~{\rm max}}$ 
is 1.12, 724, 952, 475, 126 and 861 times larger than 
$\rho_{*~{\rm max}}$ at $t=0$.}
\end{figure}

\clearpage
\begin{figure}[t]
\begin{center}
\epsfxsize=1.8in
\leavevmode
\epsffile{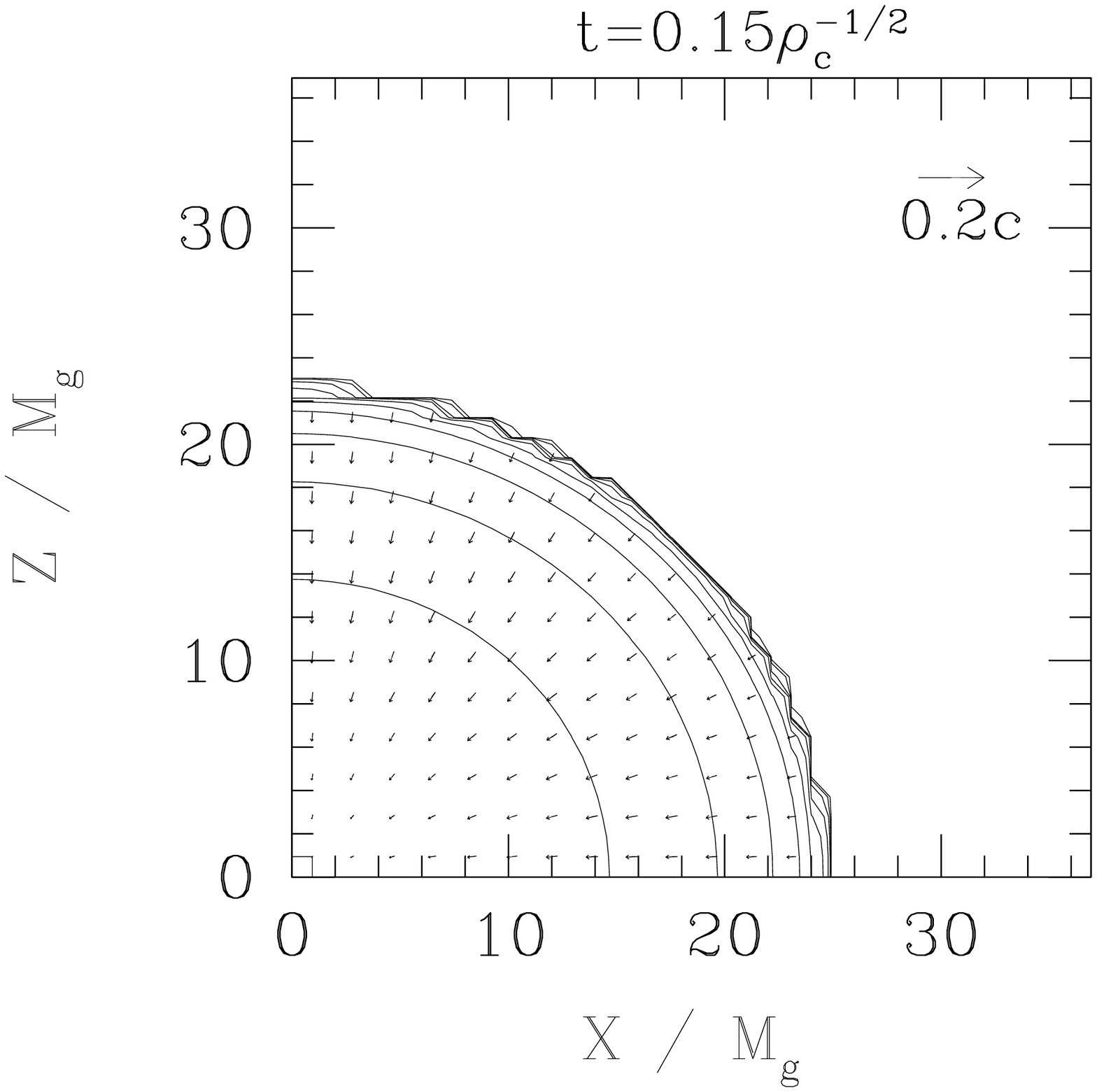}
\epsfxsize=1.8in
\leavevmode
\epsffile{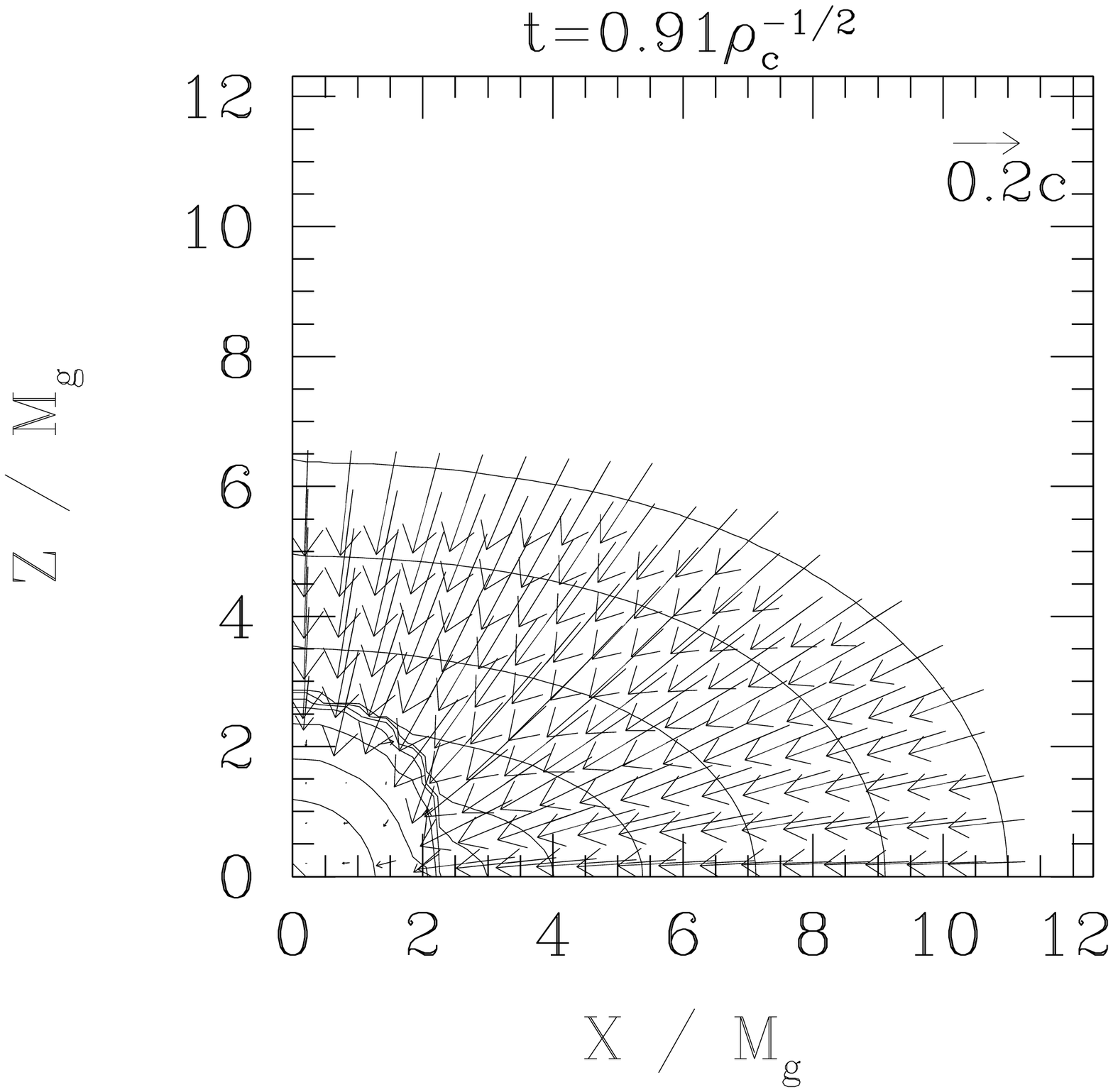}
\epsfxsize=1.8in
\leavevmode
\epsffile{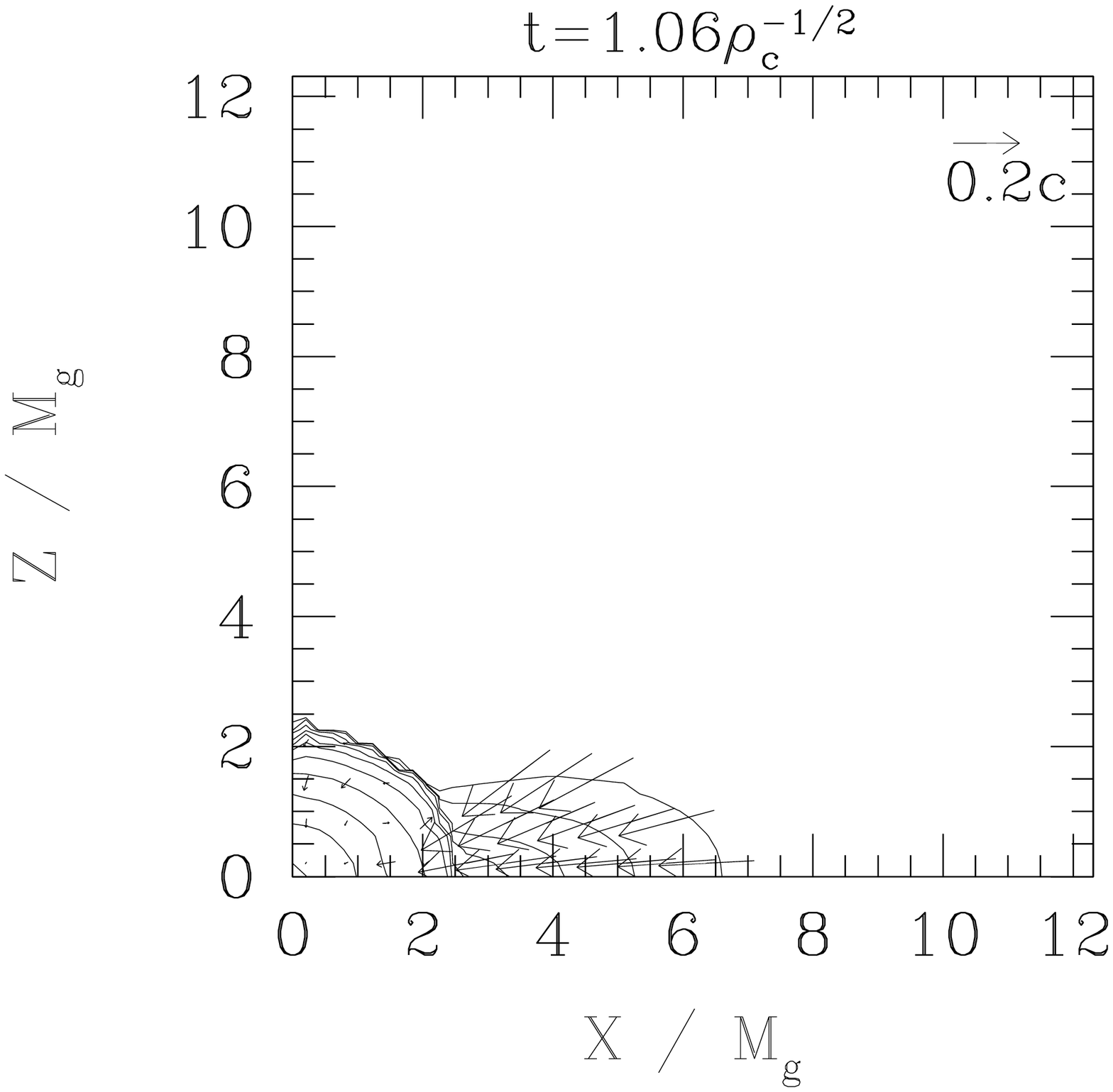}\\
\epsfxsize=1.8in
\leavevmode
\epsffile{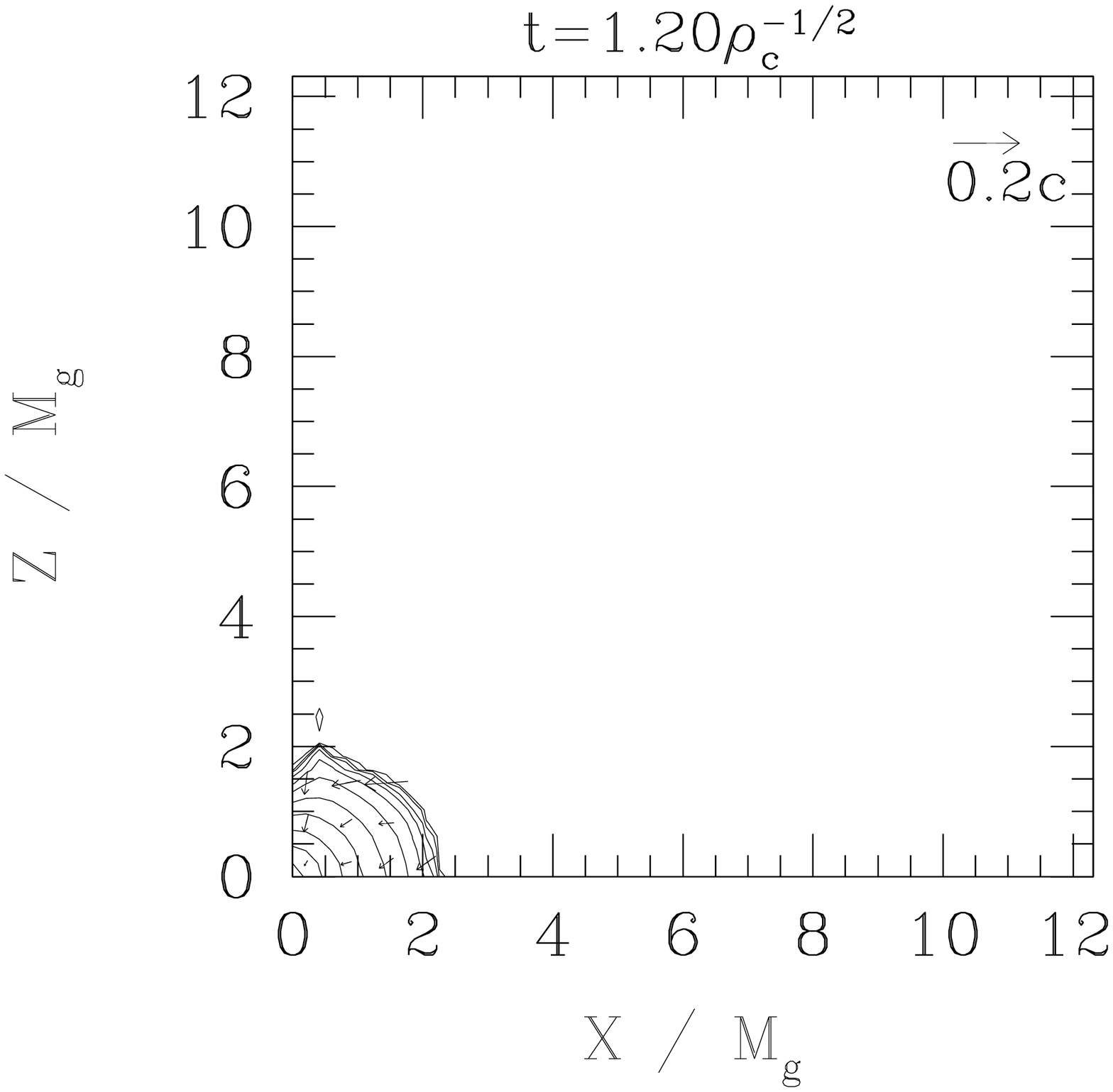}
\epsfxsize=1.8in
\leavevmode
\epsffile{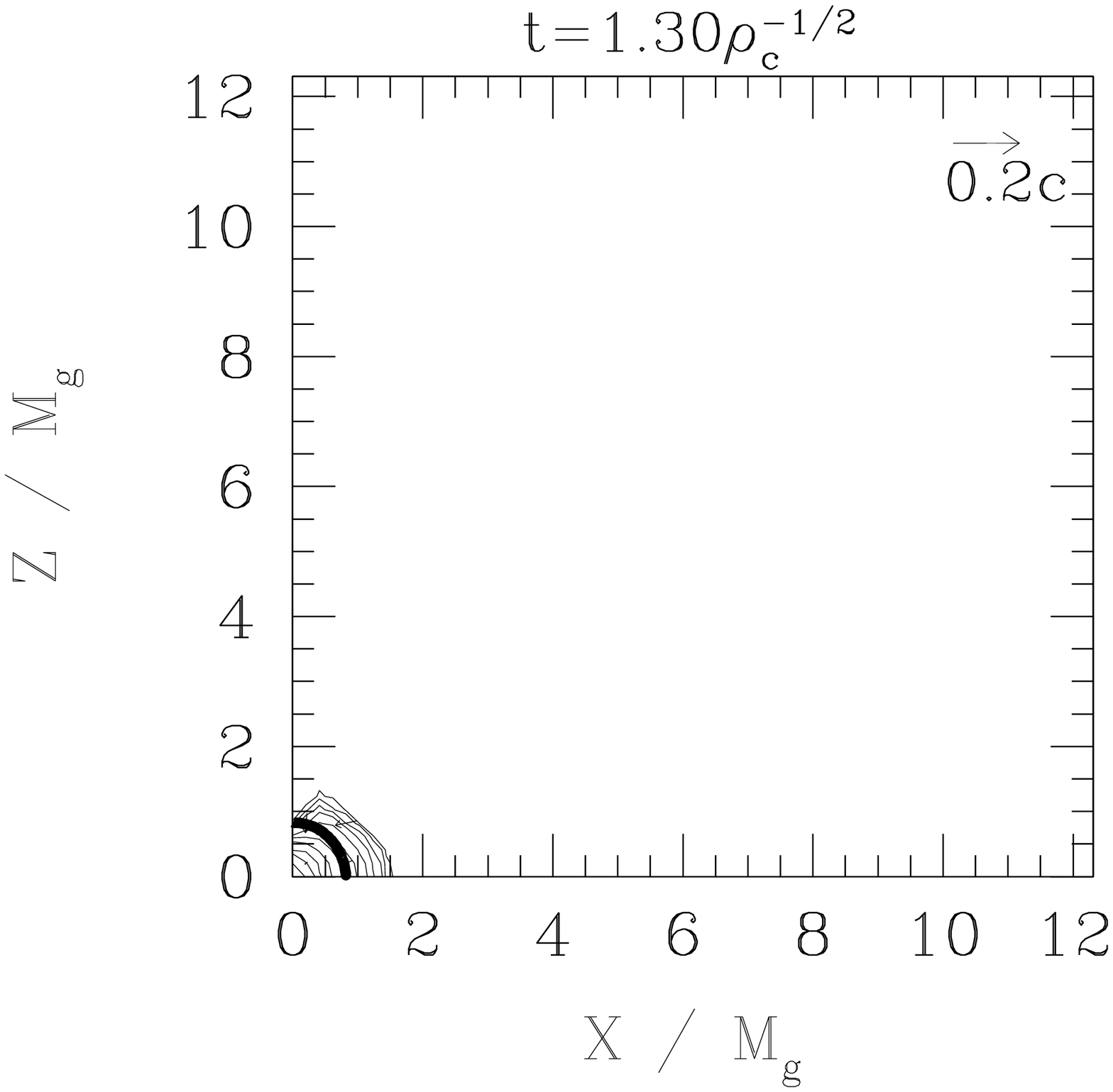}
\epsfxsize=1.8in
\leavevmode
\epsffile{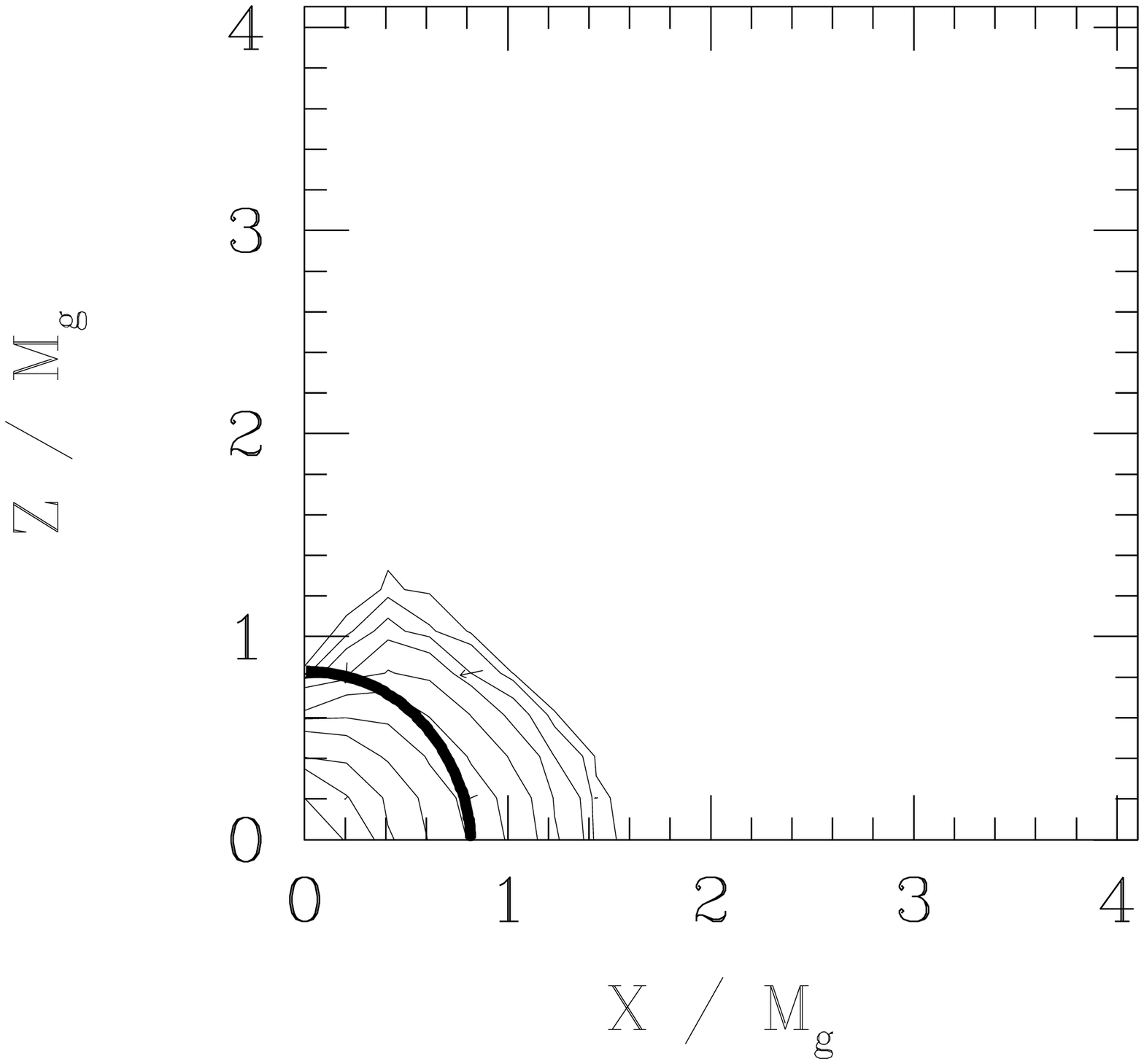}
\end{center}
\caption{The same as Fig.~4, but 
for model (E) with $K^{-1/2}=3.87$. 
The contour lines 
are drawn for $\rho_*/\rho_{*~{\rm max}}=10^{-0.4j}$ 
for $j=0,1,2,\cdots,10$, where $\rho_{*~{\rm max}}$ is $1.23$, 
1120, 2670, $1.78\times10^4$, and 
$1.82\times 10^5$ larger than the initial value. 
The last panel is the magnification of the 5th panel. 
The thick solid curve in the last two panels indicates 
the apparent horizon. }
\end{figure}

\begin{figure}[t]
\begin{center}
\epsfxsize=3.5in
\leavevmode
\epsffile{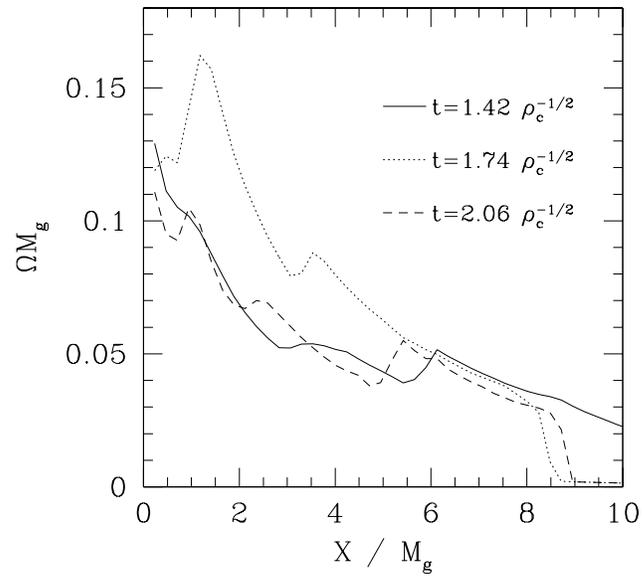}
\end{center}
\caption{$\Omega$ in units of $M_g^{-1}$ at selected time steps 
for model (C) with $K^{-1/2}=5.92$. 
Note that the rotating stars are oscillating with time, and 
at $t=1.74\rho_c^{-1/2}$ the radius is smaller than at other 
times. 
}
\end{figure}

\clearpage
\begin{figure}[t]
\begin{center}
\epsfxsize=3.5in
\leavevmode
\epsffile{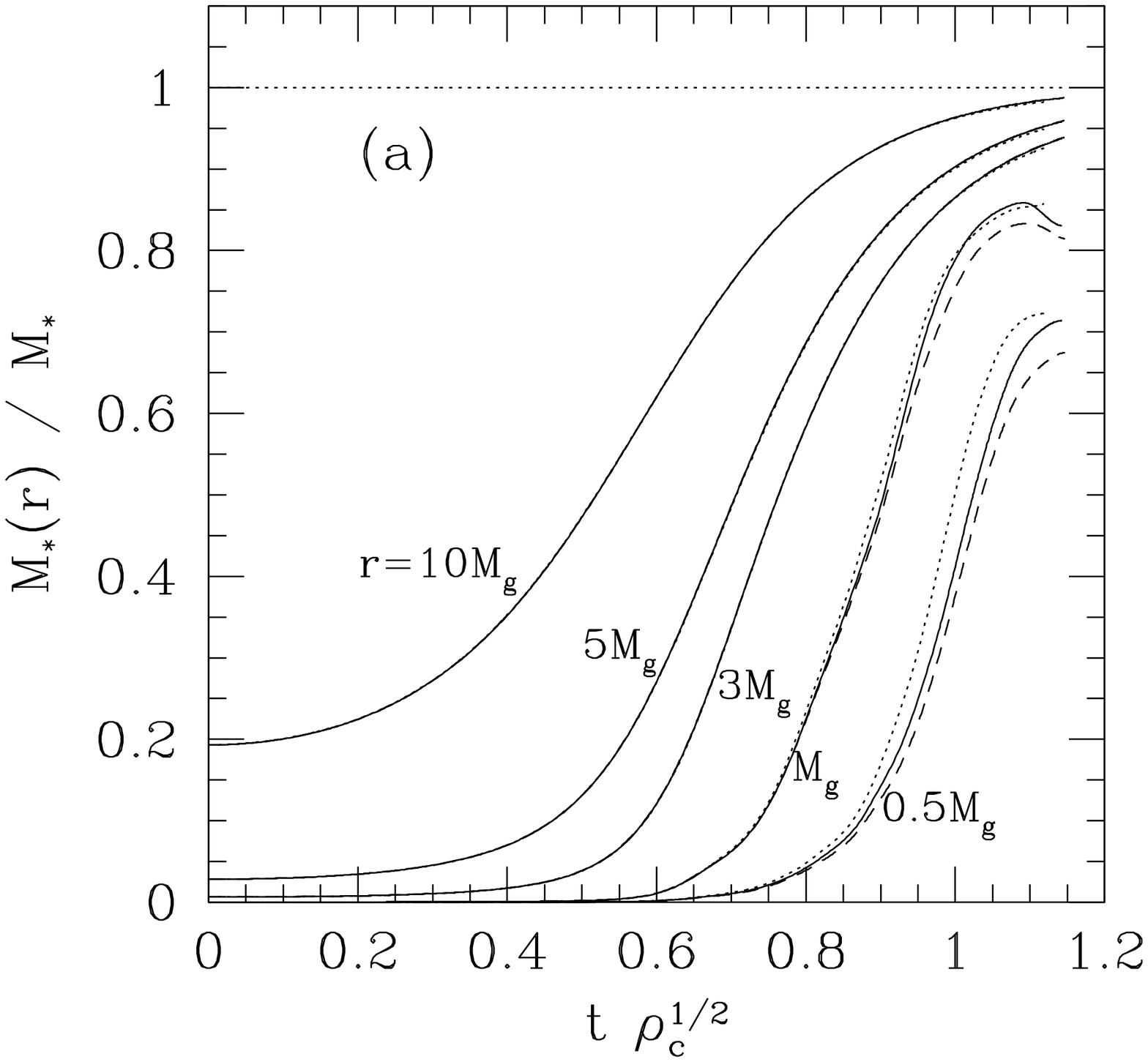}\\
\epsfxsize=3.5in
\leavevmode
\epsffile{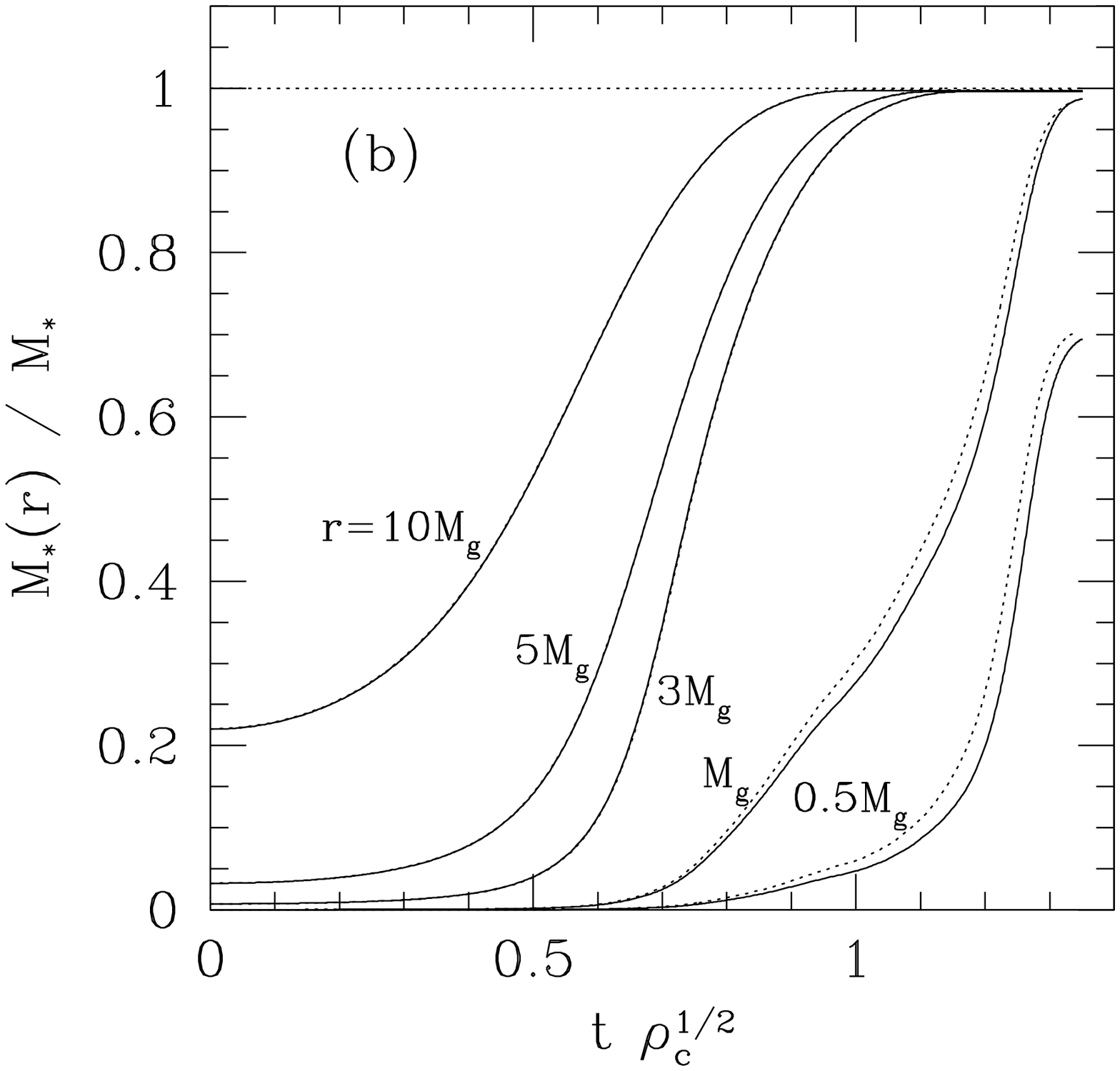}
\end{center}
\caption{Evolution of the fraction of the rest mass 
inside a coordinate radius $r$, $M_*(r)/M_*$ 
(a) for model (C) with $K^{-1/2}=6.71$ and 
(b) for model (E) with $K^{-1/2}=3.87$. 
The dotted, solid and dashed curves represent the results 
for $(271, 271)$, $(361, 361)$ and $(361, 721)$ grid resolutions, 
respectively. 
}
\end{figure}

\clearpage 
\begin{figure}[t]
\begin{center}
\epsfxsize=3.5in
\leavevmode
\epsffile{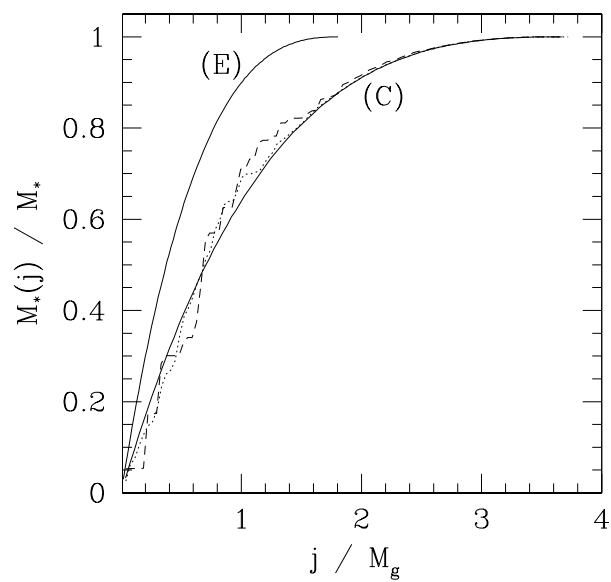}
\end{center}
\caption{$M_*(j)/M_*$ as a function of $j/M_g$ for models (C) and (E). 
The solid, dotted and dashed curves for model (C) denote 
the results at $t\rho_c^{1/2}=0$, 0.901, and 1.10 for 
$K^{-1/2}=6.71$. 
}
\end{figure}

\clearpage 

\begin{figure}[t]
\begin{center}
\epsfxsize=3.5in
\leavevmode
\epsffile{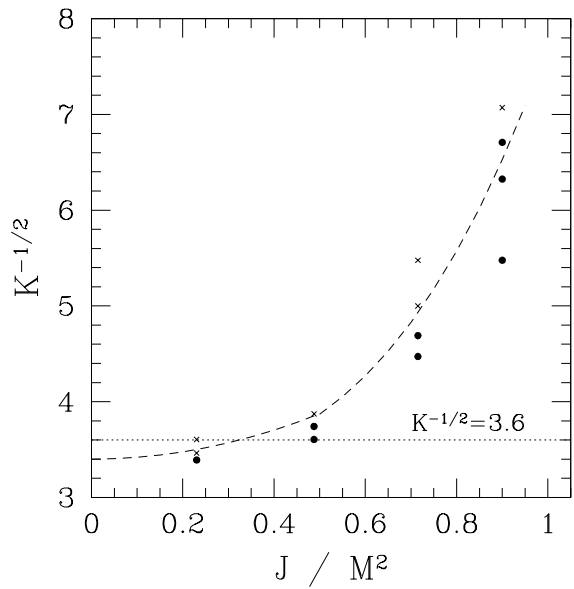}
\end{center}
\caption{The same as Fig.~1, but for differentially rotating 
initial data with $A=\varpi_e/3$. 
The dashed curve represents the approximate threshold for black hole 
formation for rigidly rotating initial data. 
}
\end{figure}

\begin{figure}[t]
\begin{center}
\epsfxsize=3.5in
\leavevmode
\epsffile{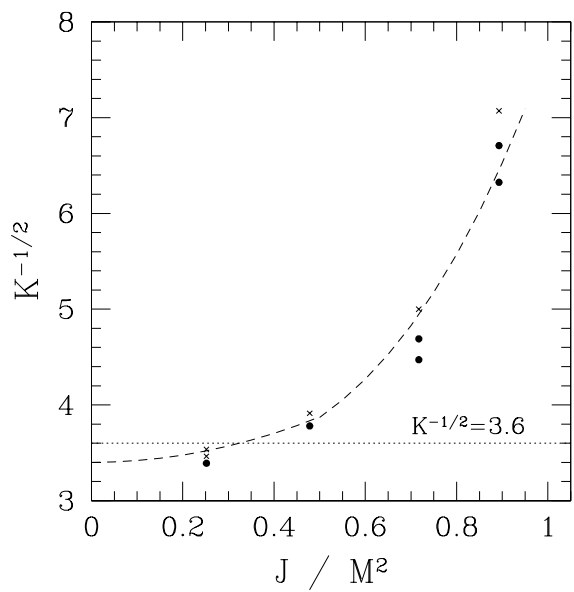}
\end{center}
\caption{The same as Fig.~11, but for differentially rotating 
initial data with $A=\varpi_e$. 
}
\end{figure}

\clearpage
\begin{figure}[t]
\begin{center}
\epsfxsize=1.8in
\leavevmode
\epsffile{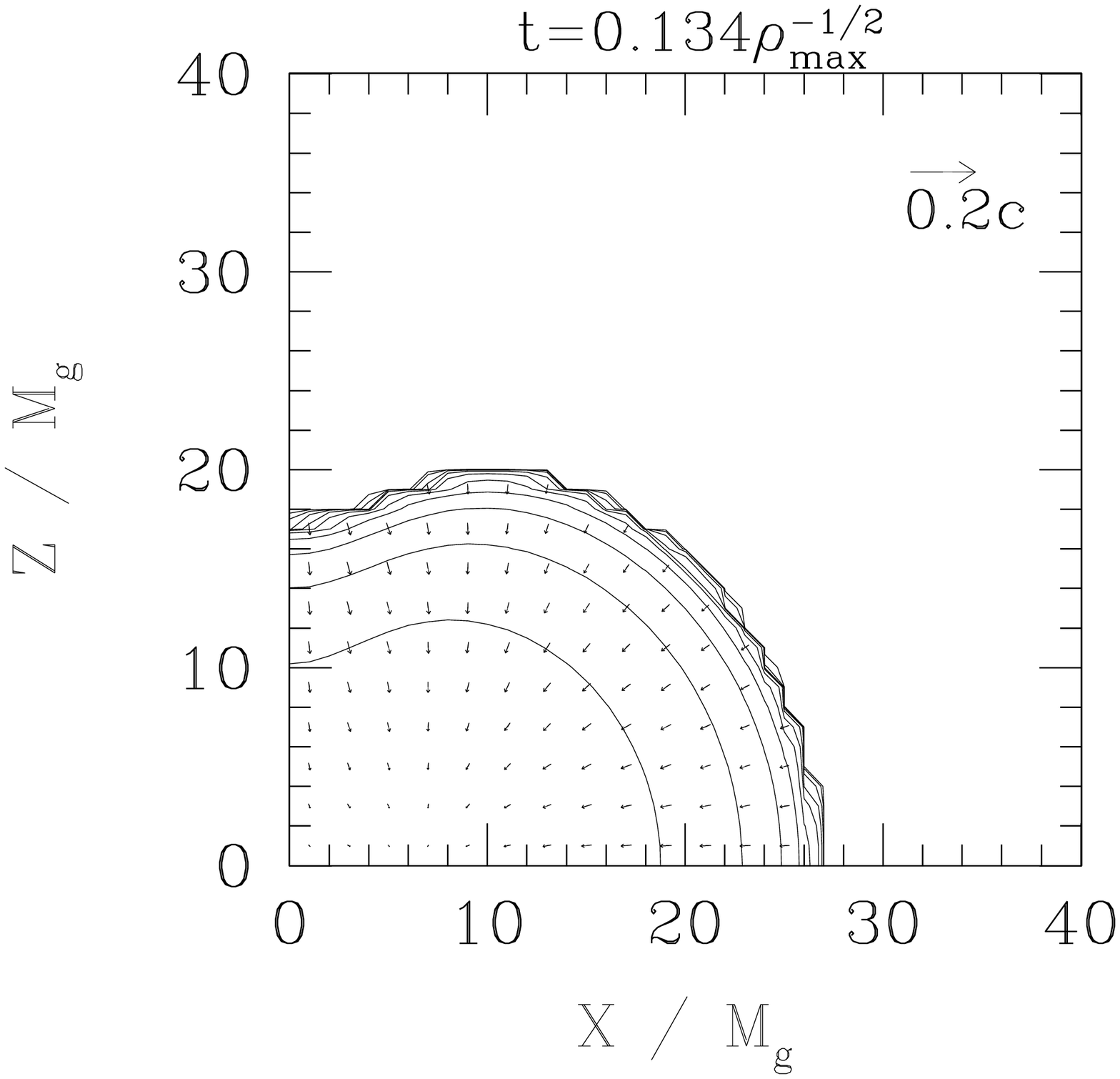}
\epsfxsize=1.8in
\leavevmode
\epsffile{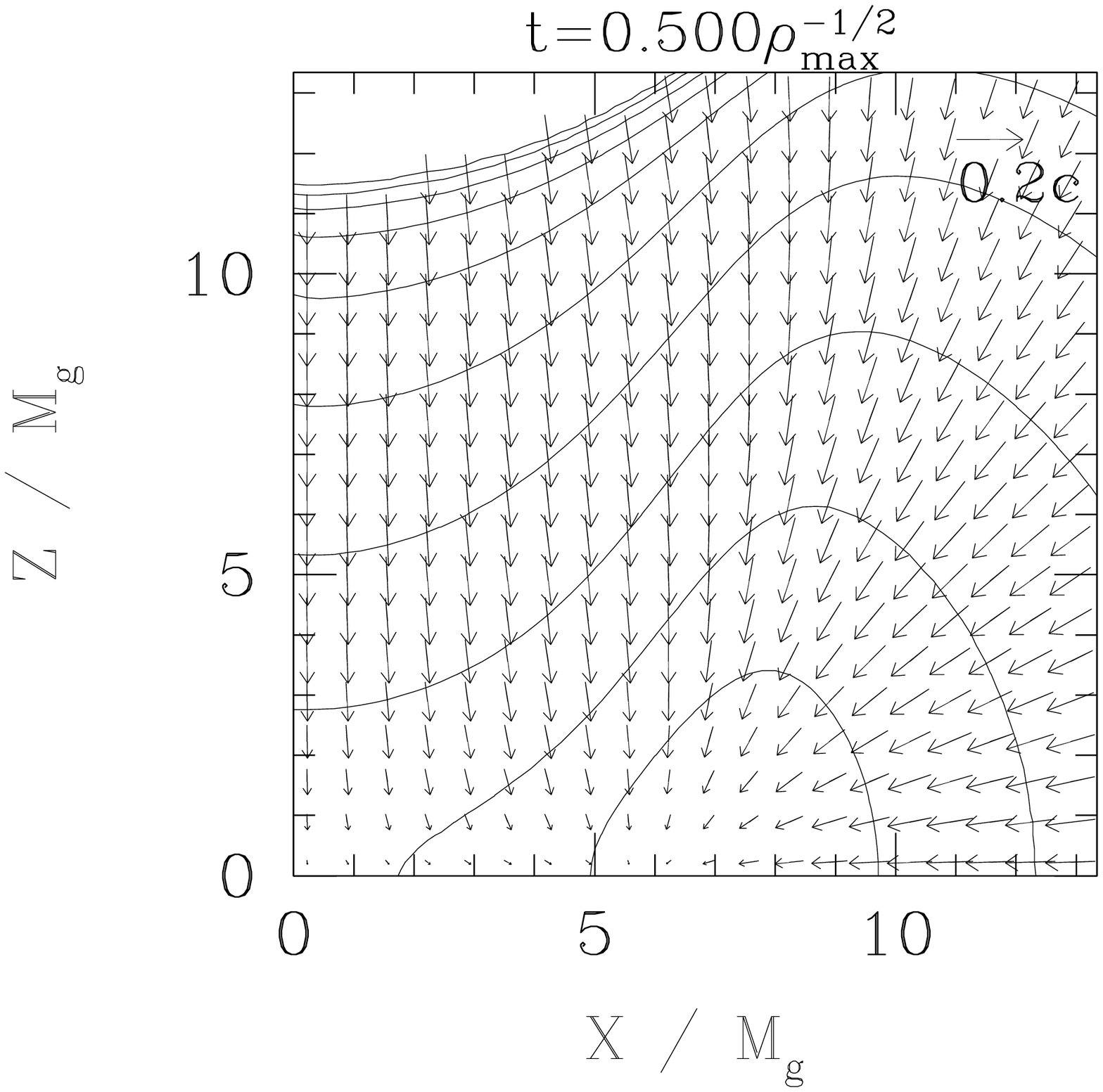}
\epsfxsize=1.8in
\leavevmode
\epsffile{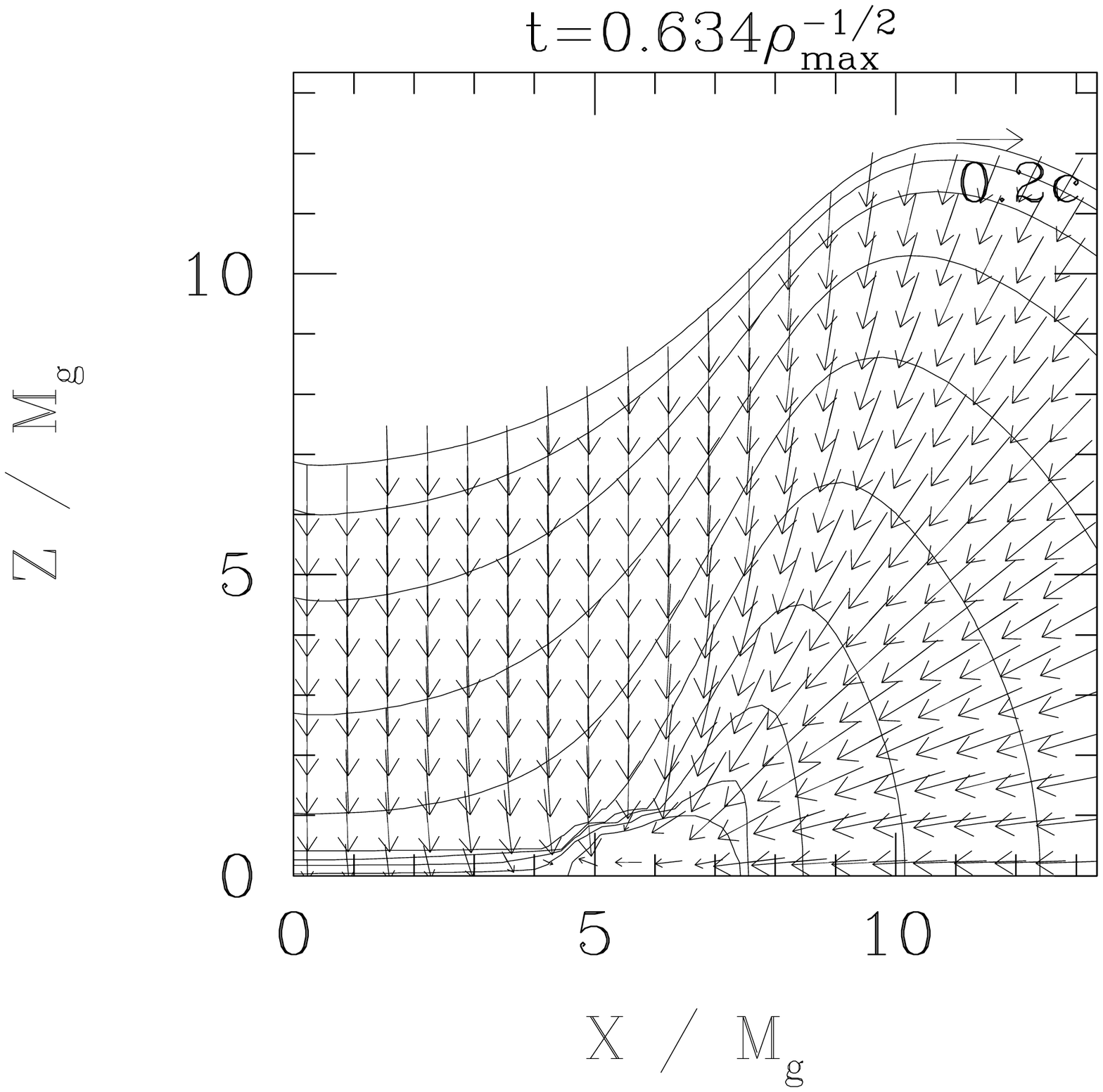} \\
\epsfxsize=1.8in
\leavevmode
\epsffile{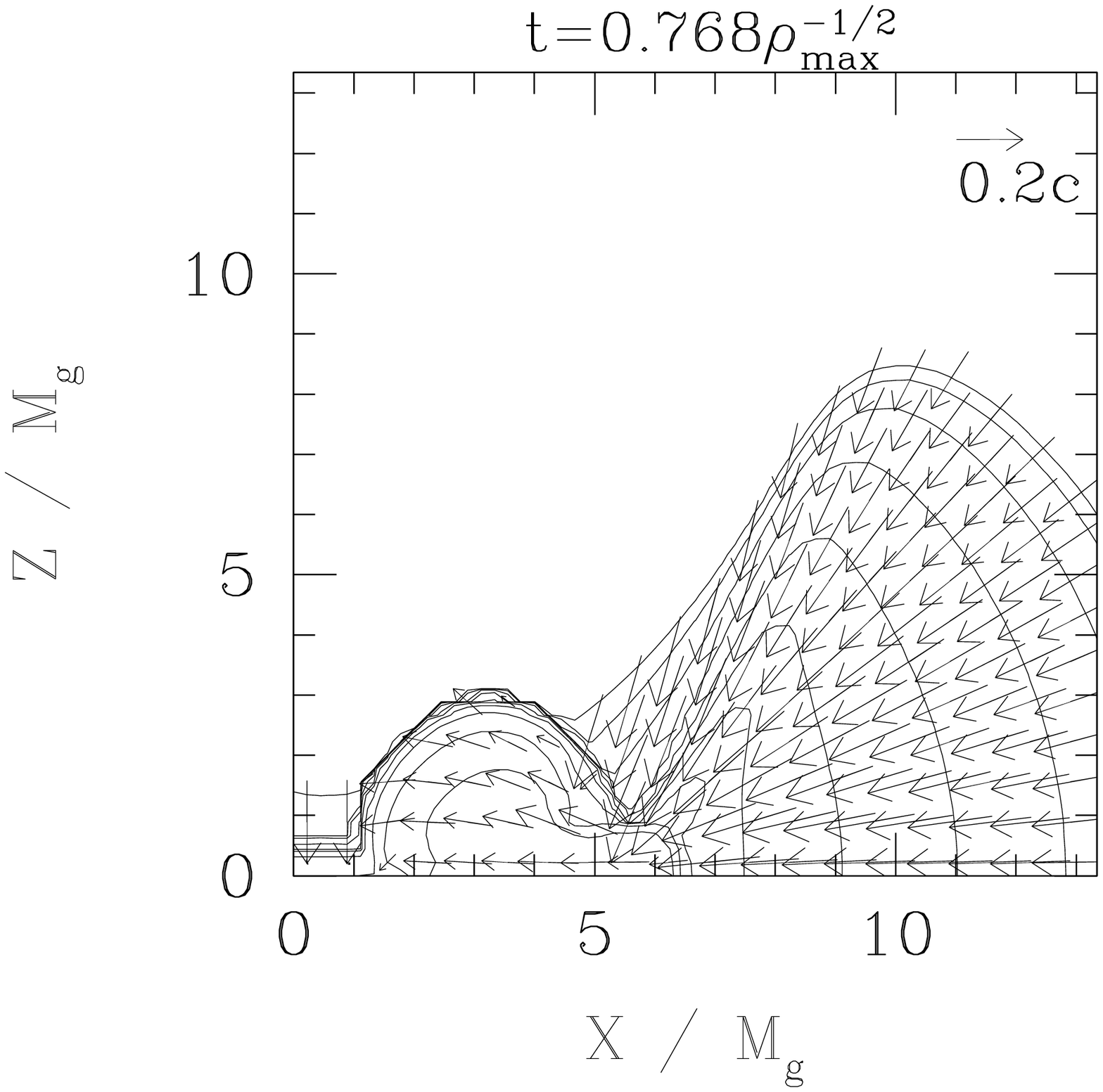}
\epsfxsize=1.8in
\leavevmode
\epsffile{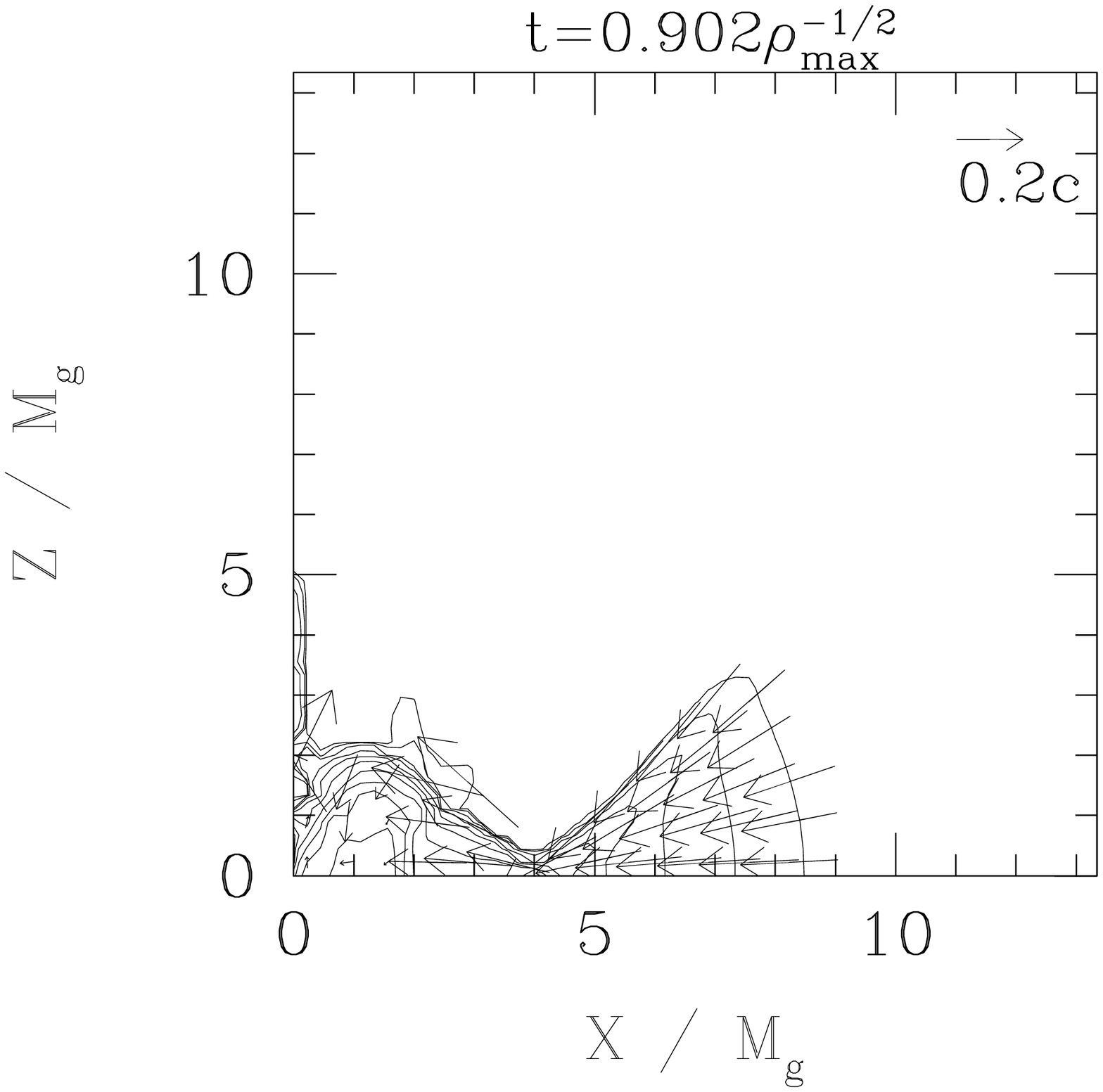}
\epsfxsize=1.8in
\leavevmode
\epsffile{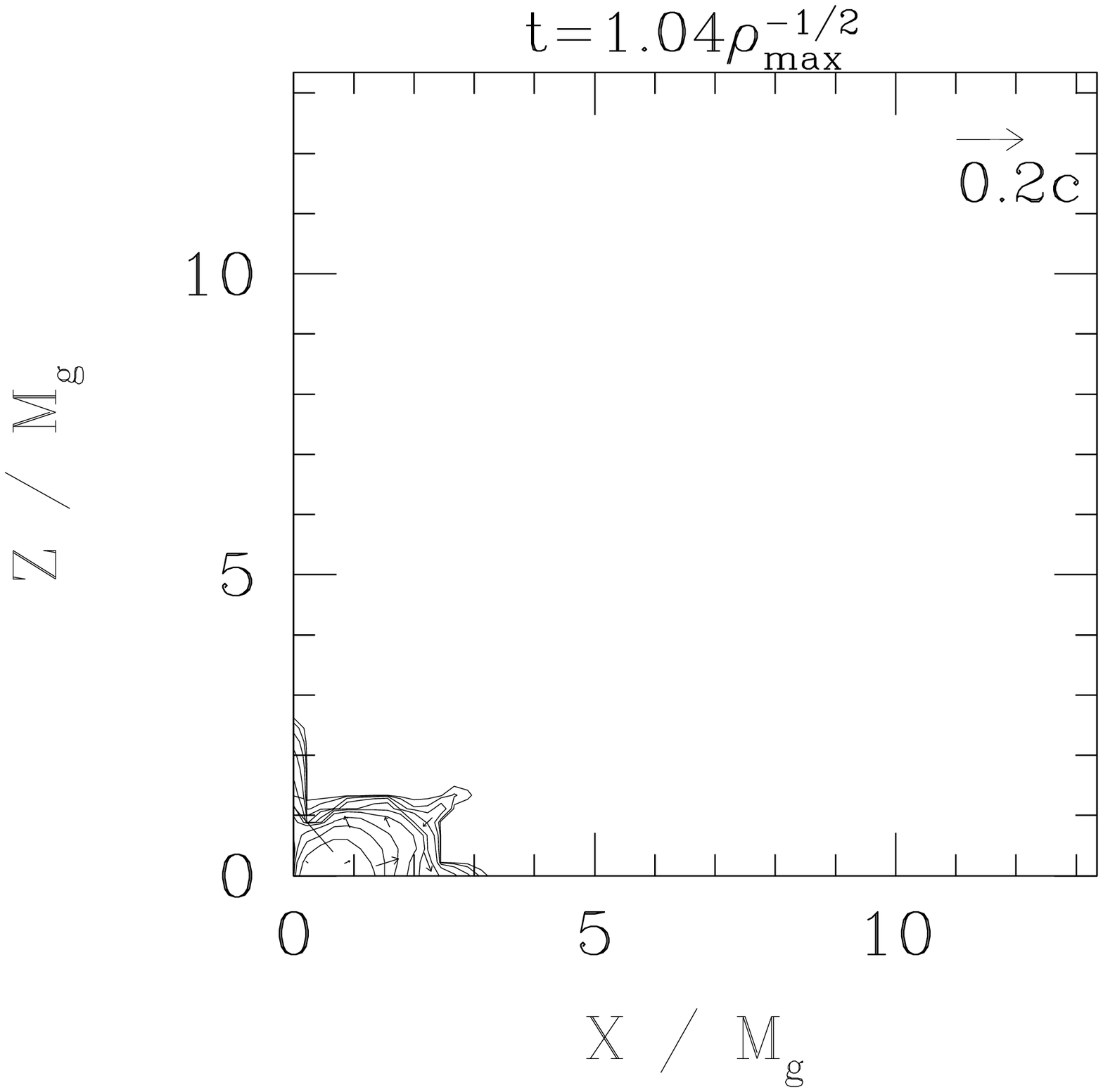} \\
\epsfxsize=1.8in
\leavevmode
\epsffile{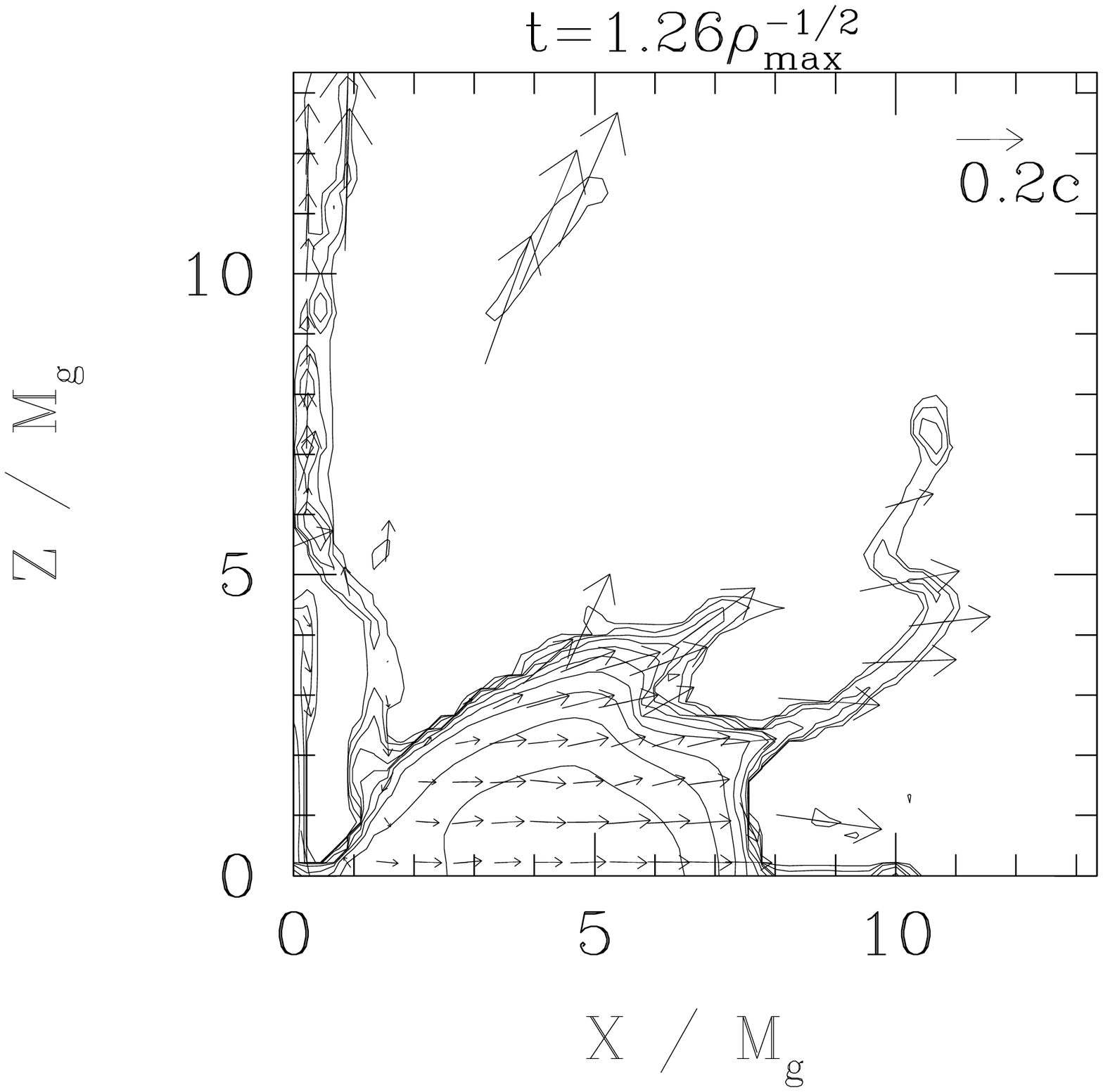}
\epsfxsize=1.8in
\leavevmode
\epsffile{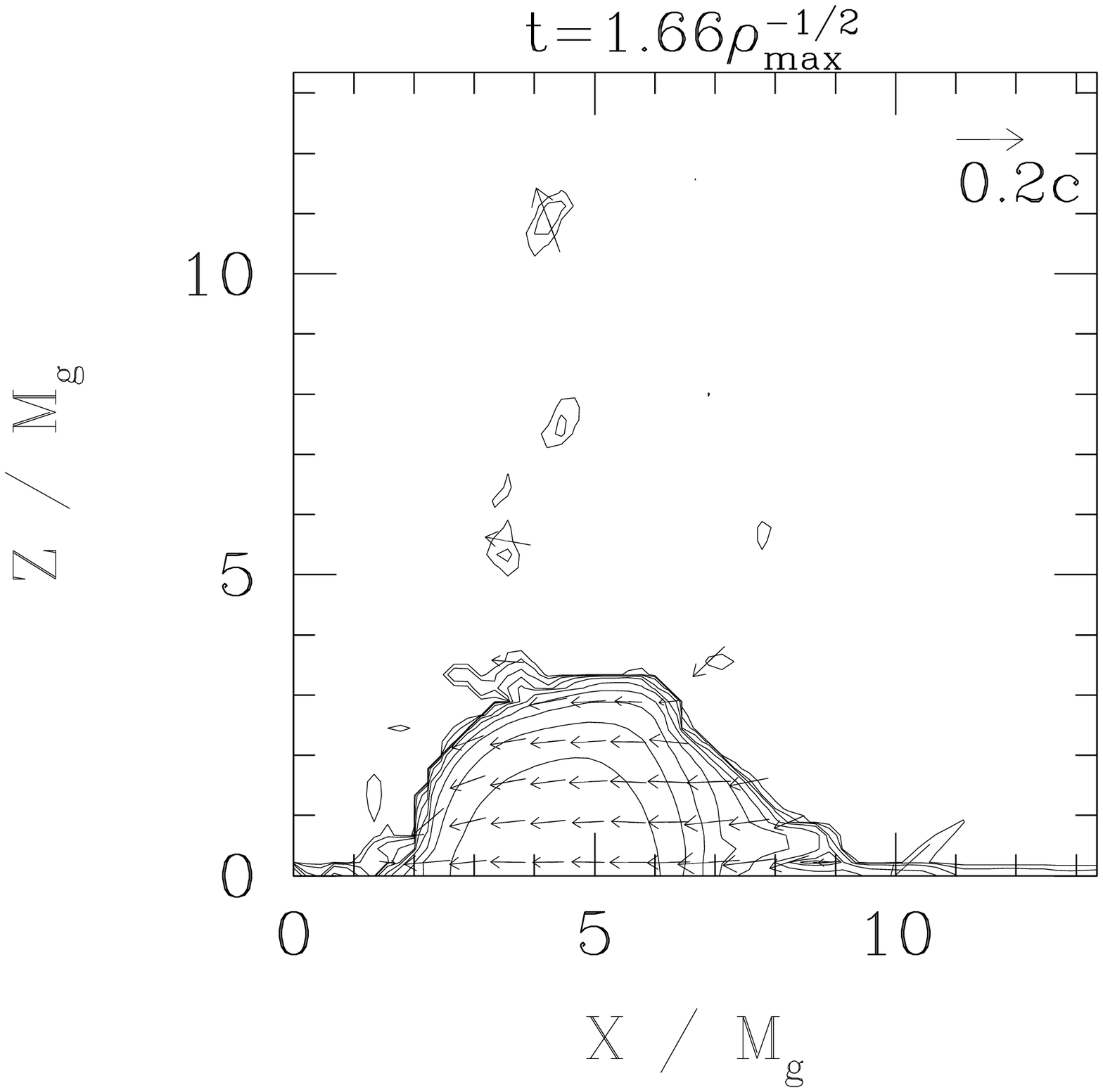}
\epsfxsize=1.8in
\leavevmode
\epsffile{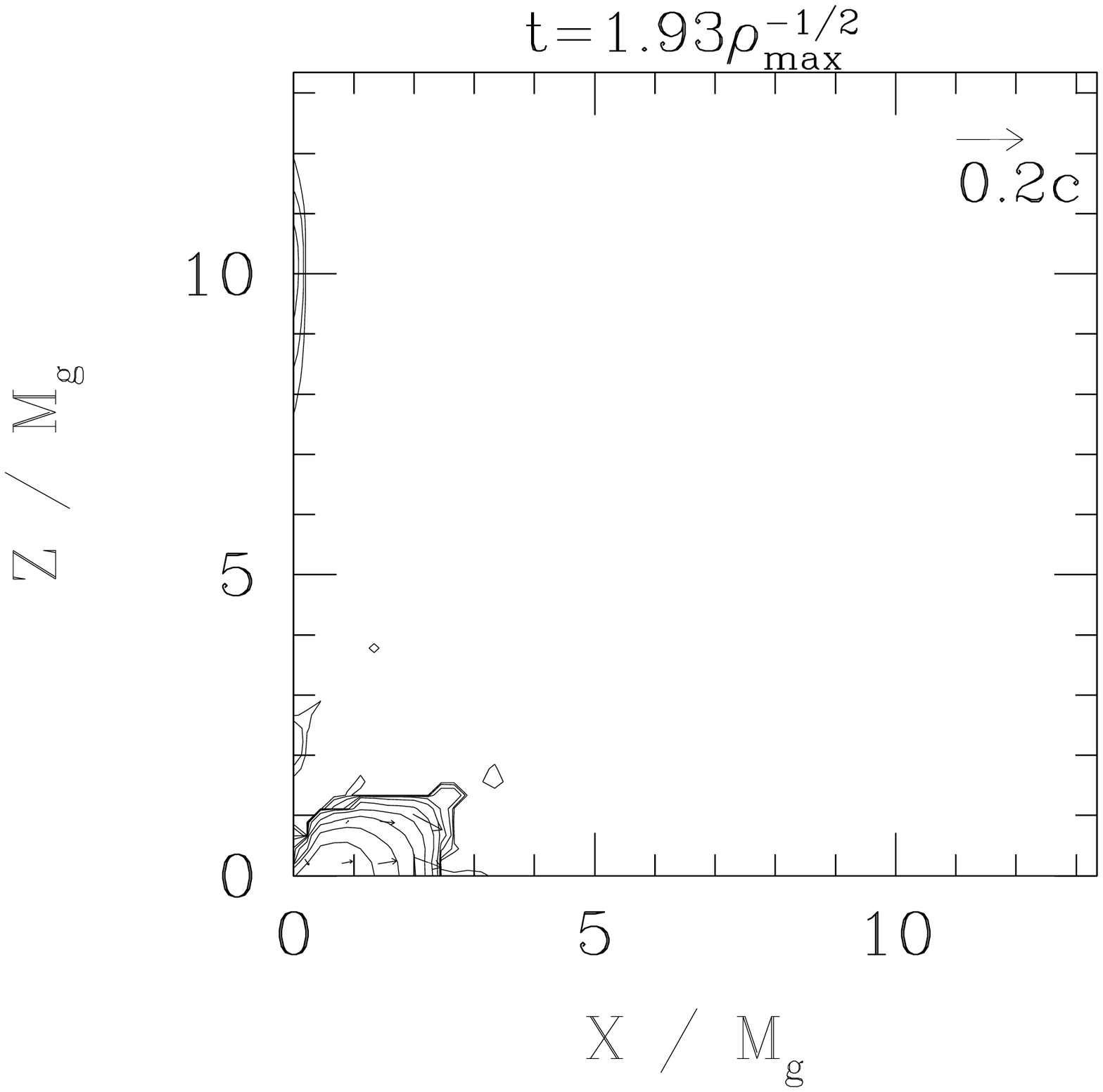}
\end{center}
\caption{The same as Fig.~4, but 
for model (H) with $K^{-1/2}=6.32$. The contour lines 
are drawn for $\rho_*/\rho_{*~{\rm max}}=10^{-0.4j}$ 
for $j=0,1,2,\cdots,10$, where $\rho_{*~{\rm max}}$ 
is 1.10, 9.22, 77.7, 99.3, 1781, 4131, 
75.1, 85.3, and 3619 times larger than $\rho_{*~{\rm max}}$ at $t=0$.
The time appears in units of $\rho_{\rm max}^{-1/2}$ at $t=0$. 
} 
\end{figure}
\clearpage
\begin{figure}[t]
\begin{center}
\epsfxsize=1.8in
\leavevmode
\epsffile{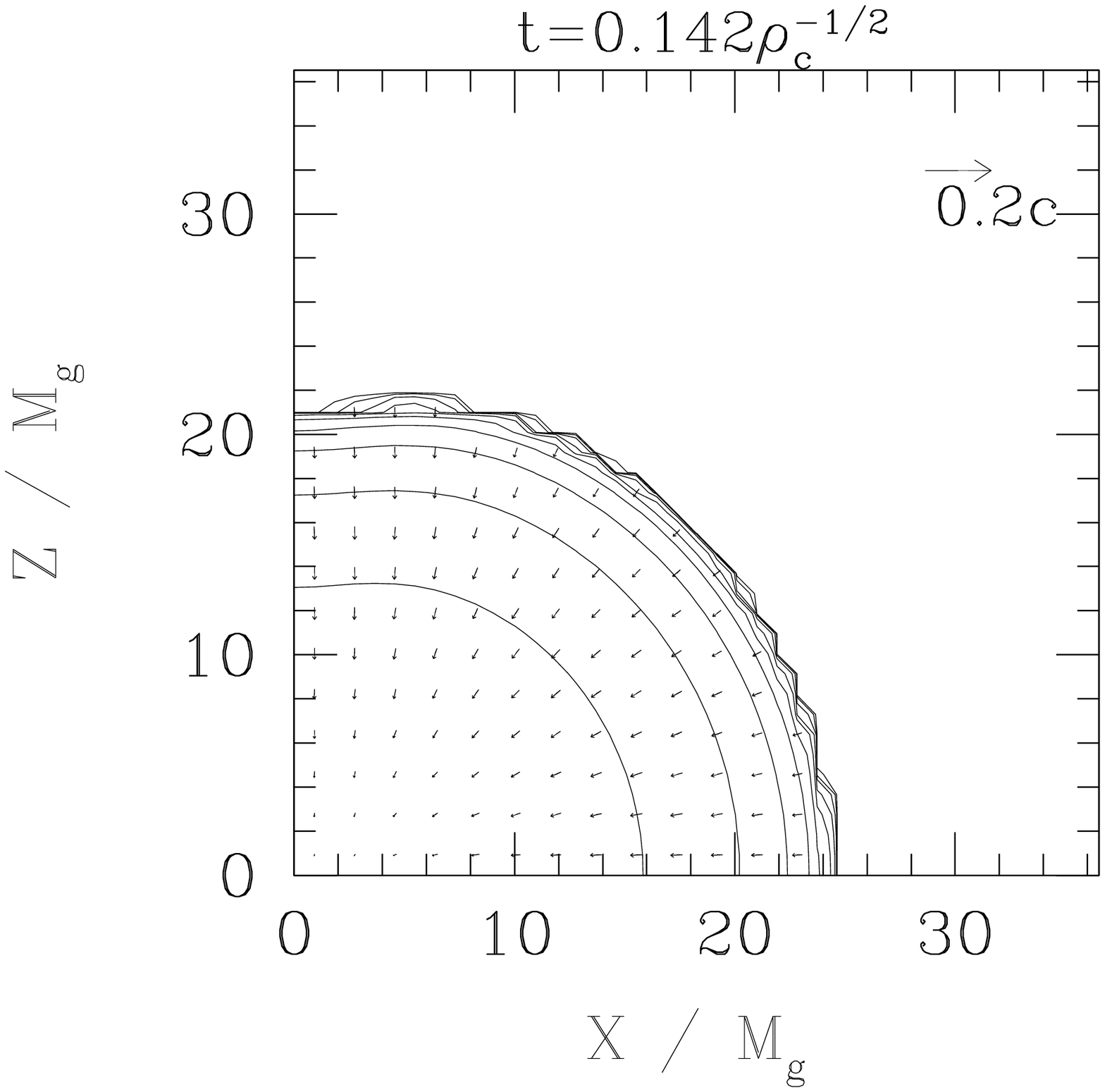}
\epsfxsize=1.8in
\leavevmode
\epsffile{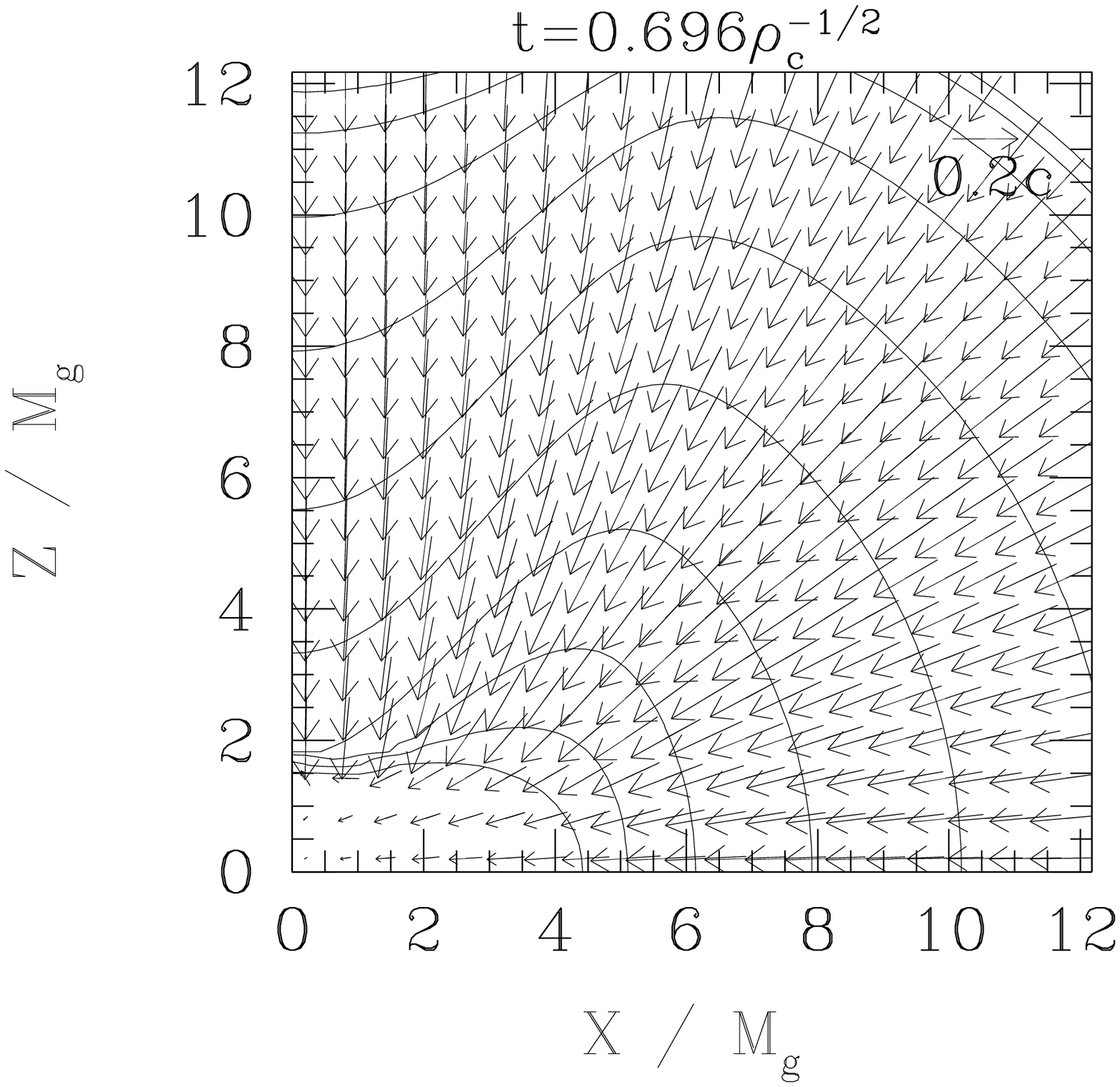}
\epsfxsize=1.8in
\leavevmode
\epsffile{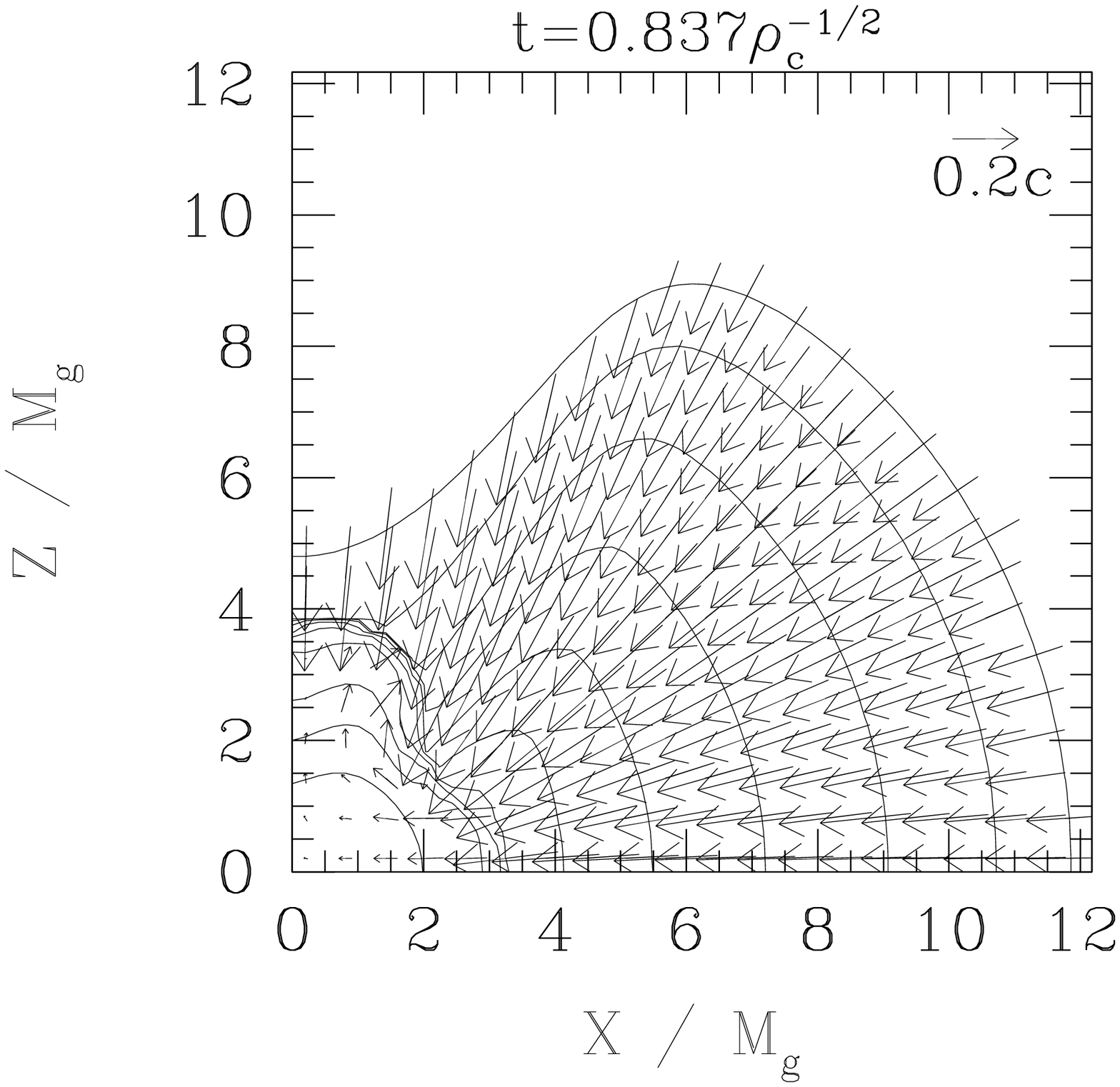} \\
\epsfxsize=1.8in
\leavevmode
\epsffile{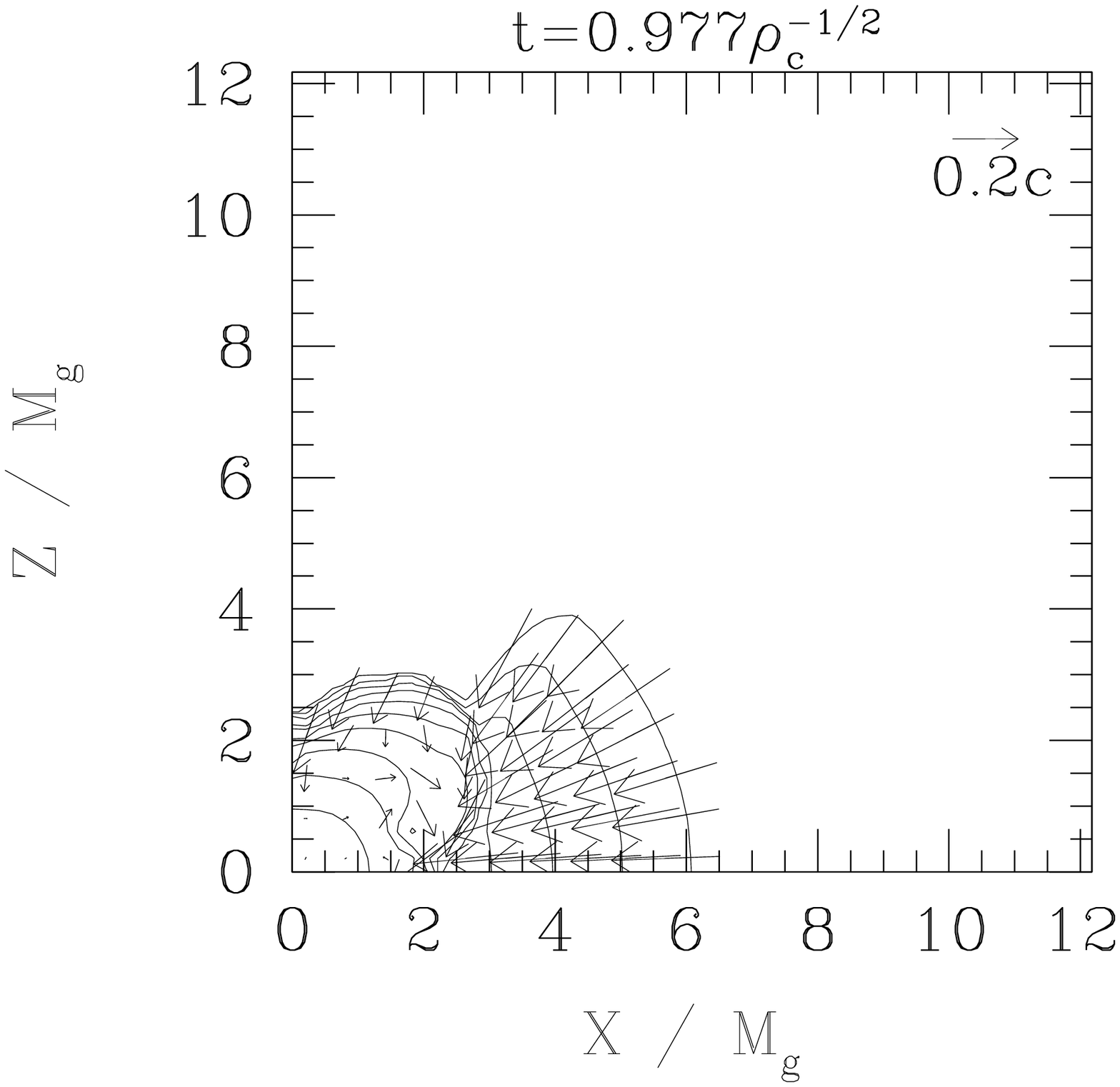}
\epsfxsize=1.8in
\leavevmode
\epsffile{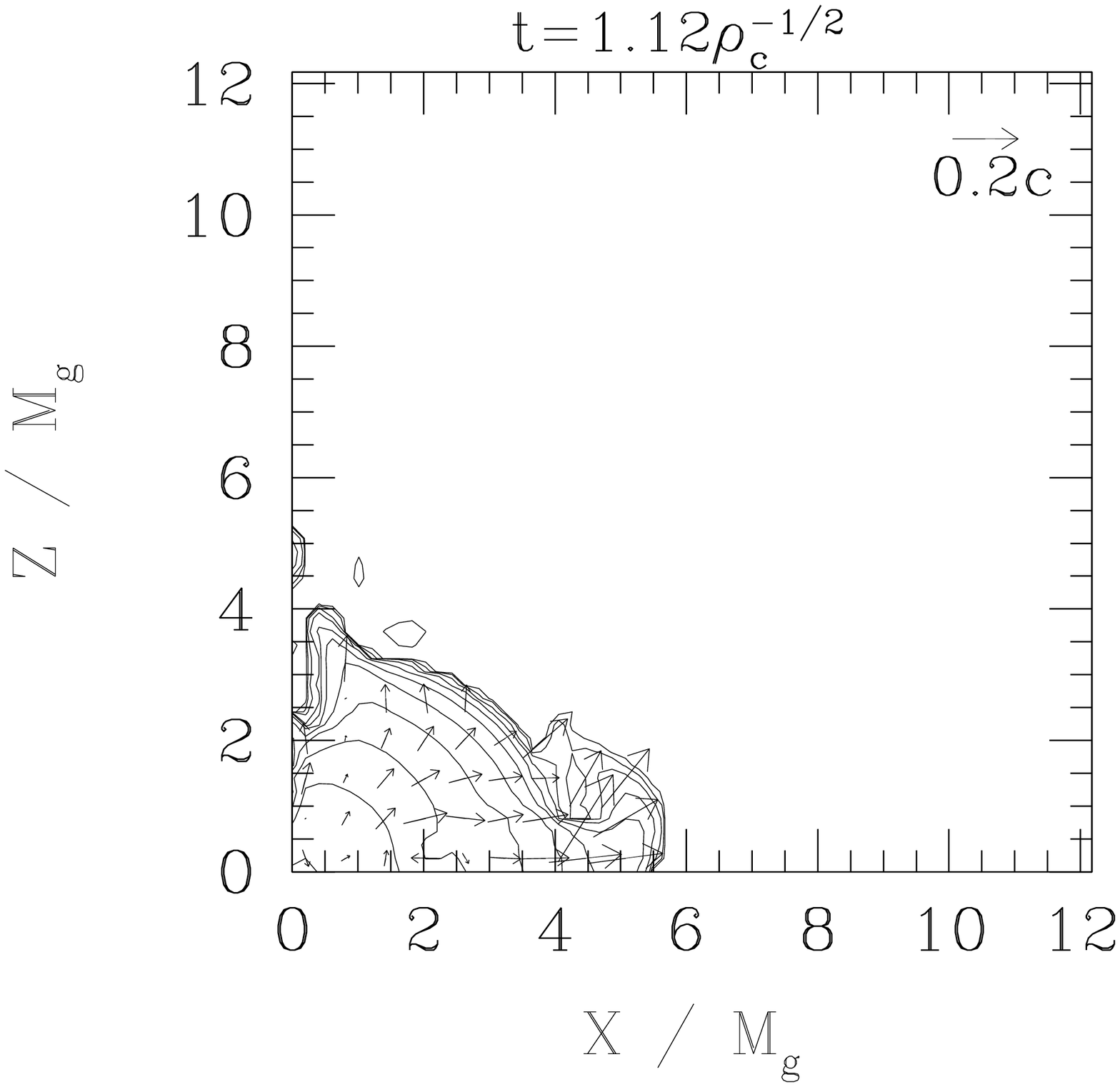}
\epsfxsize=1.8in
\leavevmode
\epsffile{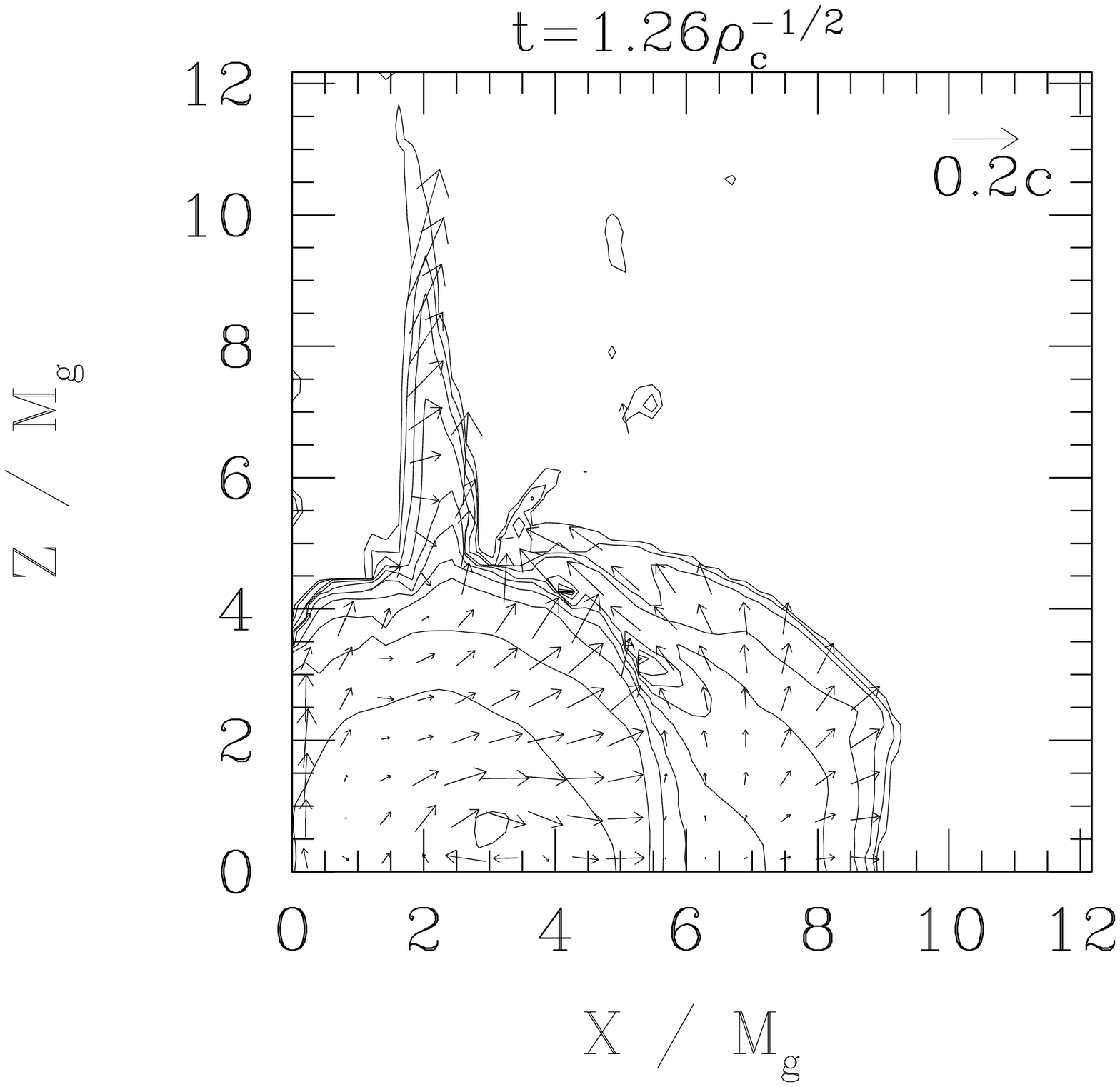} \\
\epsfxsize=1.8in
\leavevmode
\epsffile{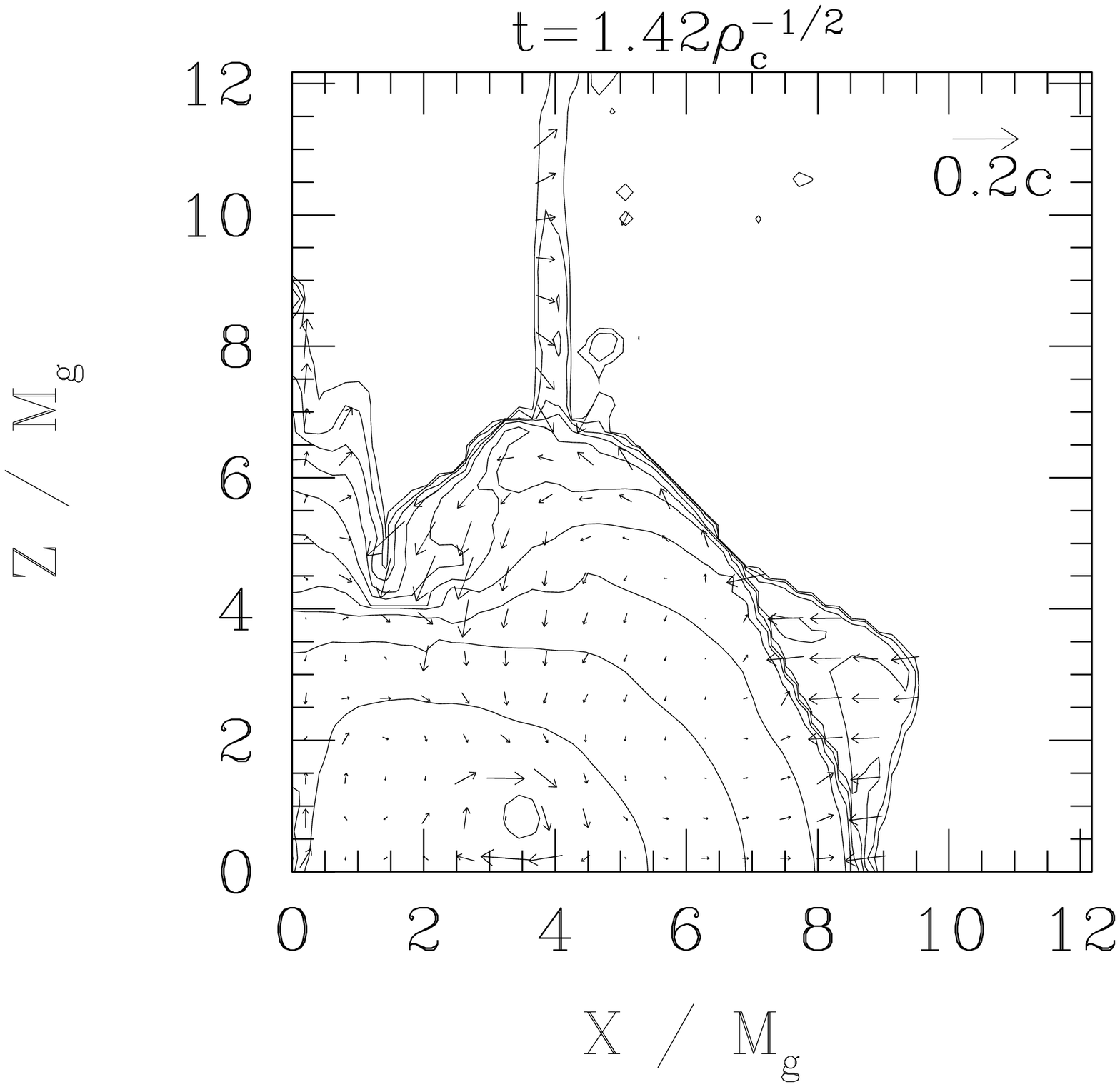}
\epsfxsize=1.8in
\leavevmode
\epsffile{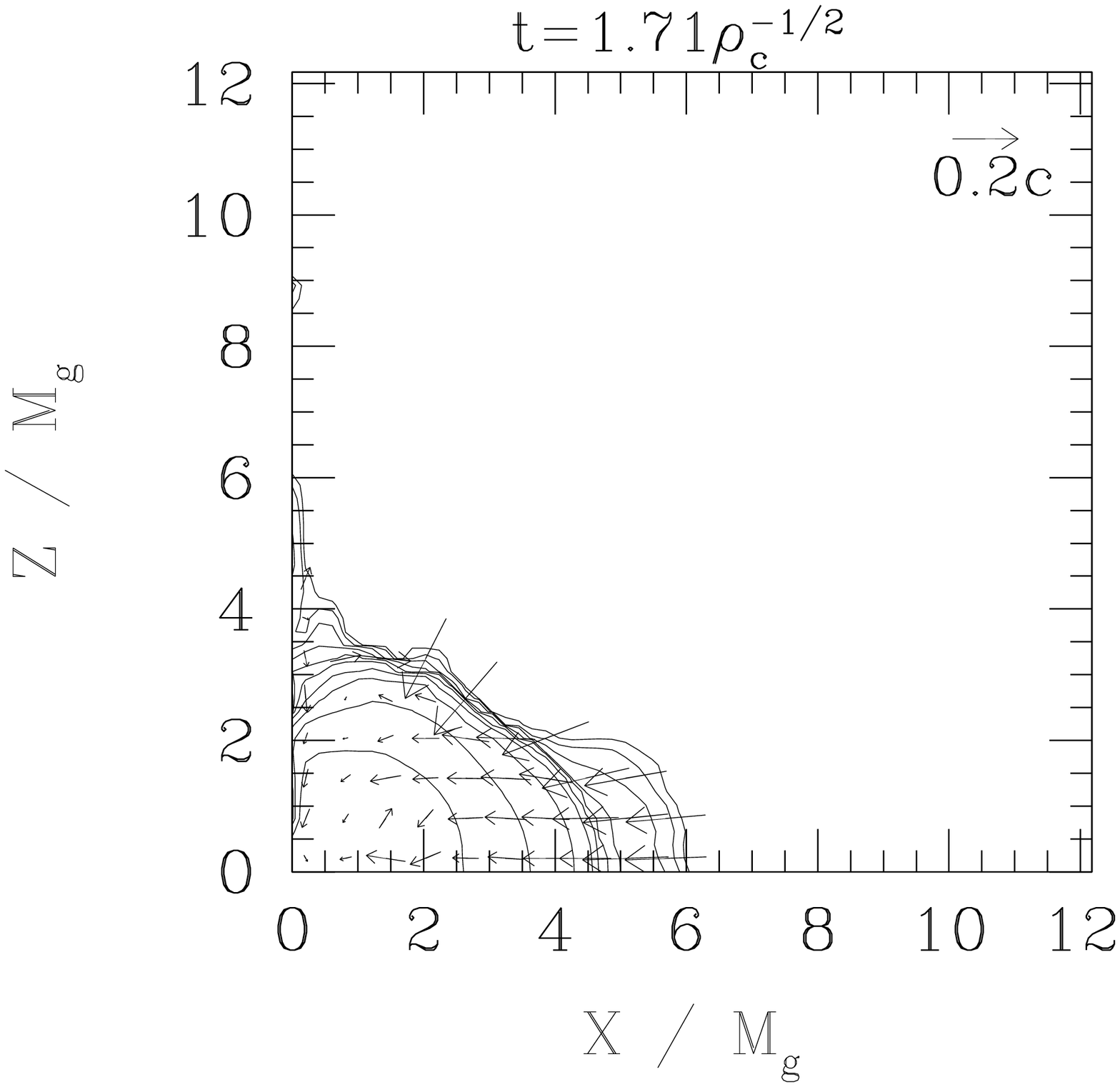}
\epsfxsize=1.8in
\leavevmode
\epsffile{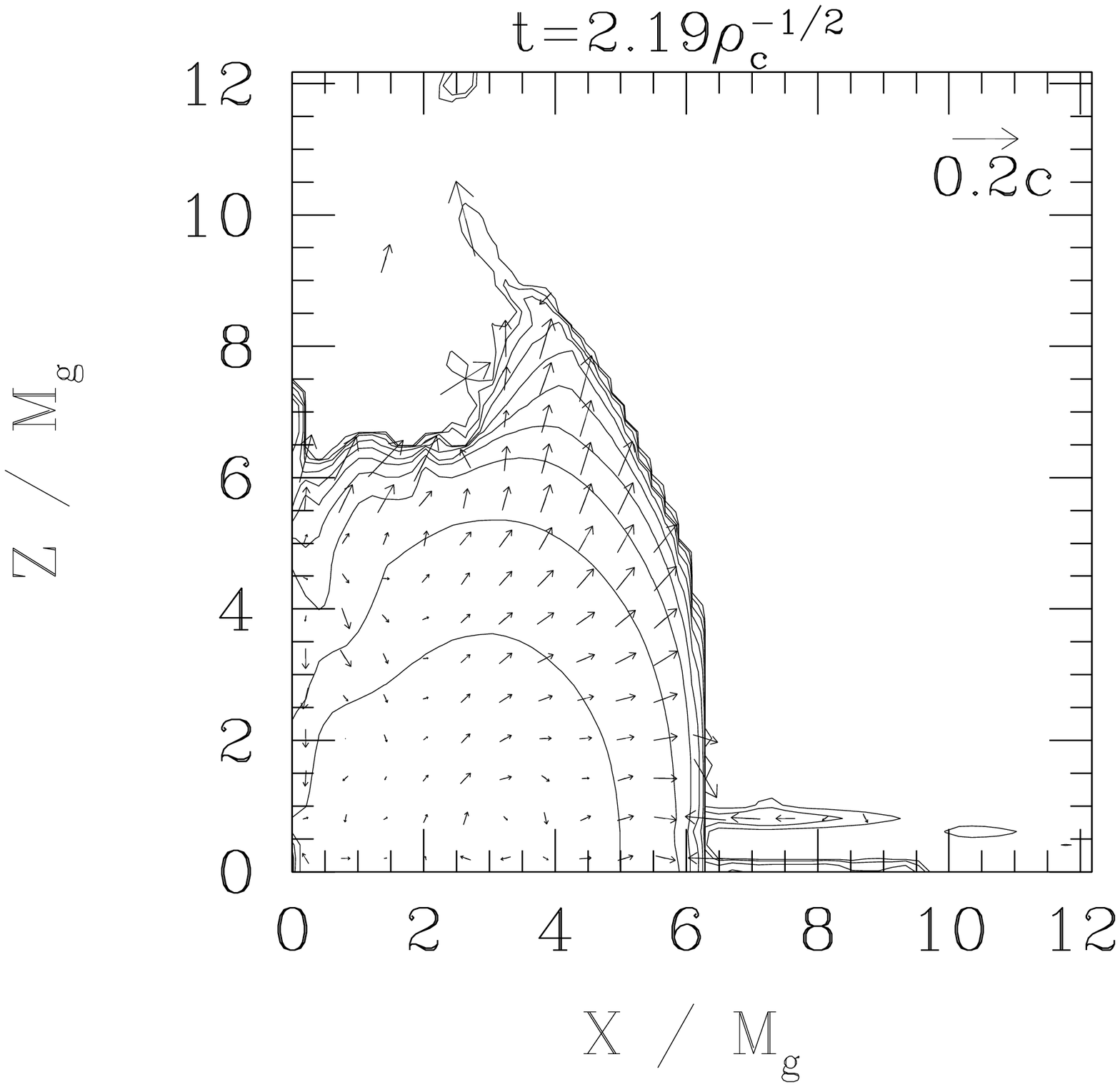}
\end{center}
\caption{The same as Fig.~13, but 
for model (J) with $K^{-1/2}=3.74$. The contour lines 
are drawn for $\rho_*/\rho_{*~{\rm max}}=10^{-0.4j}$ 
for $j=0,1,2,\cdots,10$, where $\rho_{*~{\rm max}}$ 
is 1.06, 61.5, 436, 1554, 610, 63.7, 41.8, 241, and 36.5 
times larger than $\rho_{*~{\rm max}}$ at $t=0$. 
Note that $\rho_c=\rho_{\rm max}$ at $t=0$ in this case. 
} 
\end{figure}

\clearpage
\begin{figure}[t]
\begin{center}
\epsfxsize=3.5in
\leavevmode
\epsffile{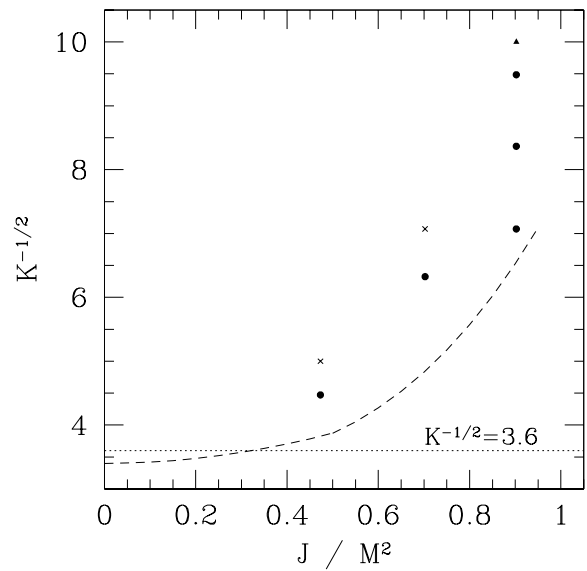}
\end{center}
\caption{The same as Fig.~1, but for 
non-adiabatic collapse with rigidly rotating initial data. 
The dashed curve represents the approximate threshold for black hole 
formation in adiabatic collapse. 
}
\end{figure}

\begin{figure}[t]
\begin{center}
\epsfxsize=3.5in
\leavevmode
\epsffile{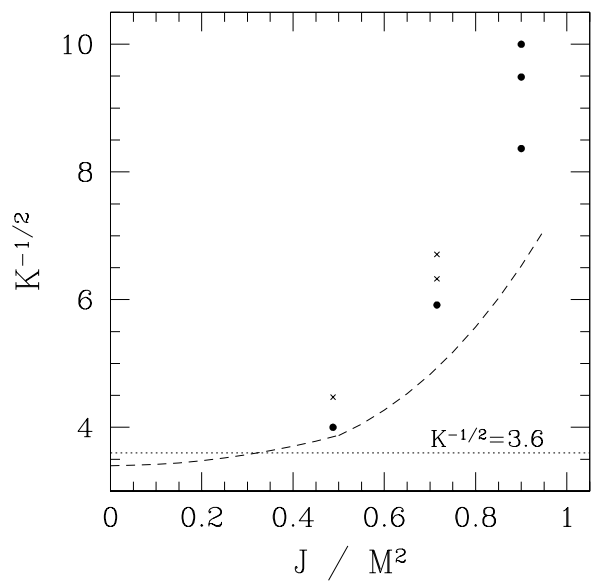}
\end{center}
\caption{The same as Fig.~1, but for 
non-adiabatic collapse for differentially rotating initial data 
with $A=\varpi_e/3$. 
The dashed curve represents the approximate threshold for black hole 
formation in adiabatic collapse. 
}
\end{figure}

\clearpage
\begin{figure}[t]
\begin{center}
\epsfxsize=1.8in
\leavevmode
\epsffile{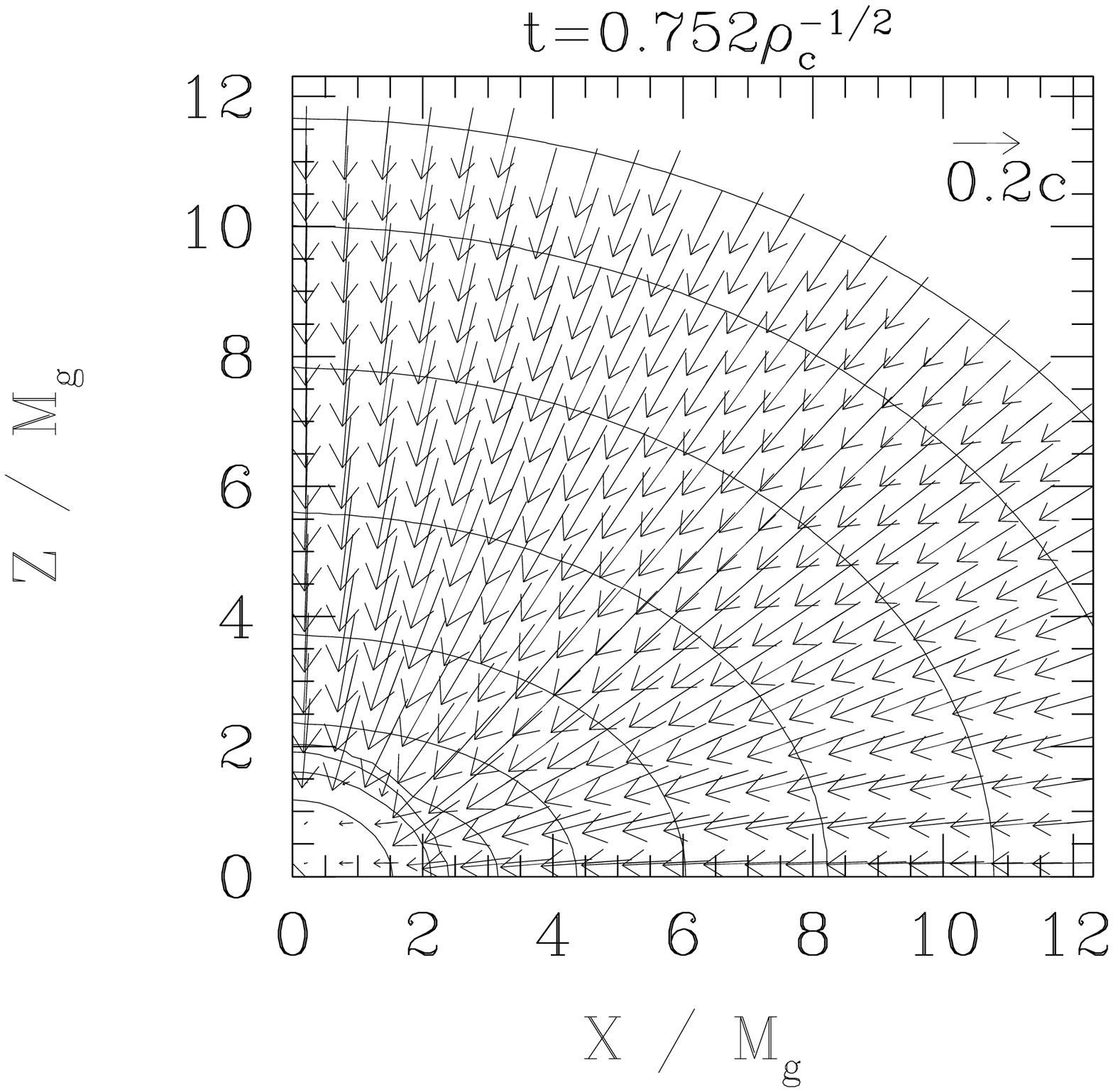}
\epsfxsize=1.8in
\leavevmode
\epsffile{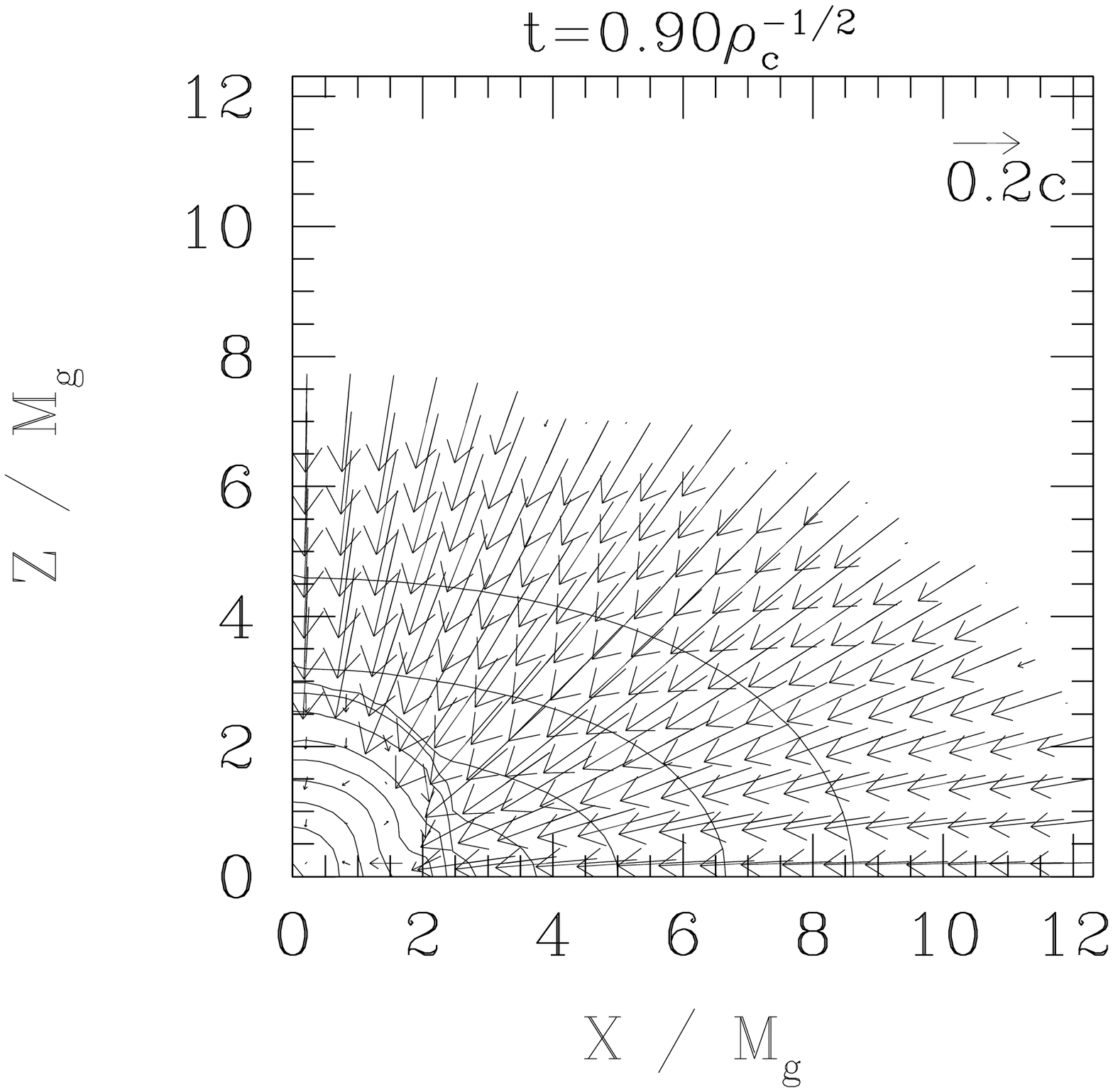}
\epsfxsize=1.8in
\leavevmode
\epsffile{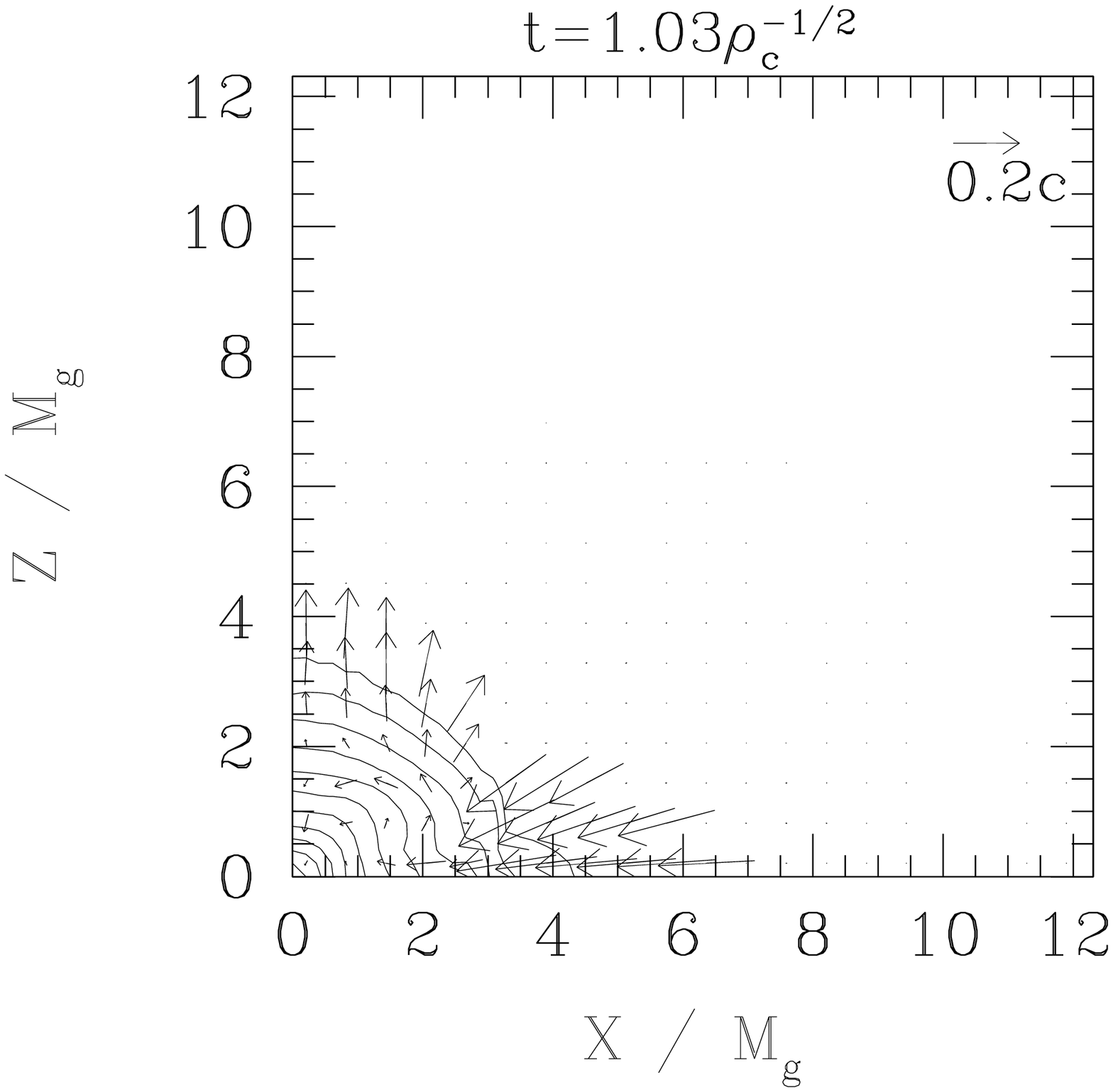}\\
\epsfxsize=1.8in
\leavevmode
\epsffile{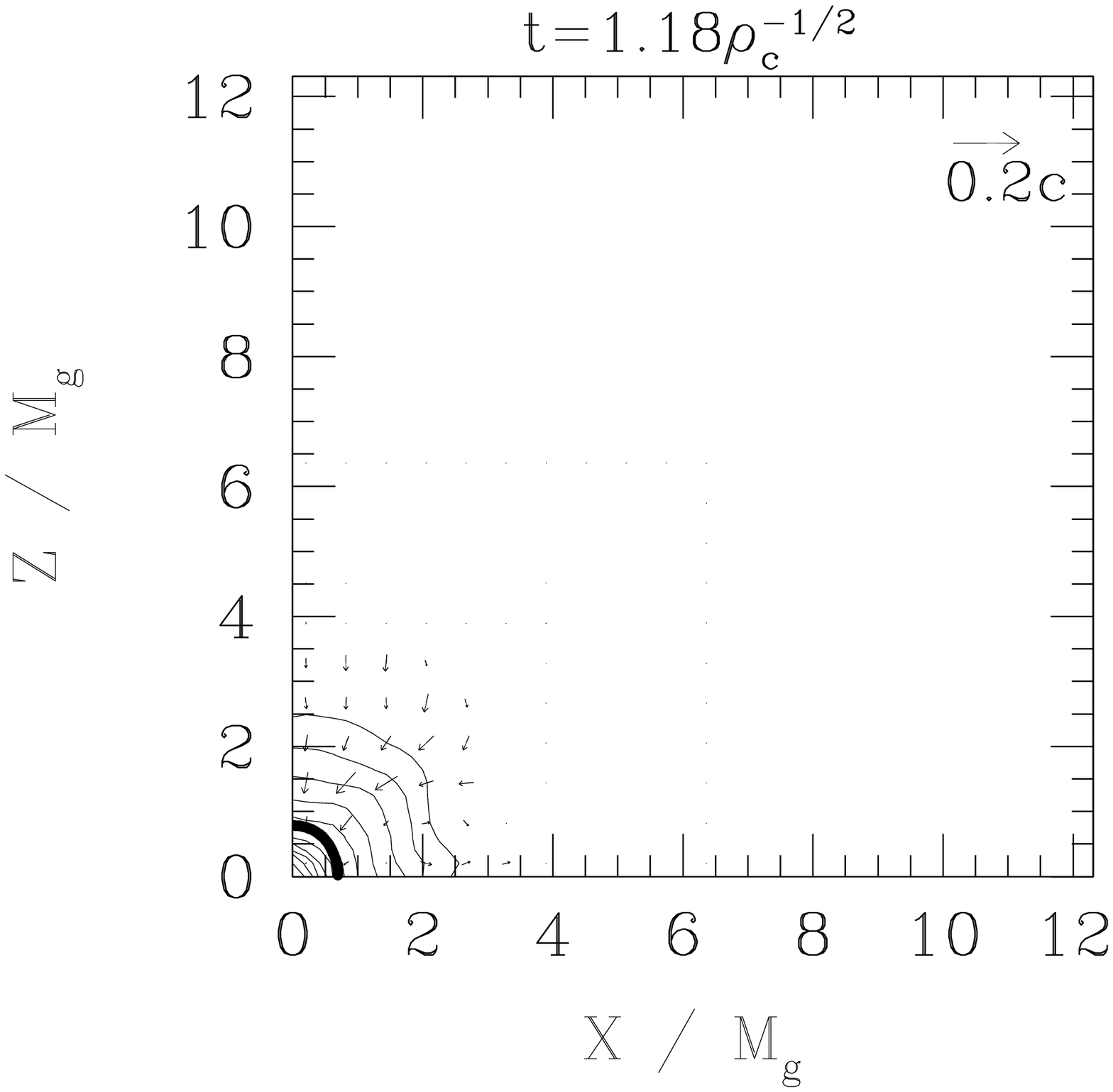}
\epsfxsize=1.8in
\leavevmode
\epsffile{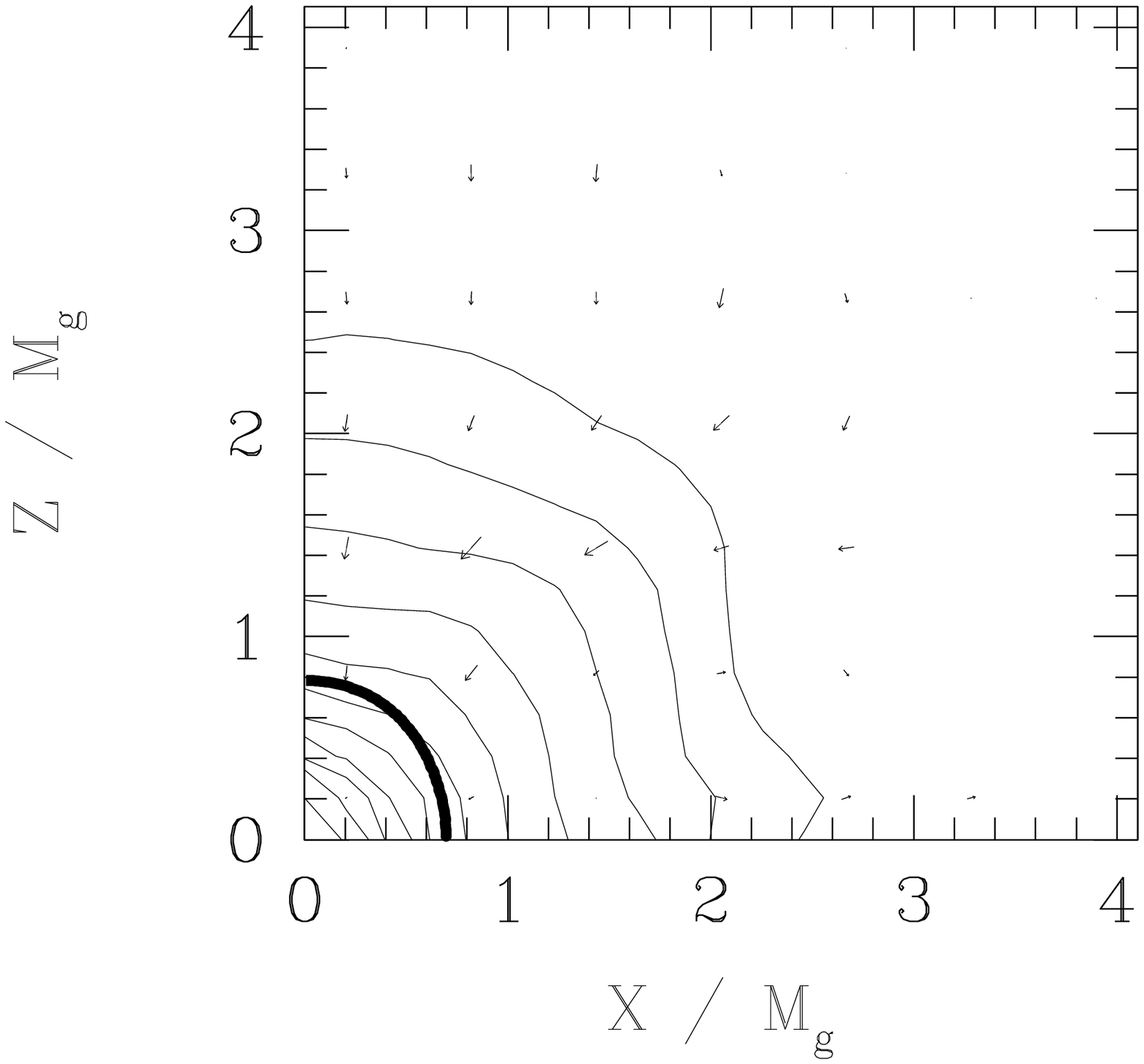}
\end{center}
\caption{The same as Fig.~4, but 
for non-adiabatic collapse for 
model (E) with $K^{-1/2}=5.00$. The contour lines 
are drawn for $\rho_*/\rho_{*~{\rm max}}=10^{-0.4j}$ 
for $j=0,1,2,\cdots,10$, where $\rho_{*~{\rm max}}$ 
is 643, 3814, $2.65\times 10^4$, and $2.26\times 10^5$ times larger than 
$\rho_{*~{\rm max}}$ at $t=0$.
The last panel is the magnification of the 4th panel. 
The thick solid curve in the last two panels indicates 
the apparent horizon. }
\end{figure}

\clearpage
\begin{figure}[t]
\begin{center}
\epsfxsize=1.8in
\leavevmode
\epsffile{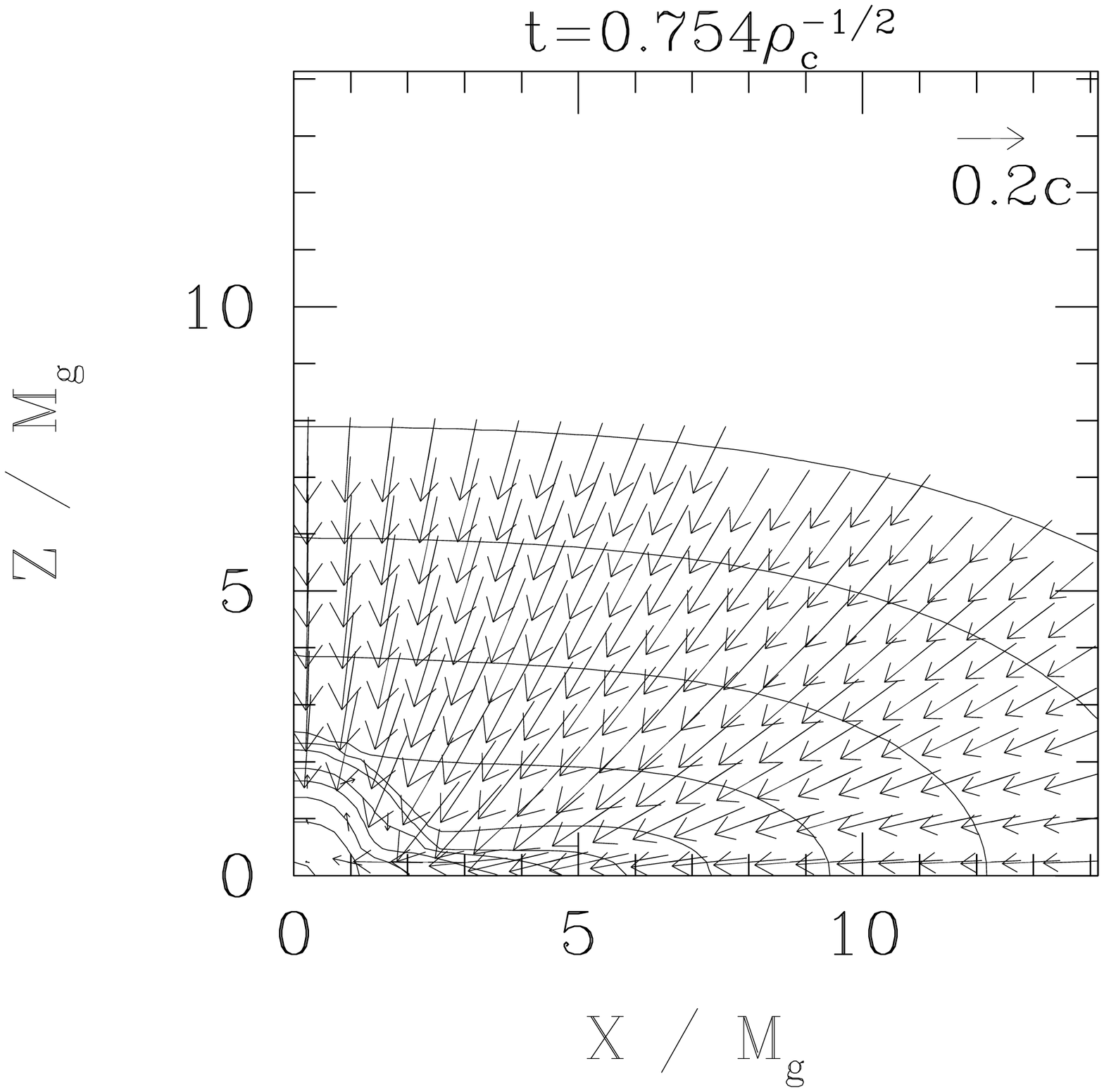}
\epsfxsize=1.8in
\leavevmode
\epsffile{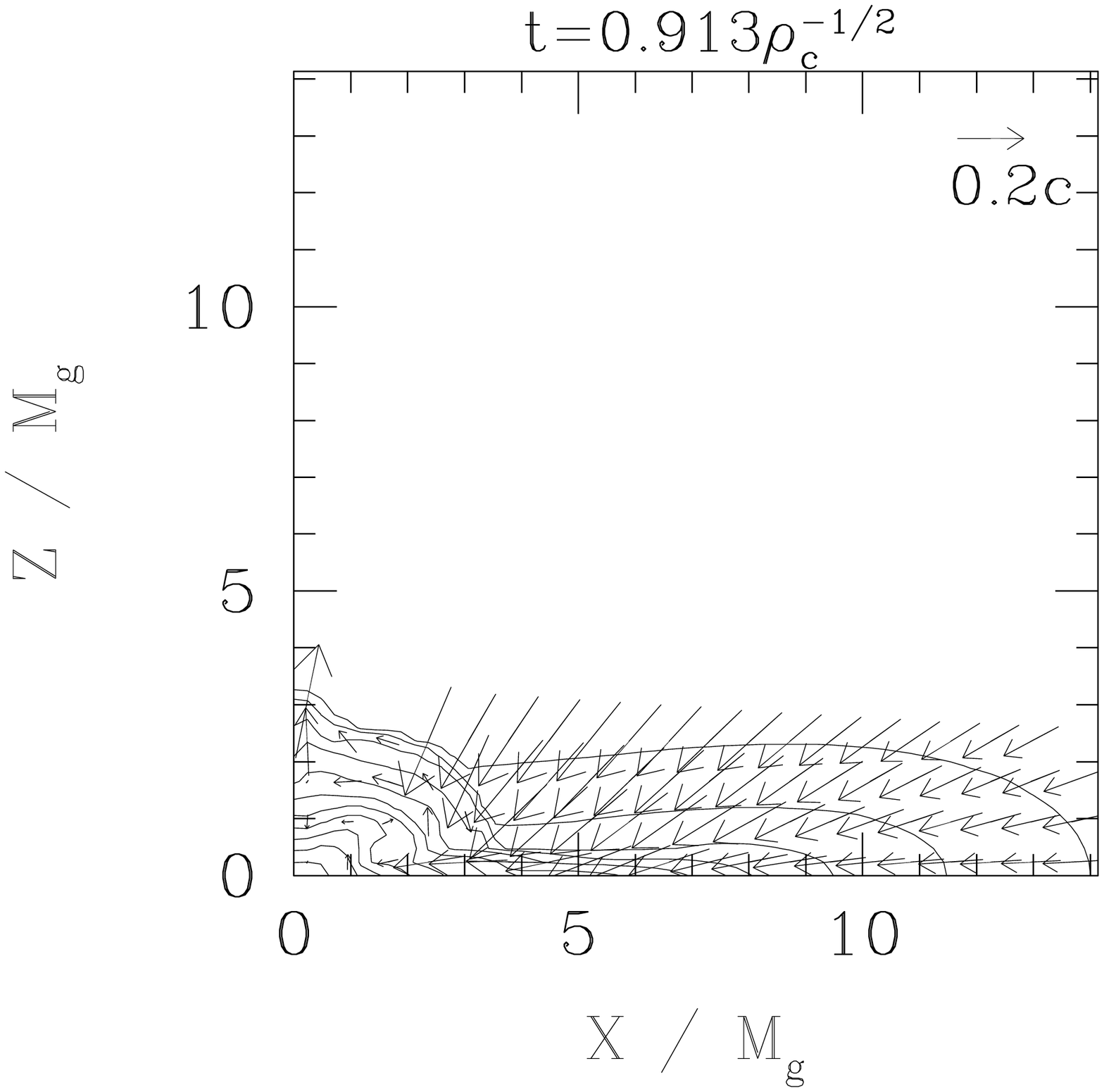}
\epsfxsize=1.8in
\leavevmode
\epsffile{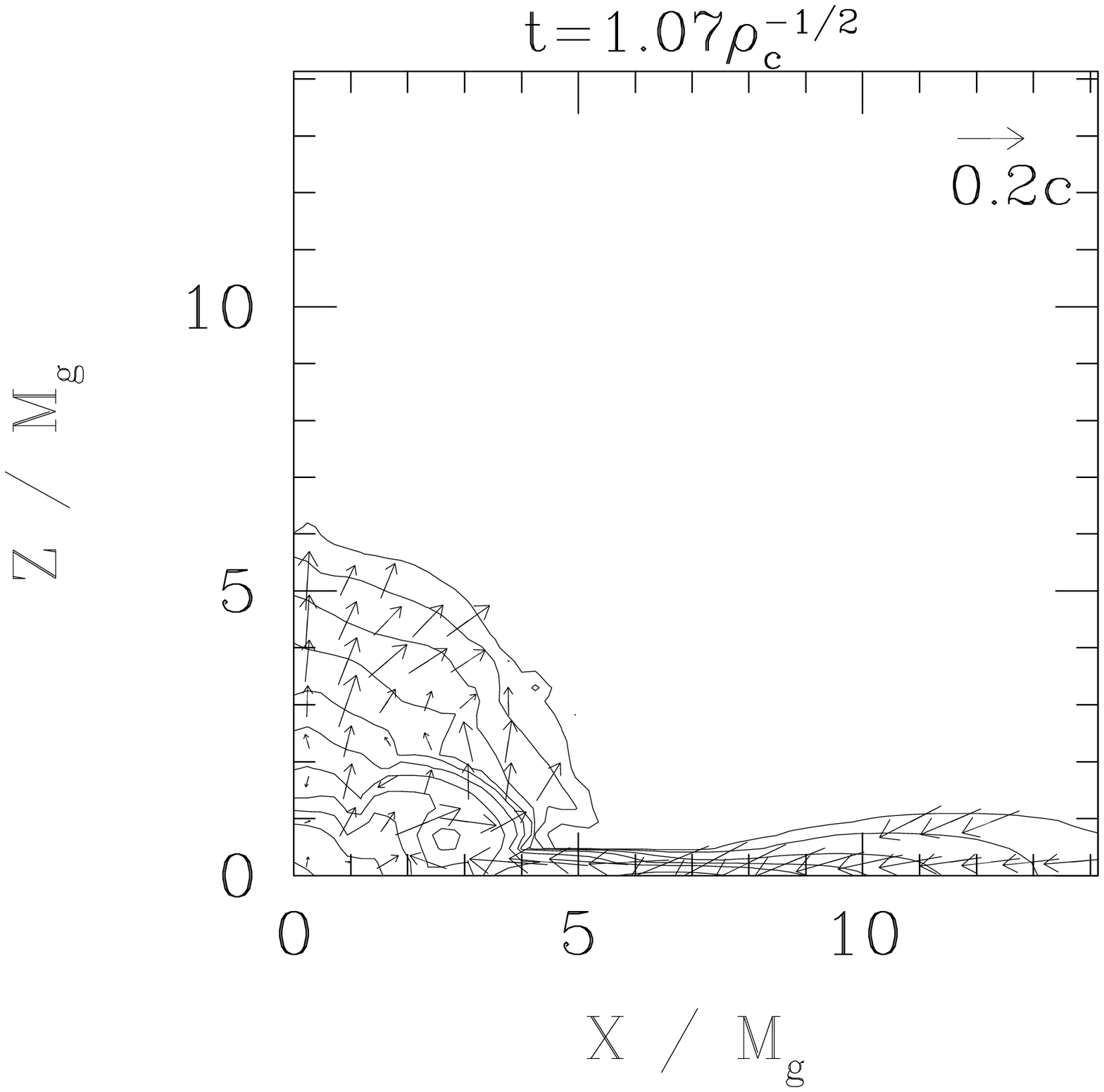}\\
\epsfxsize=1.8in
\leavevmode
\epsffile{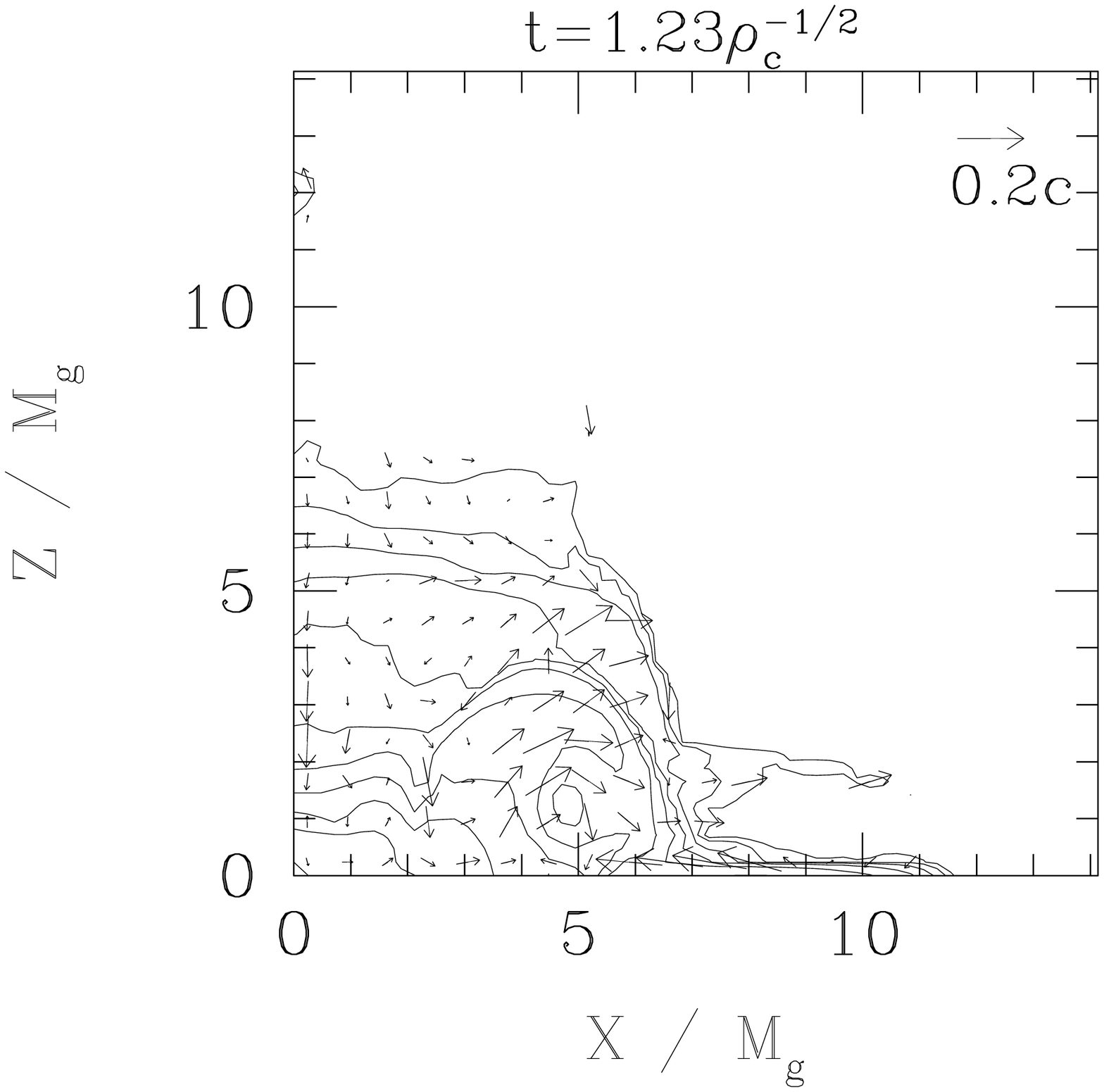}
\epsfxsize=1.8in
\leavevmode
\epsffile{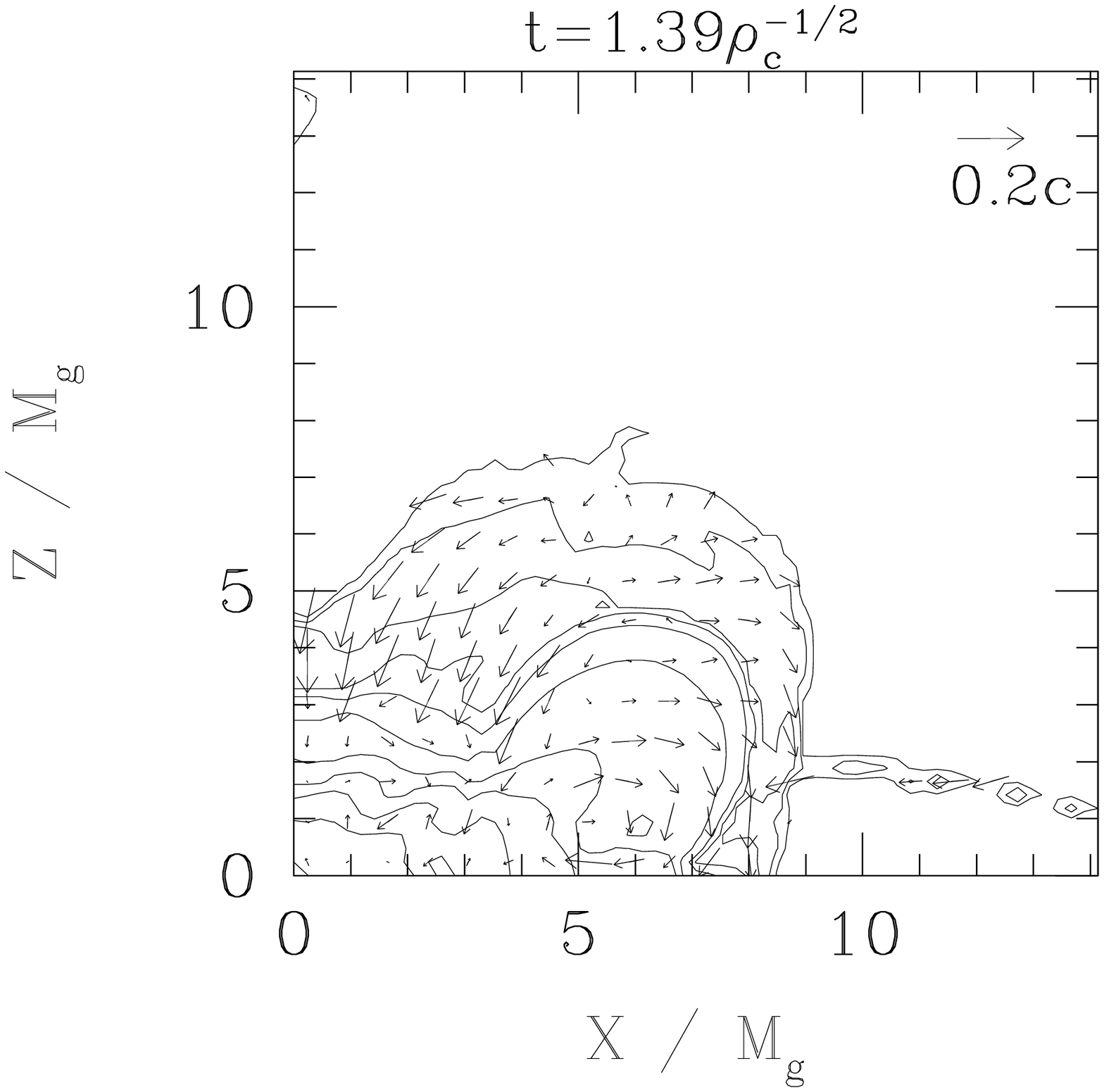}
\epsfxsize=1.8in
\leavevmode
\epsffile{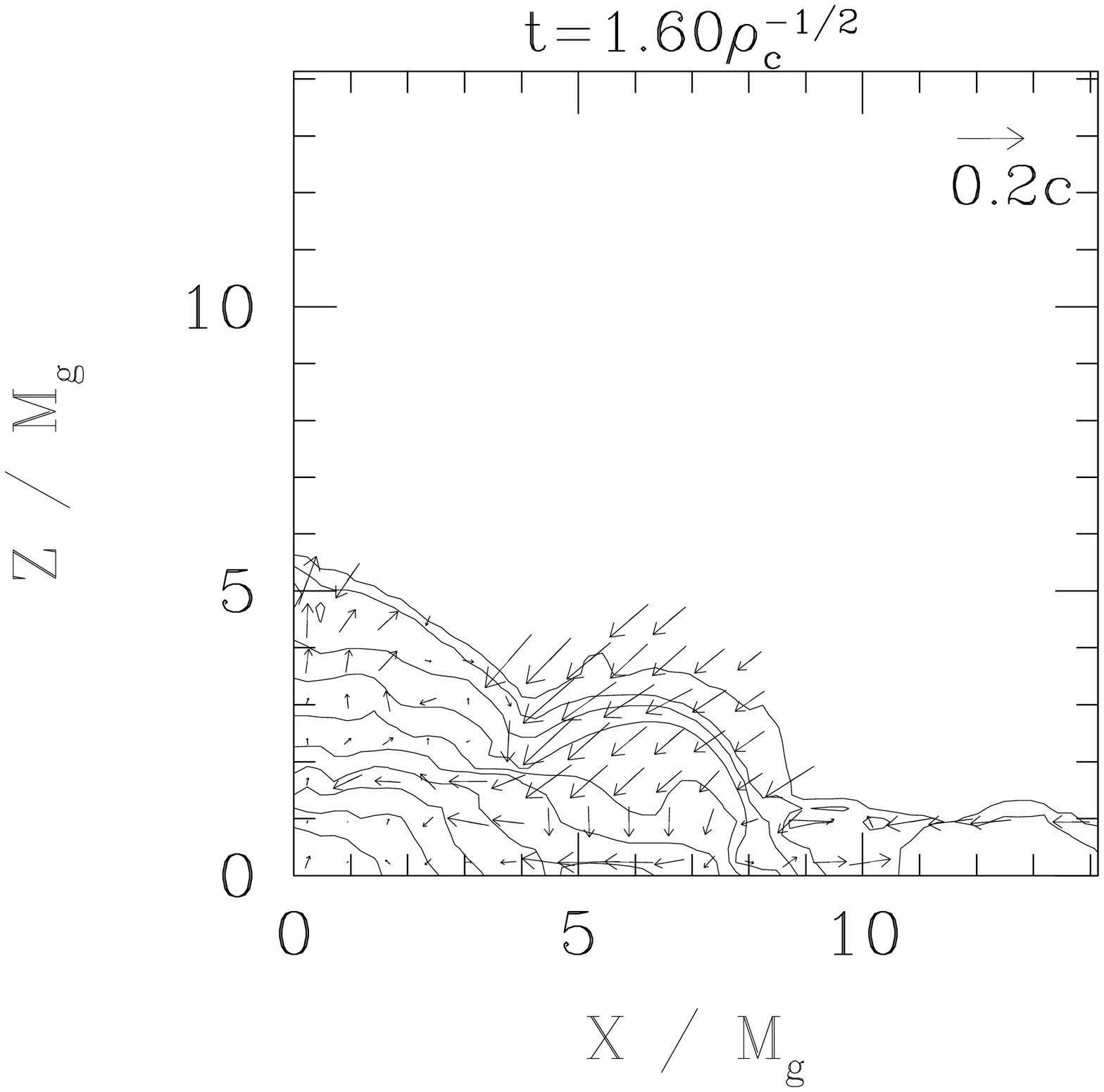}\\
\epsfxsize=1.8in
\leavevmode
\epsffile{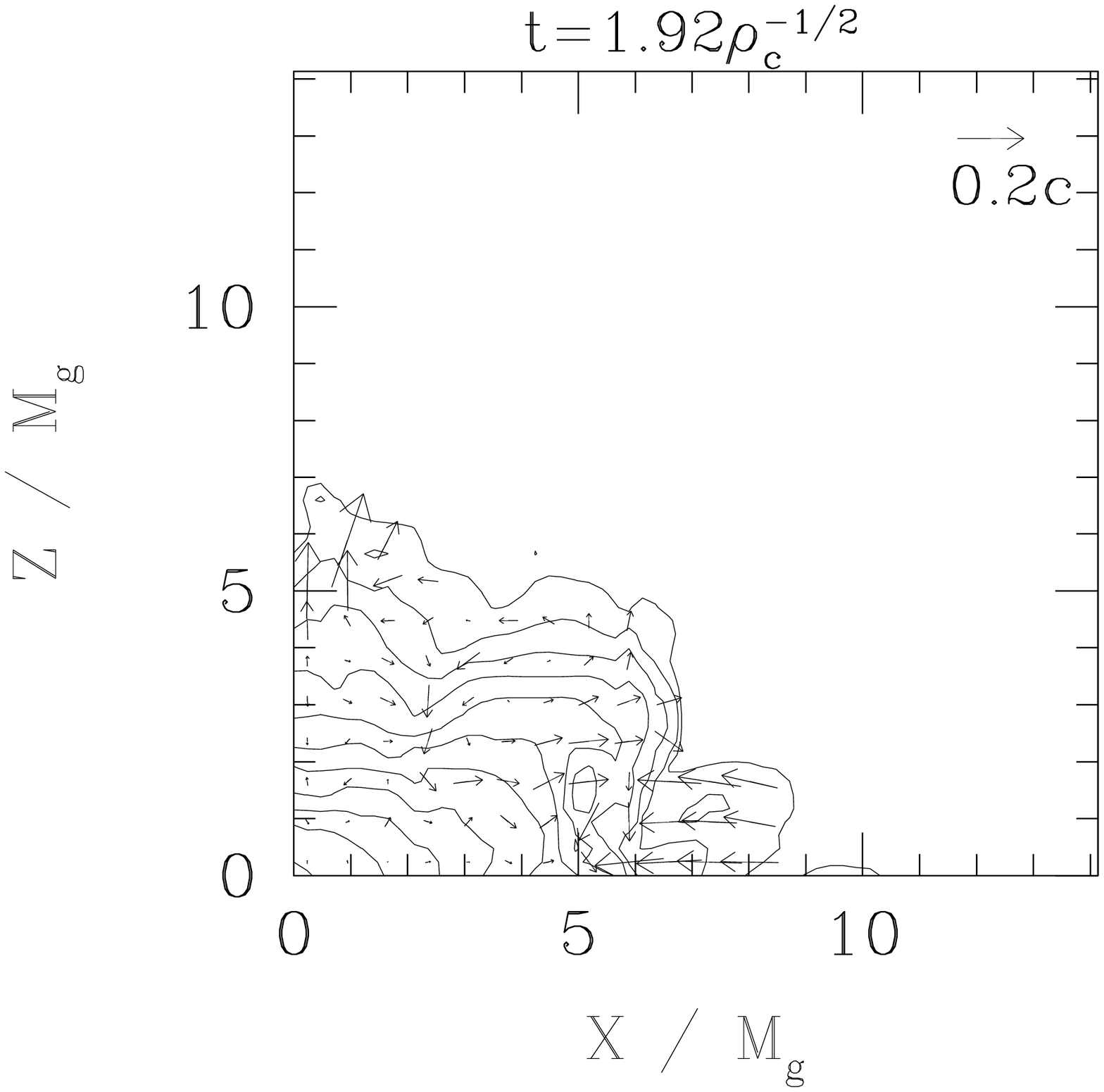}
\epsfxsize=1.8in
\leavevmode
\epsffile{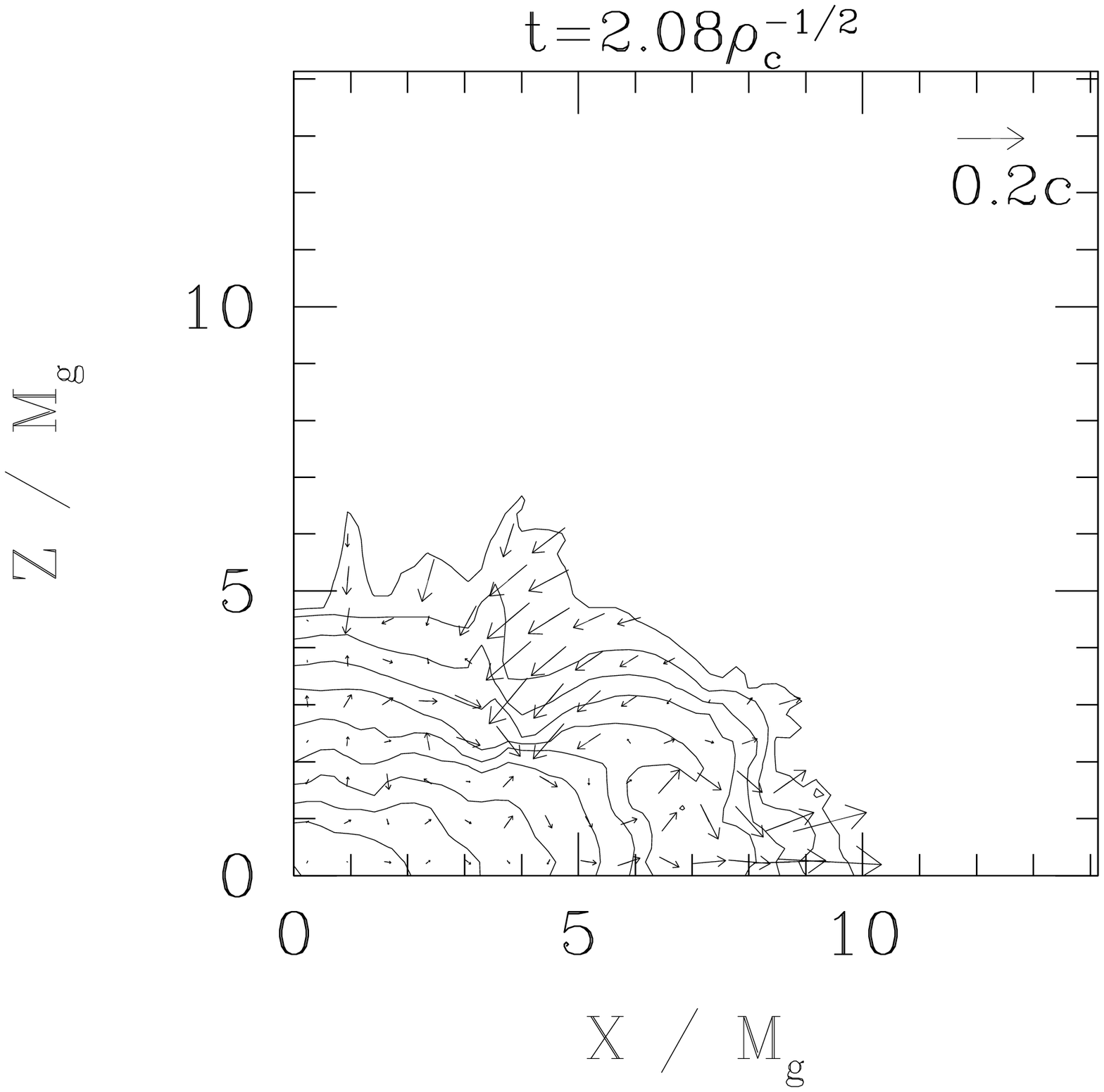}
\end{center}
\caption{The same as Fig.~4, but for non-adiabatic collapse 
for model (C) with $K^{-1/2}=8.37$. The contour lines 
are drawn for $\rho_*/\rho_{*~{\rm max}}=10^{-0.4j}$ 
for $j=0,1,2,\cdots,10$, where $\rho_{*~{\rm max}}$ 
is 1178, $5015$, 1186, 403, 317, 619, 814, and 
436 times larger than $\rho_{*~{\rm max}}$ at $t=0$.
} 
\end{figure}

\clearpage
\begin{figure}[t]
\begin{center}
\epsfxsize=1.8in
\leavevmode
\epsffile{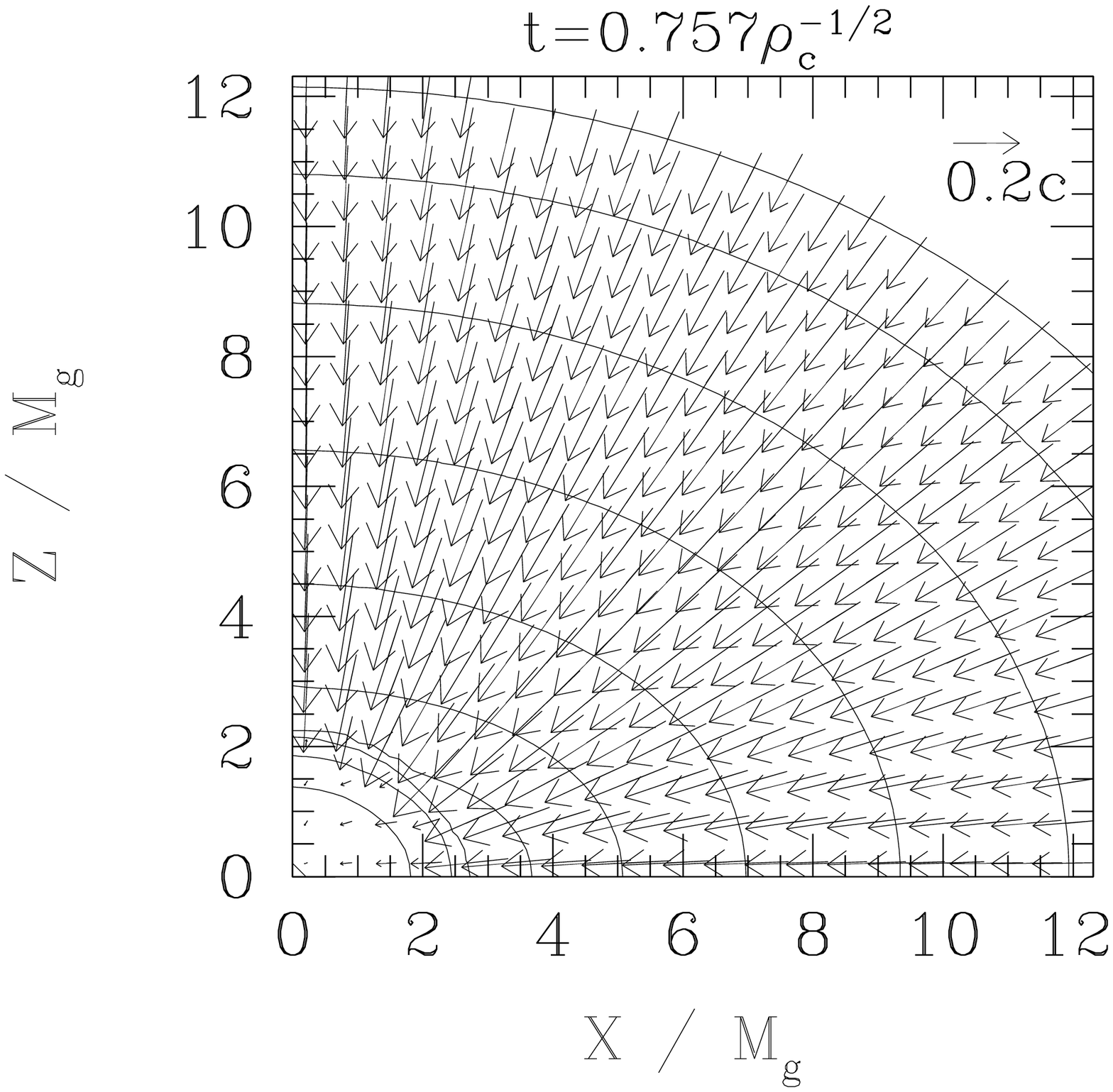}
\epsfxsize=1.8in
\leavevmode
\epsffile{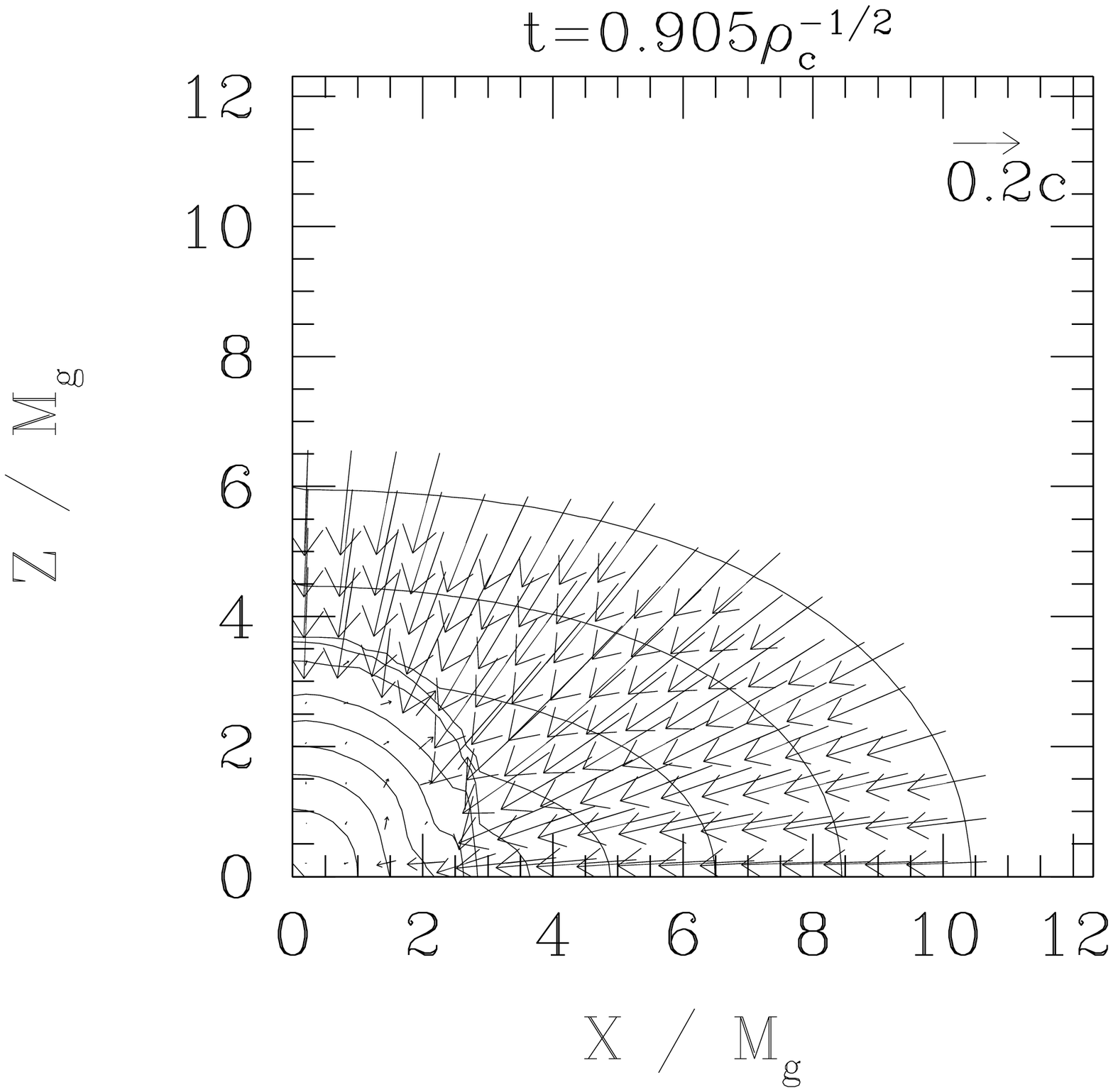}
\epsfxsize=1.8in
\leavevmode
\epsffile{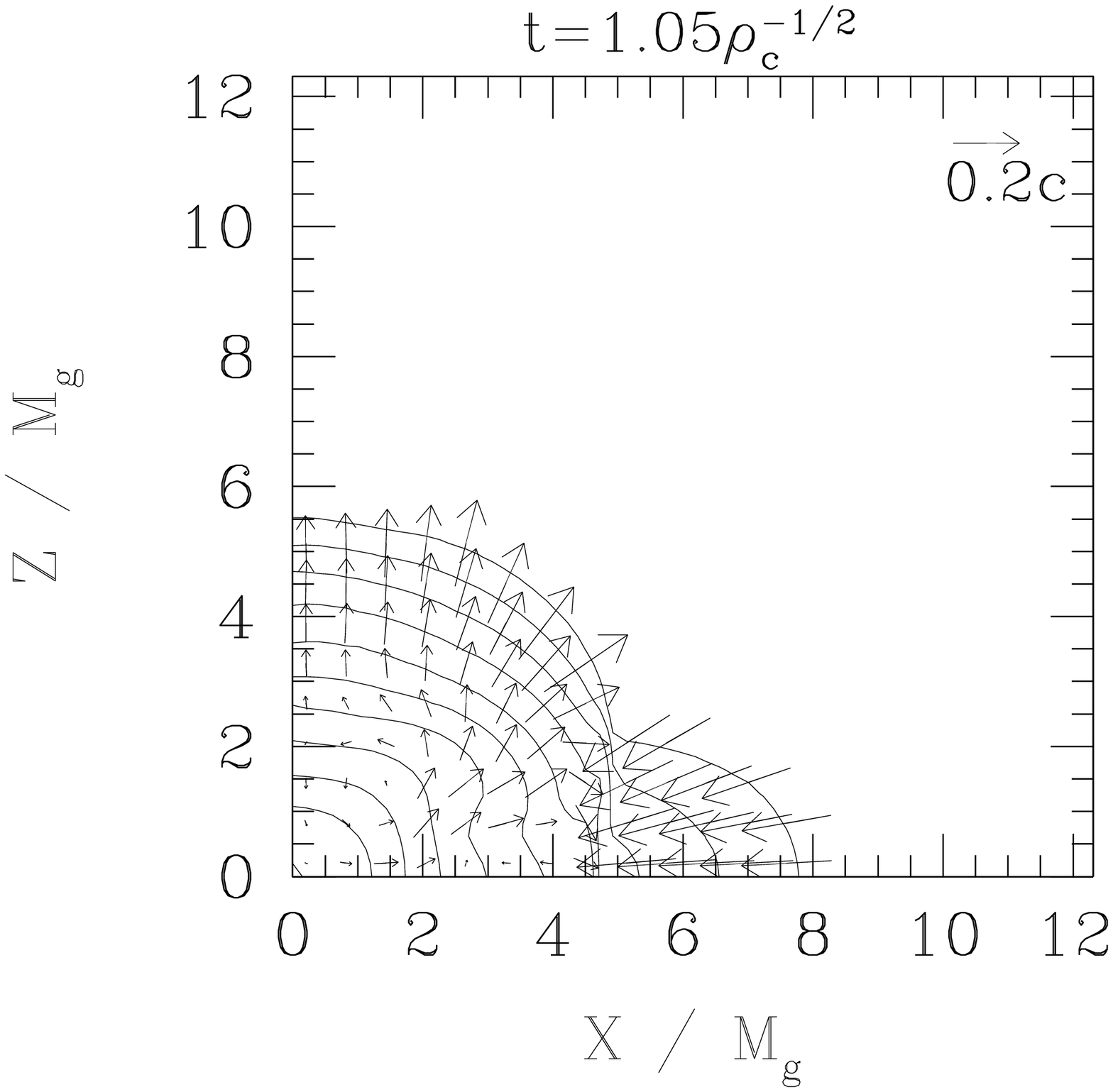}\\
\epsfxsize=1.8in
\leavevmode
\epsffile{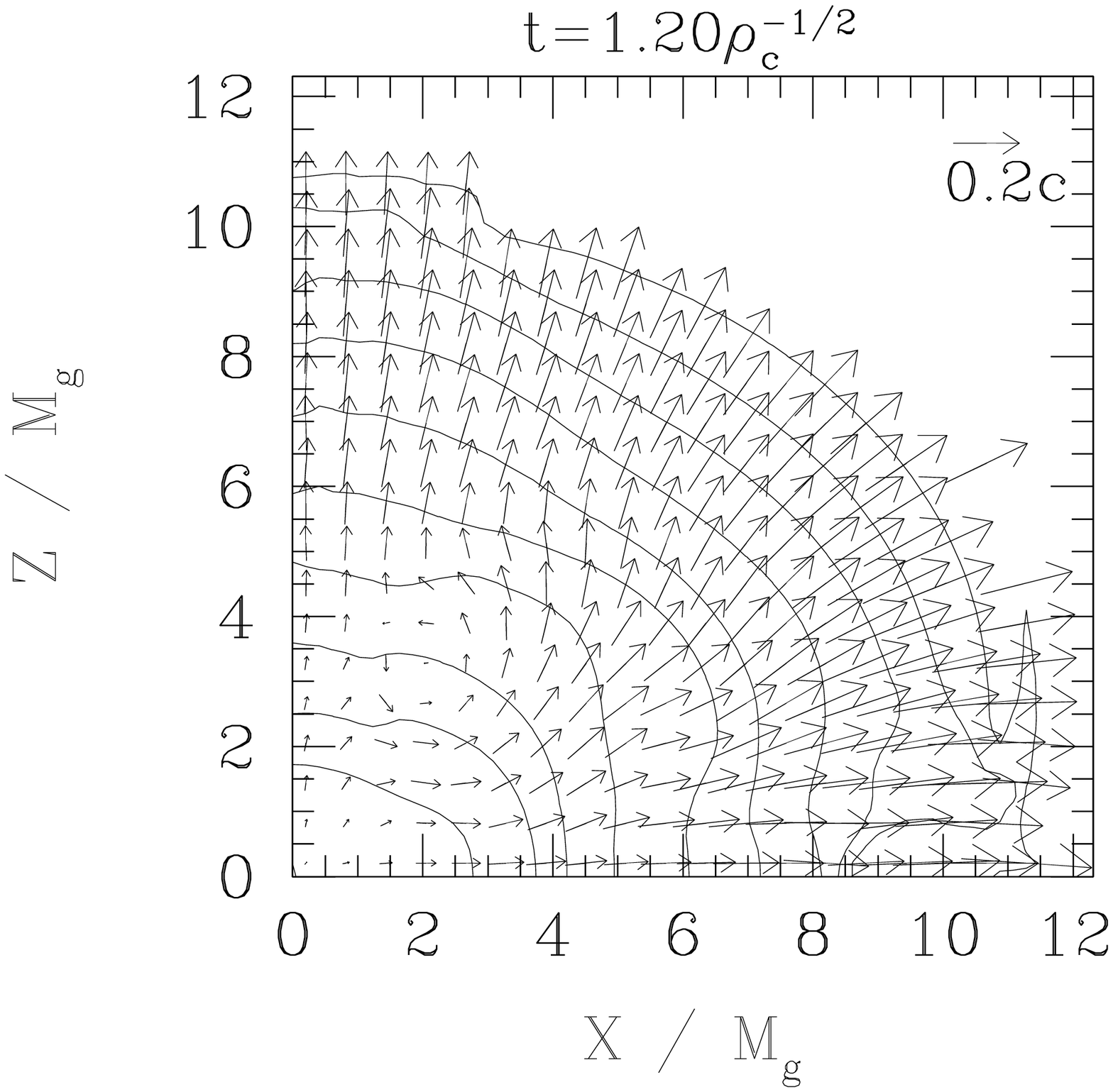}
\epsfxsize=1.8in
\leavevmode
\epsffile{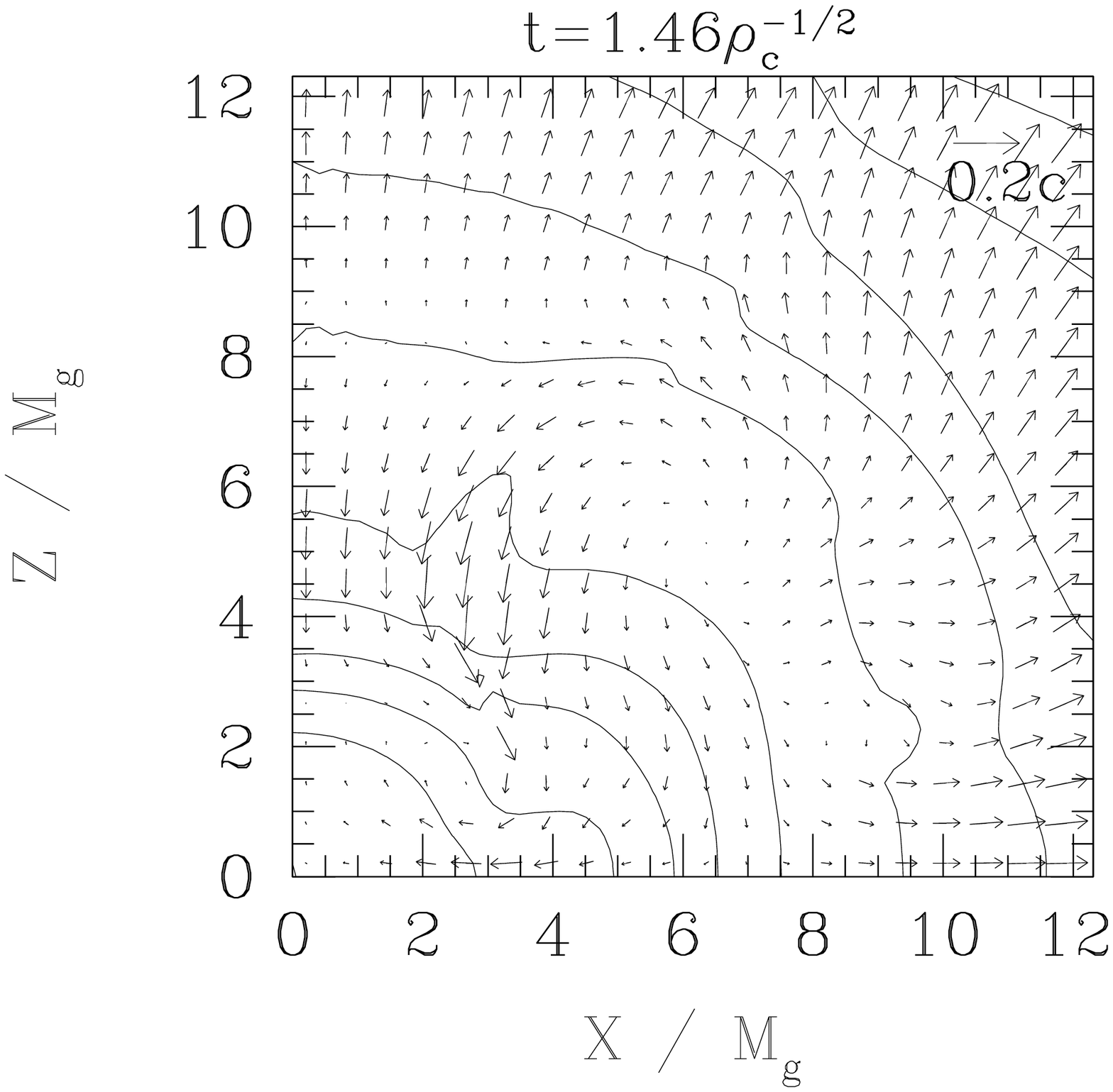}
\epsfxsize=1.8in
\leavevmode
\epsffile{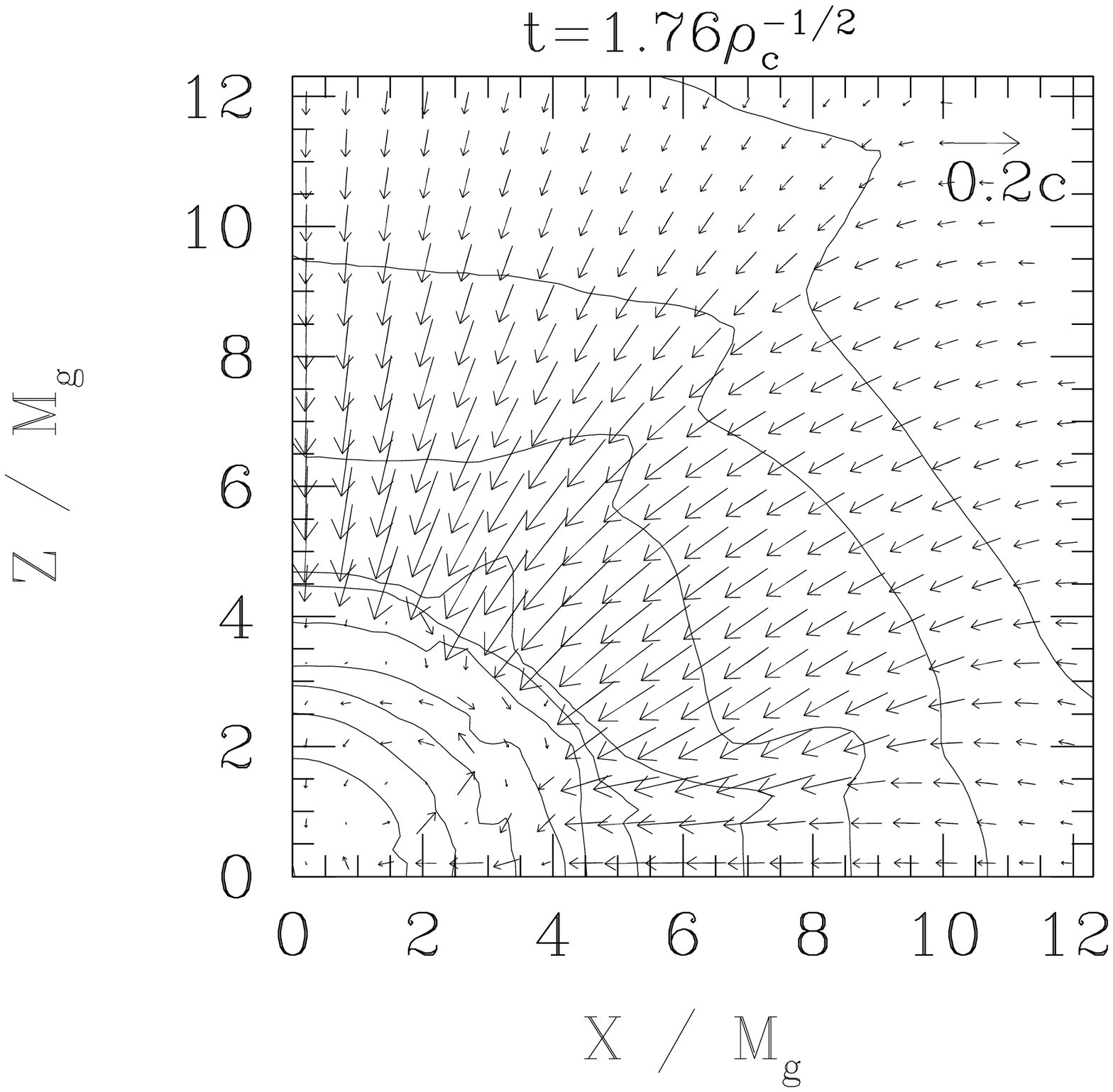}\\
\epsfxsize=1.8in
\leavevmode
\epsffile{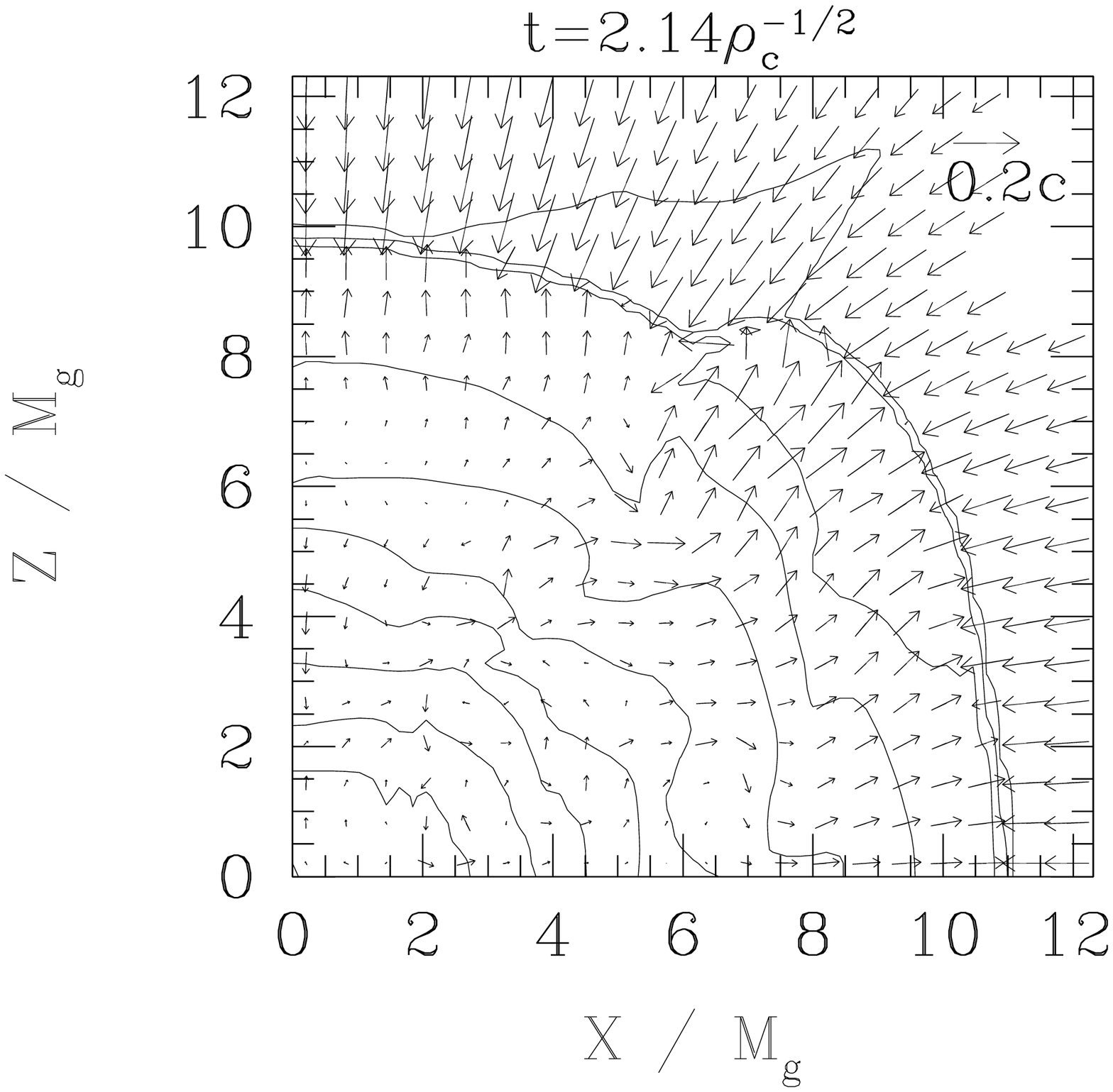}
\epsfxsize=1.8in
\leavevmode
\epsffile{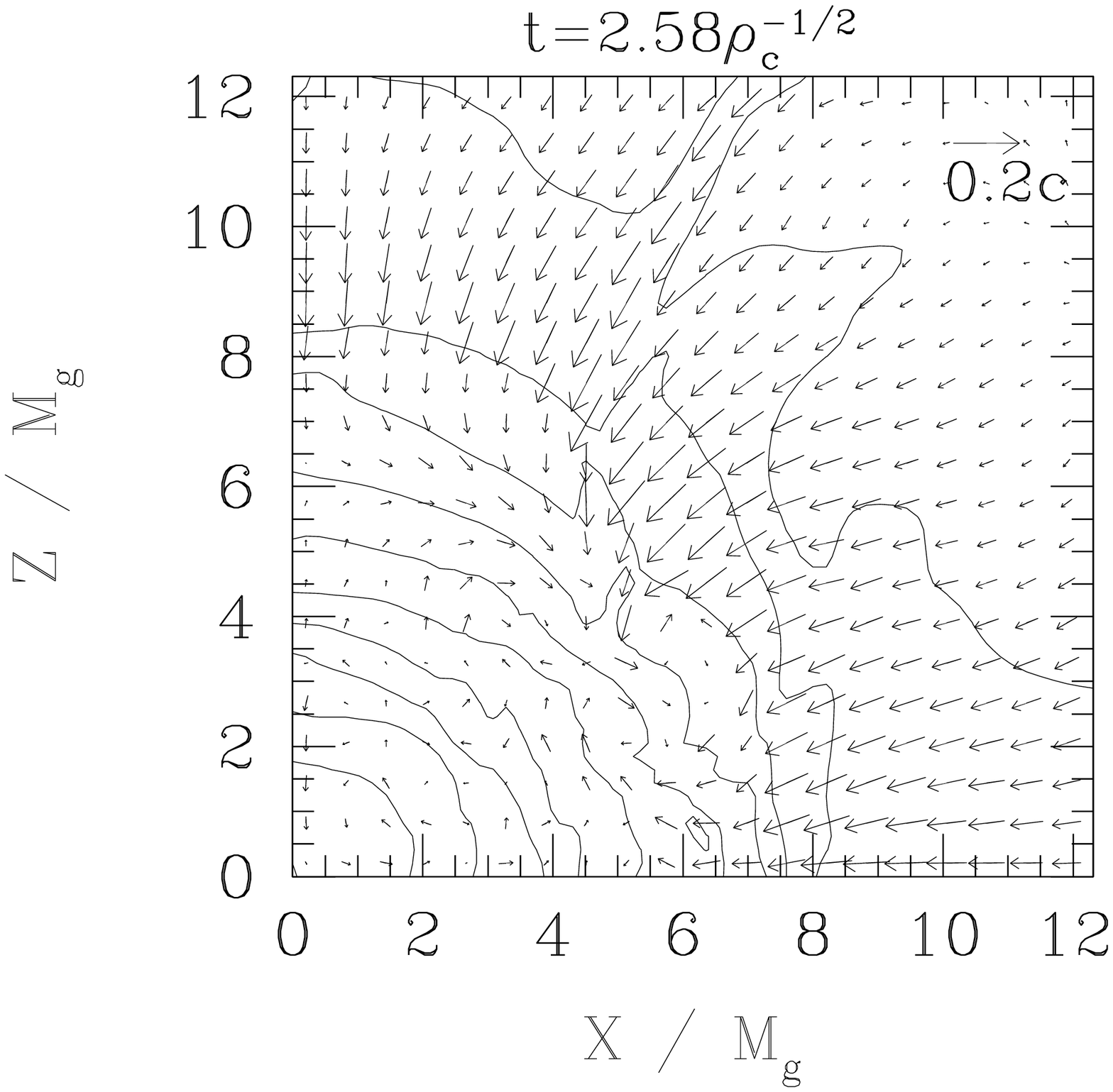}
\epsfxsize=1.8in
\leavevmode
\epsffile{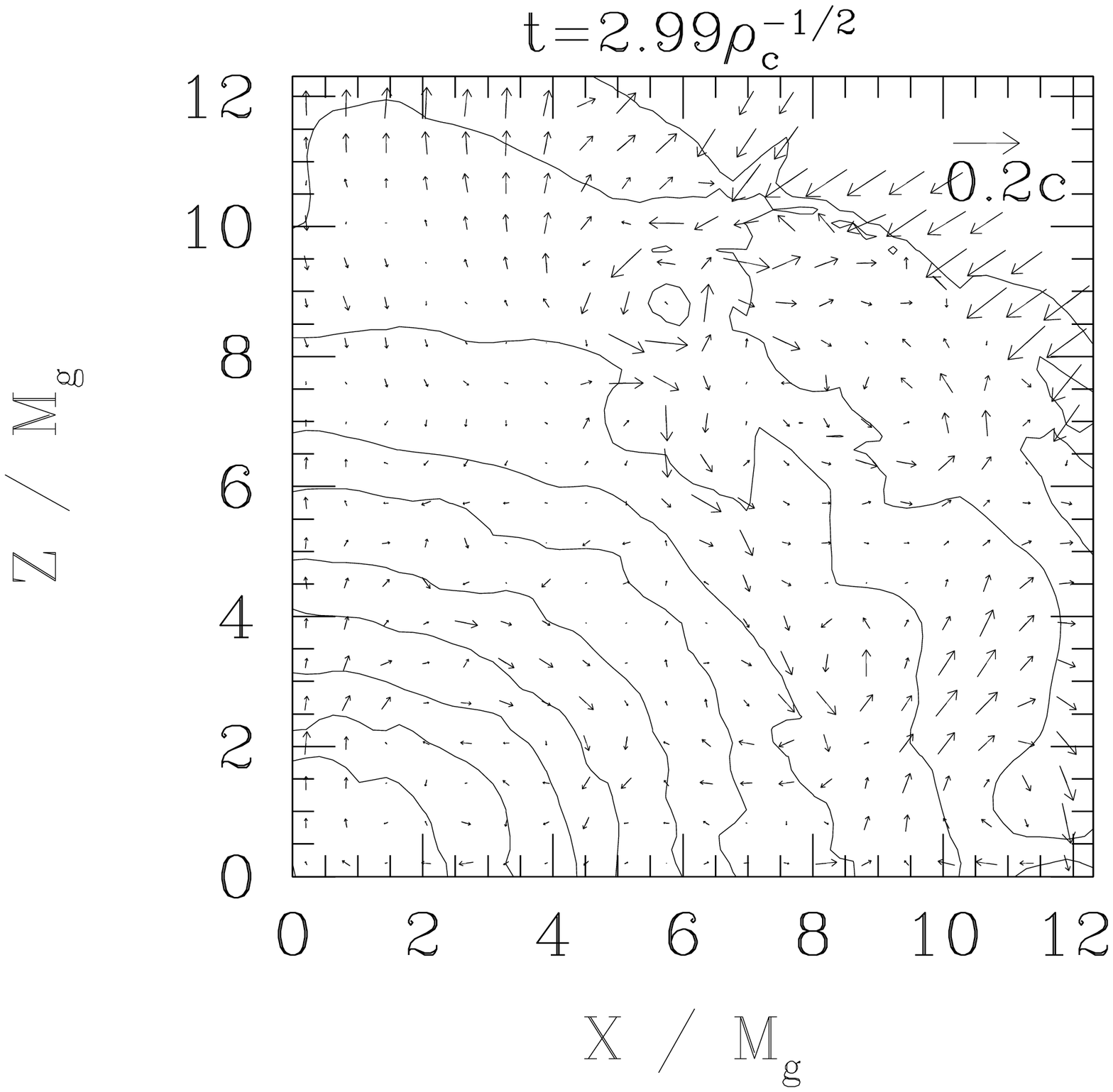}
\end{center}
\caption{The same as Fig.~4, but for non-adiabatic collapse 
for model (E) with $K^{-1/2}=4.47$. The contour lines 
are drawn for $\rho_*/\rho_{*~{\rm max}}=10^{-0.4j}$ 
for $j=0,1,2,\cdots,10$, where $\rho_{*~{\rm max}}$ 
is 424, 1566, 1027, 151, 100, 360, 162, 289, and 187 times larger than 
$\rho_{*~{\rm max}}$ at $t=0$.
}
\end{figure}

\clearpage
\begin{figure}[t]
\begin{center}
\epsfxsize=1.8in
\leavevmode
\epsffile{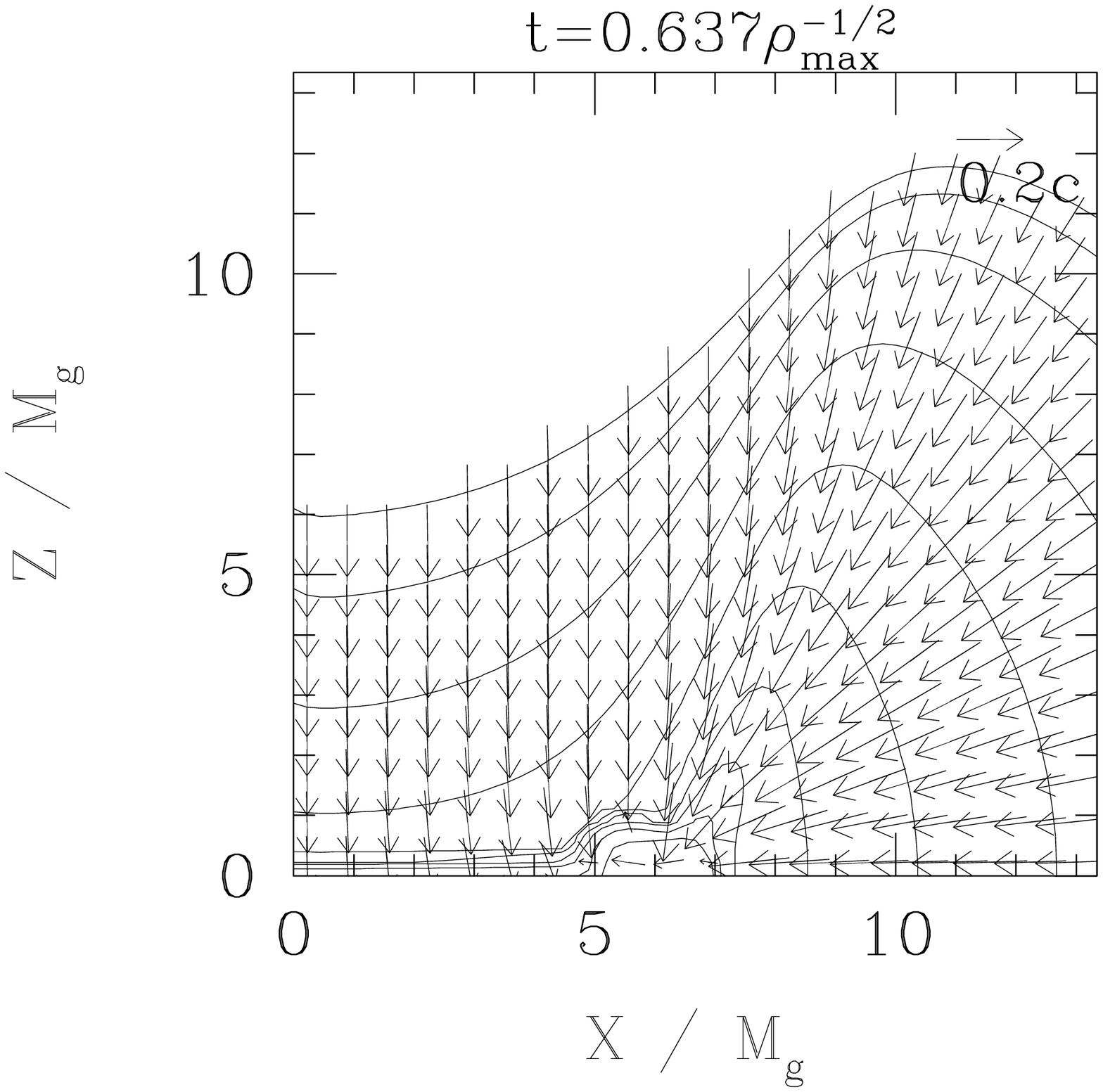}
\epsfxsize=1.8in
\leavevmode
\epsffile{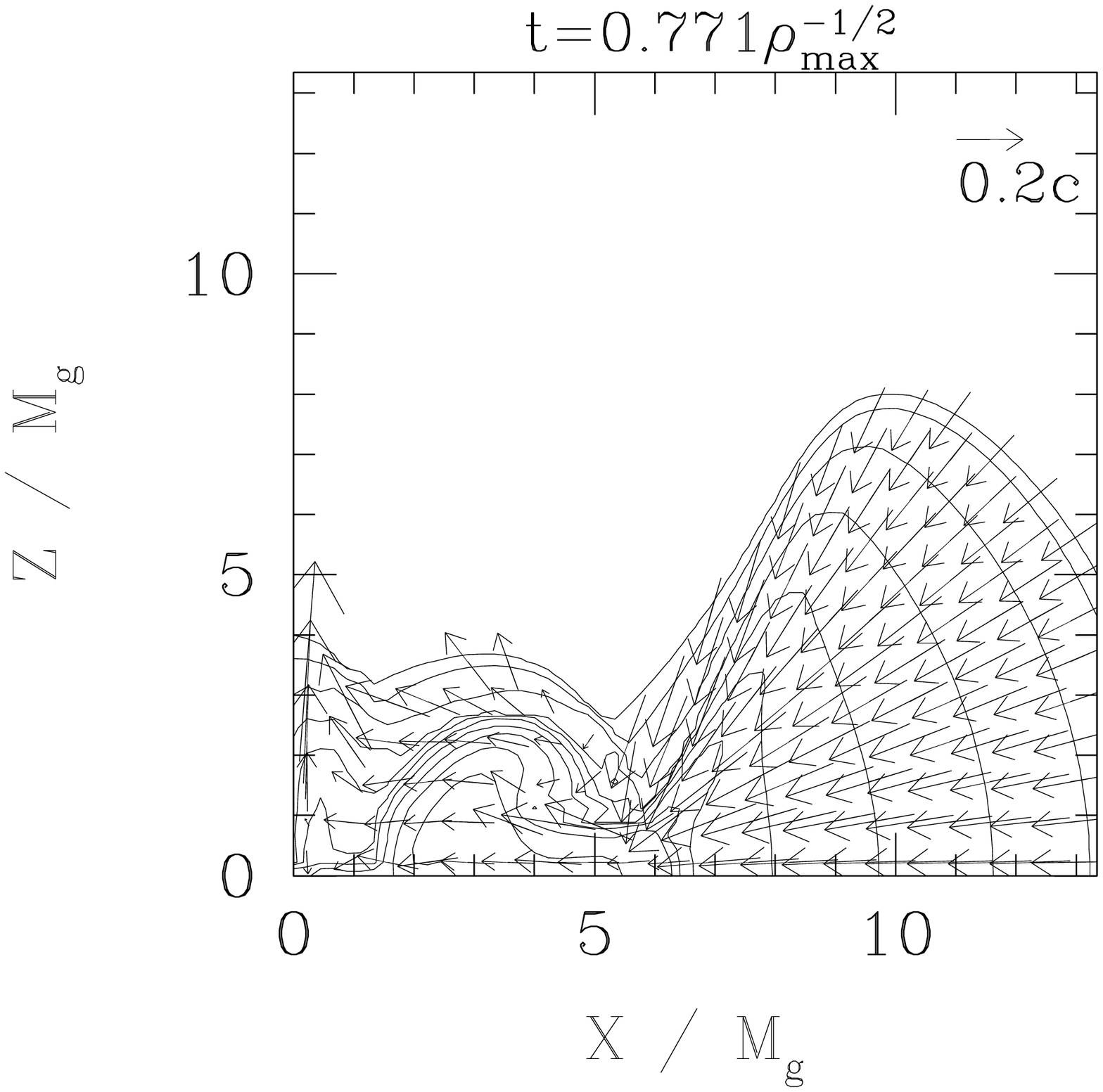}
\epsfxsize=1.8in
\leavevmode
\epsffile{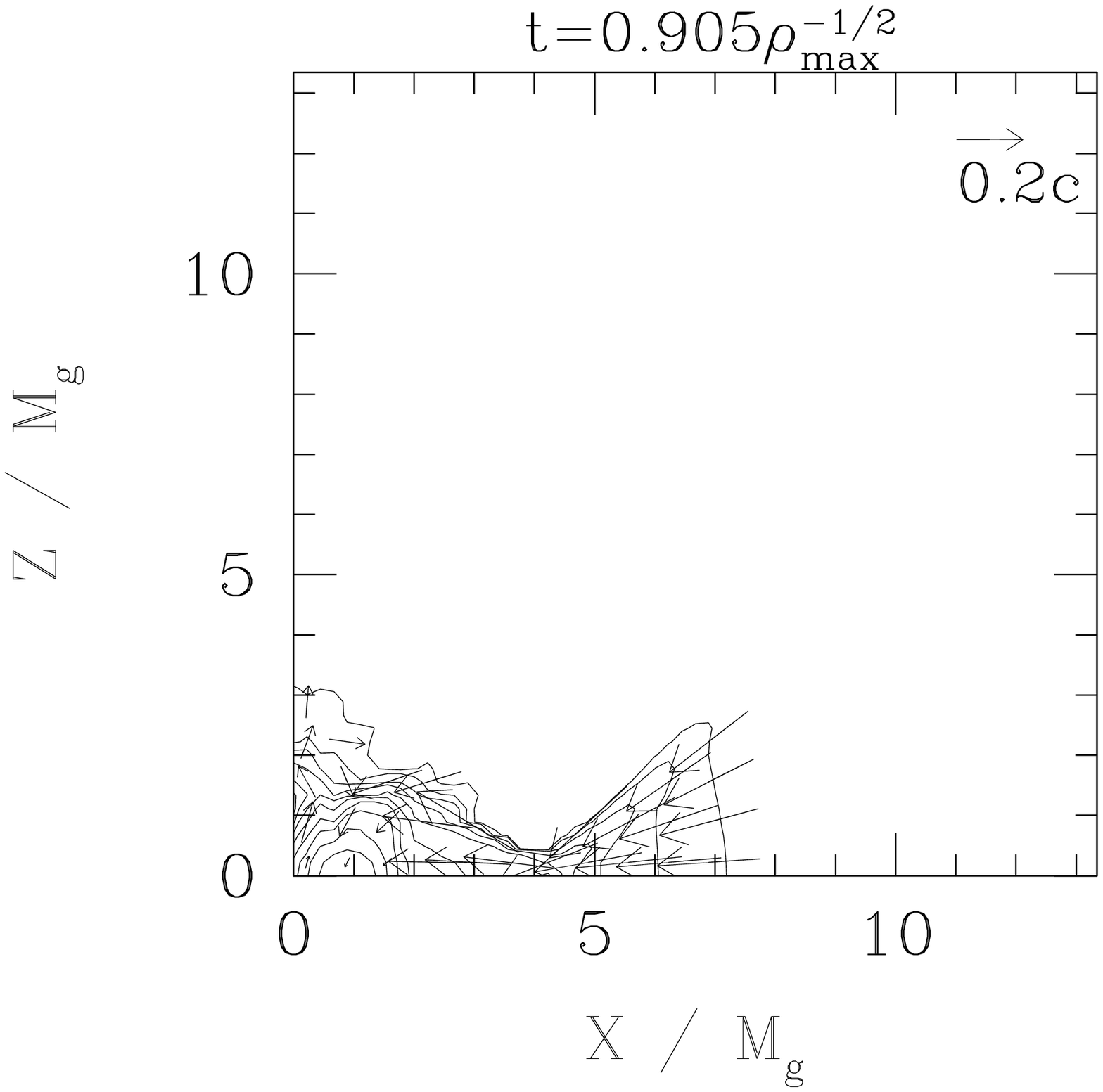}\\
\epsfxsize=1.8in
\leavevmode
\epsffile{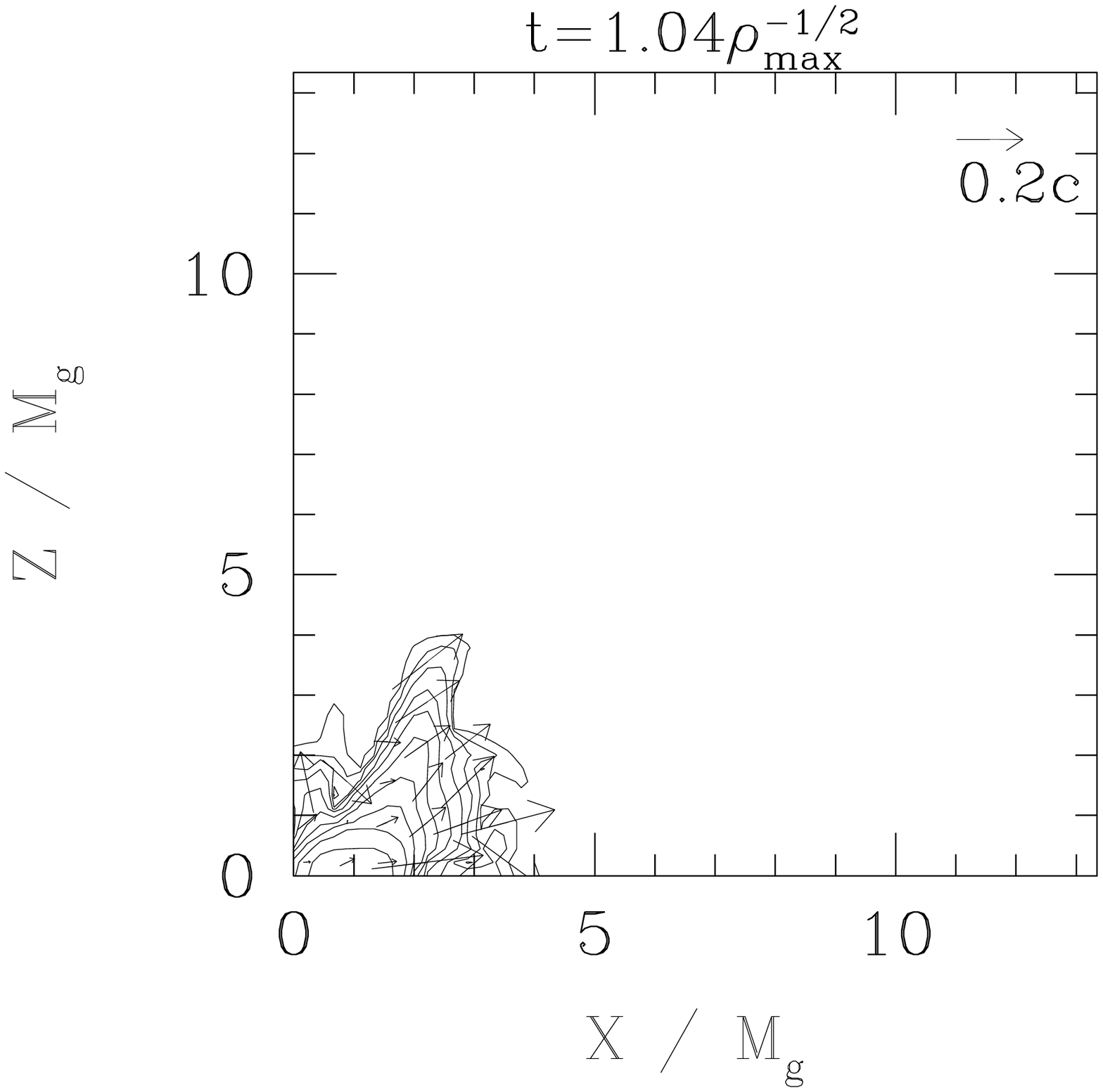}
\epsfxsize=1.8in
\leavevmode
\epsffile{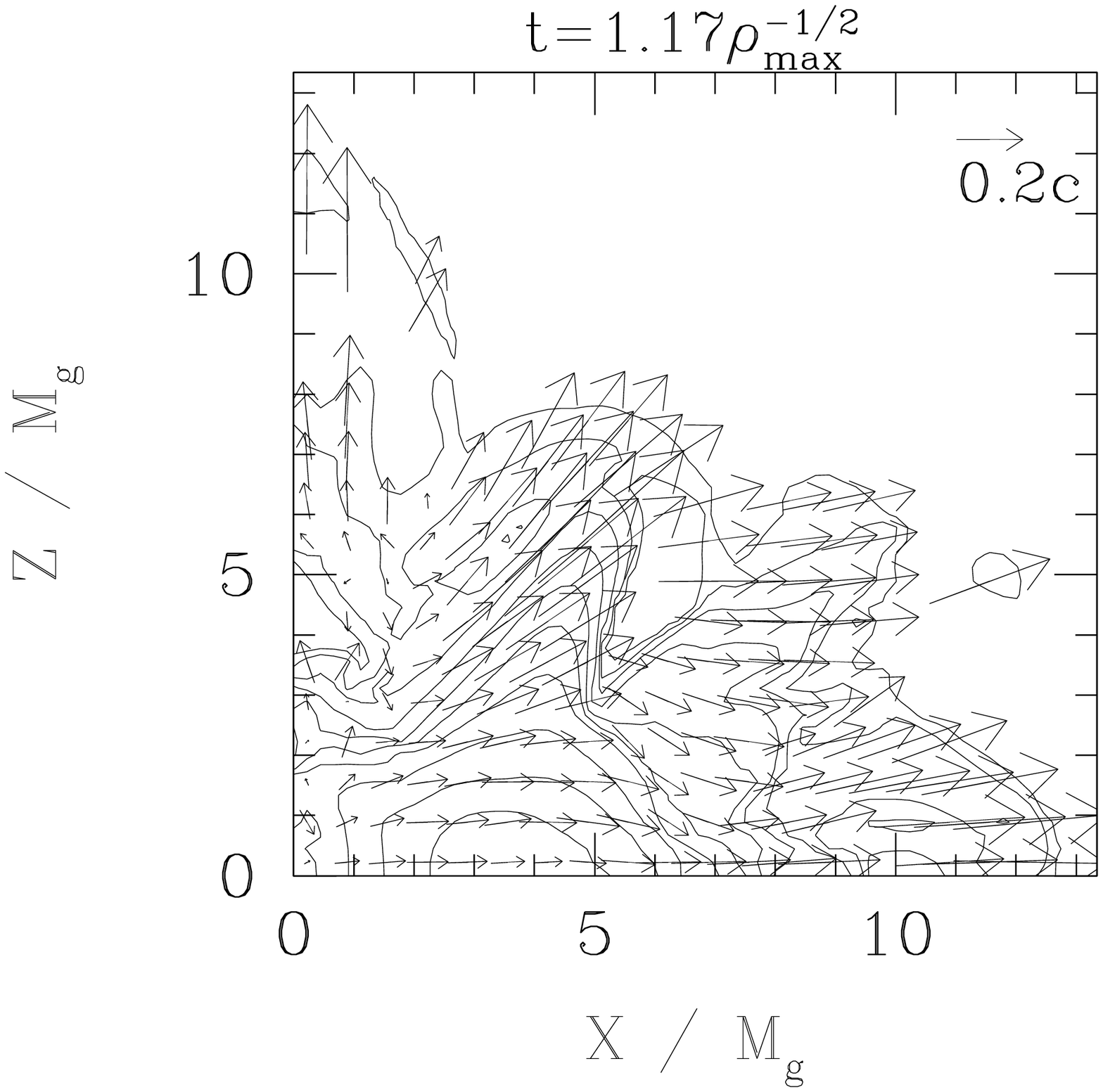}
\epsfxsize=1.8in
\leavevmode
\epsffile{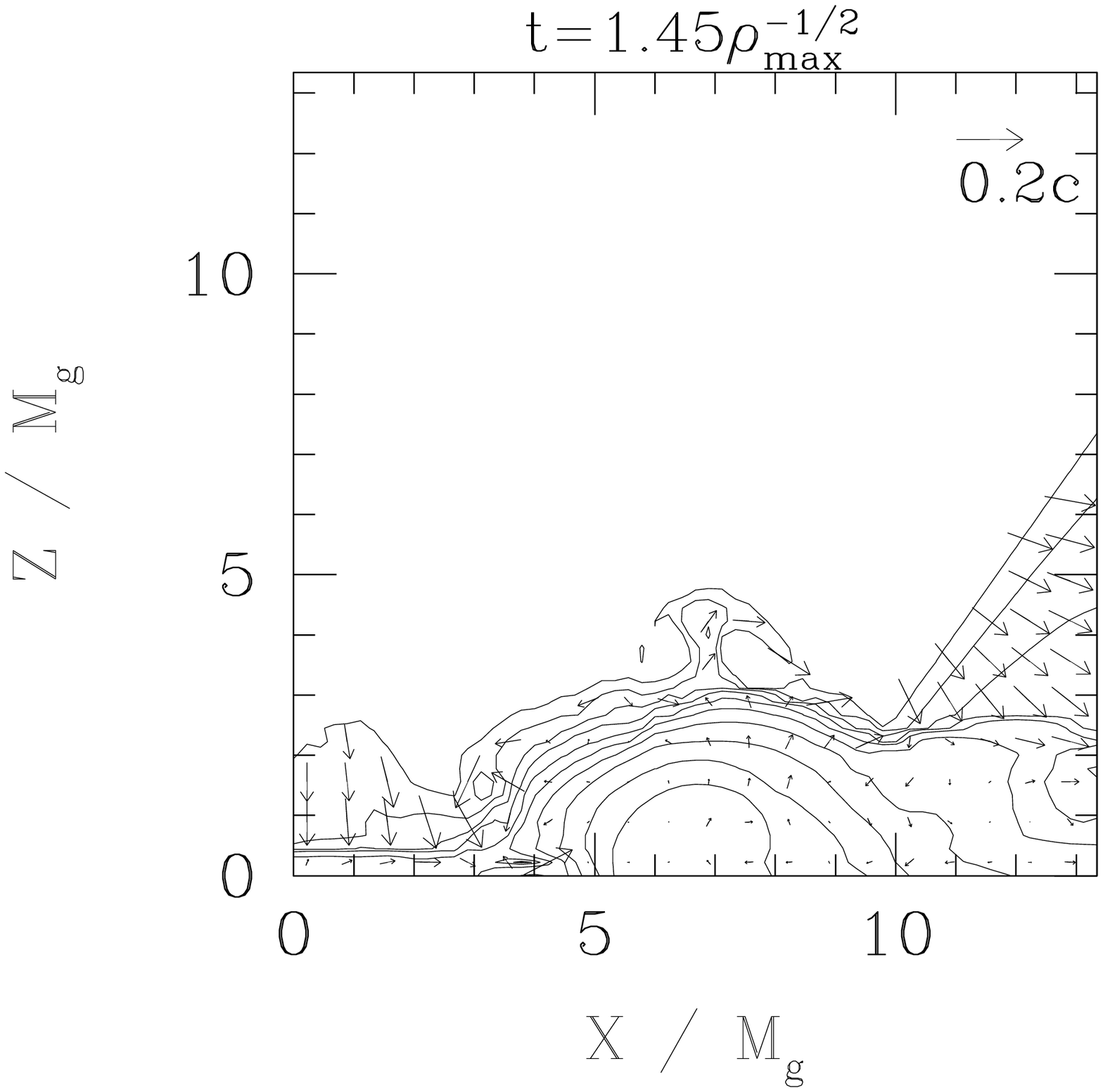}\\
\epsfxsize=1.8in
\leavevmode
\epsffile{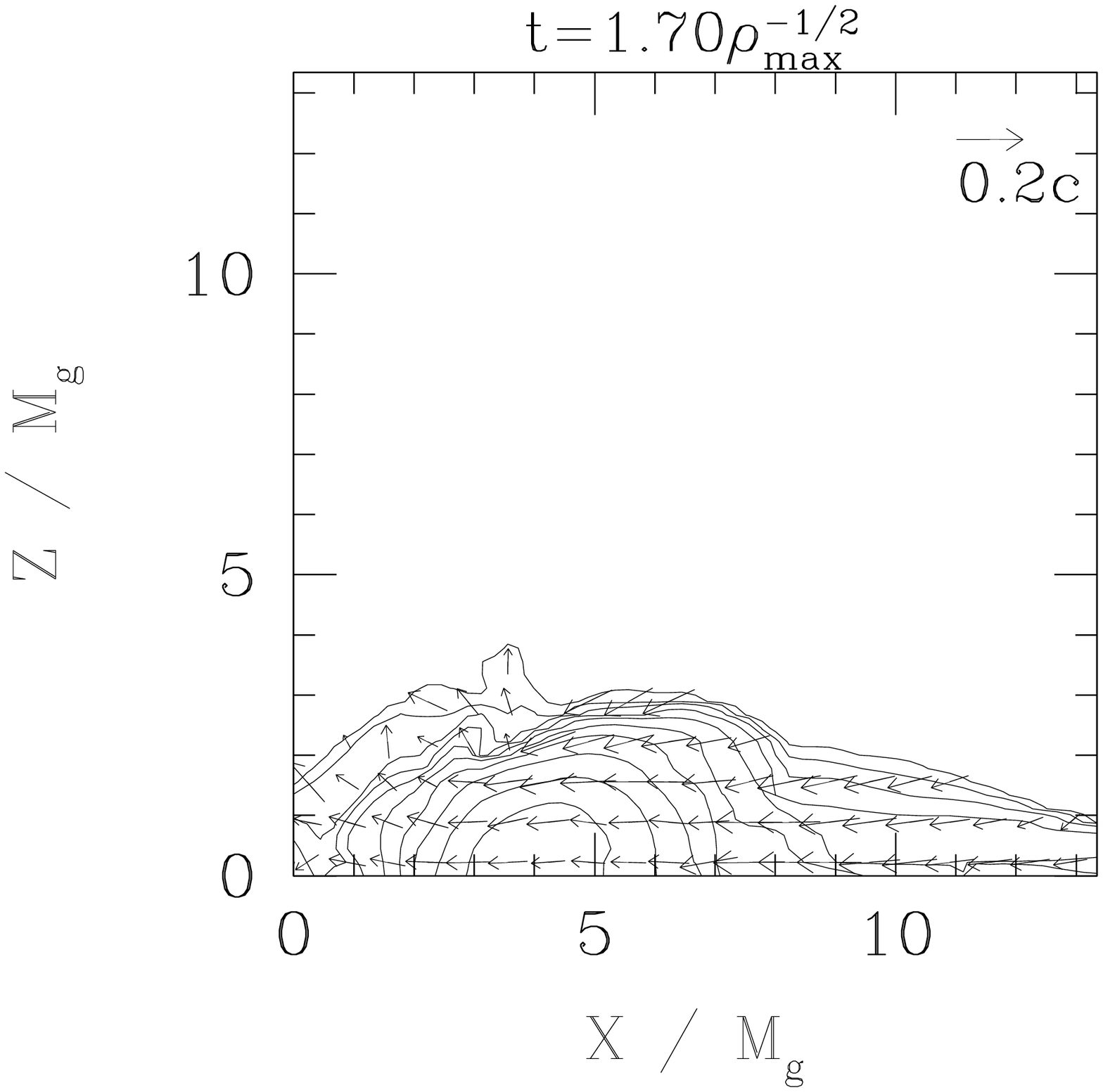}
\epsfxsize=1.8in
\leavevmode
\epsffile{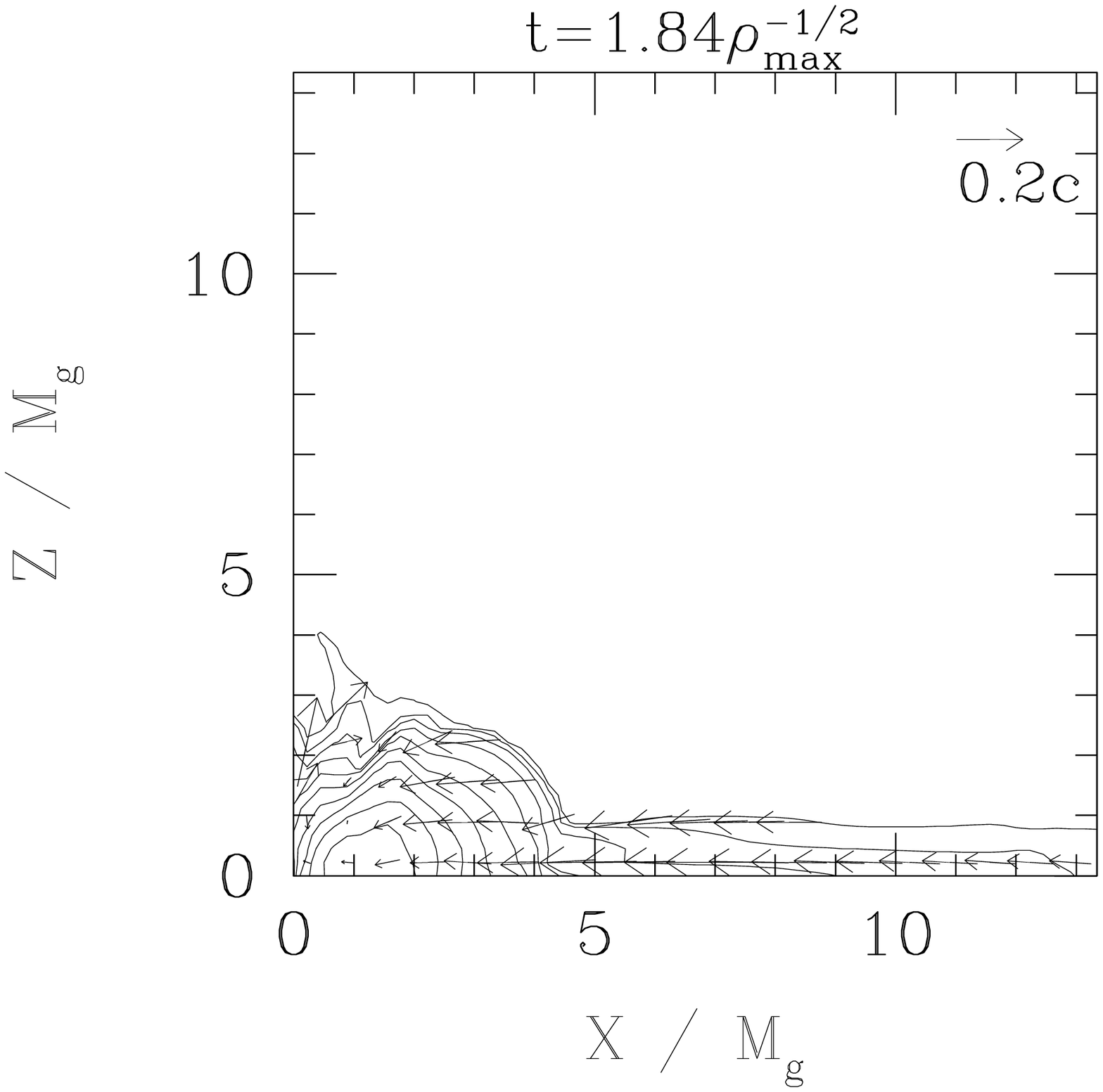}
\epsfxsize=1.8in
\leavevmode
\epsffile{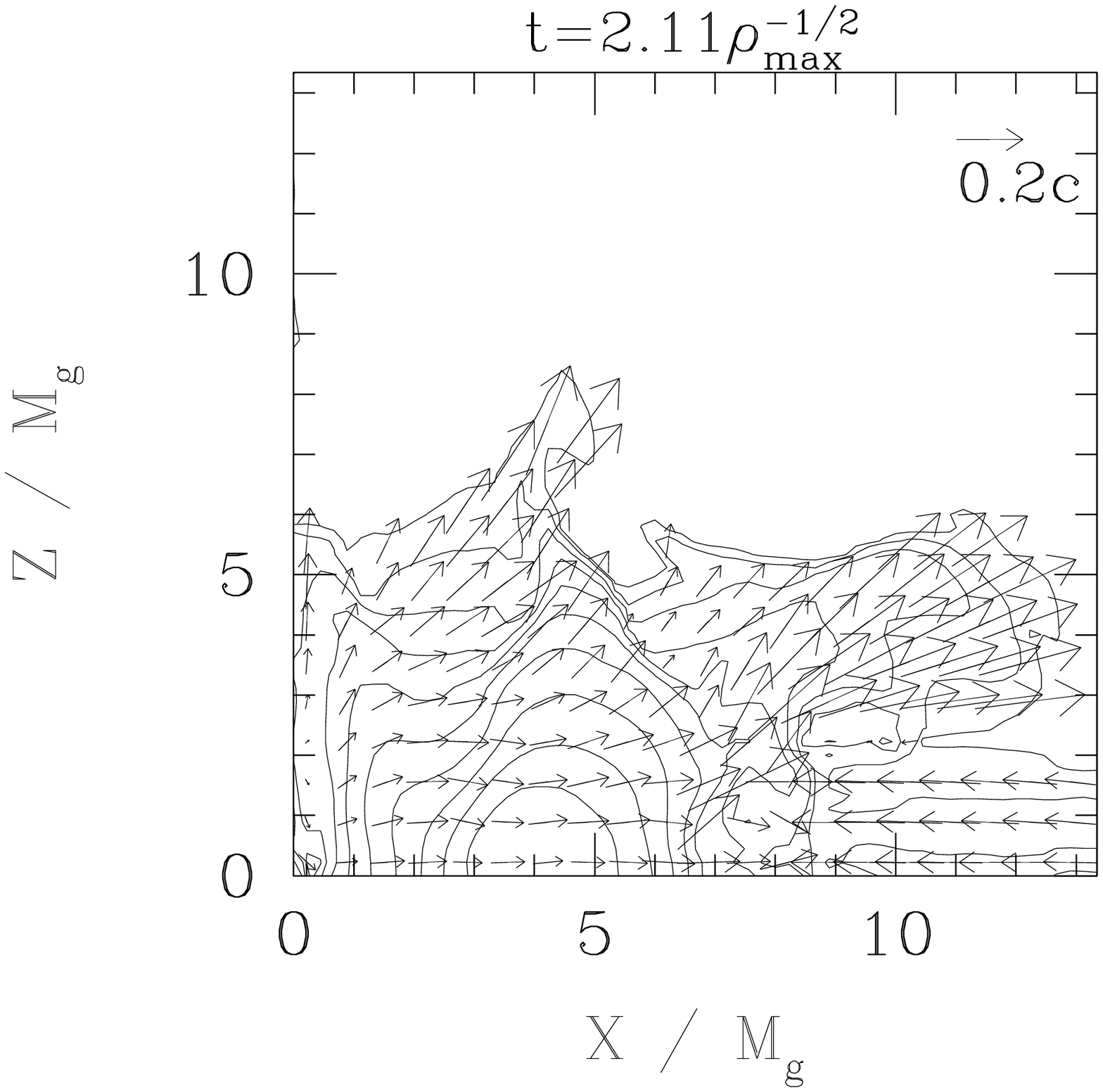}
\end{center}
\caption{The same as Fig.~4, but for non-adiabatic collapse 
for model (H) with $K^{-1/2}=9.49$. The contour lines 
are drawn for $\rho_*/\rho_{*~{\rm max}}=10^{-0.4j}$ 
for $j=0,1,2,\cdots,10$, where $\rho_{*~{\rm max}}$ 
is 165, 168, 4276, 2684, 165, 73.6, 
161, 1136, 1192, and 124 times larger than $\rho_{*~{\rm max}}$ at $t=0$.
The time appears in units of $\rho_{\rm max}^{-1/2}$ at $t=0$. 
} 
\end{figure}

\clearpage
\begin{figure}[t]
\begin{center}
\epsfxsize=1.8in
\leavevmode
\epsffile{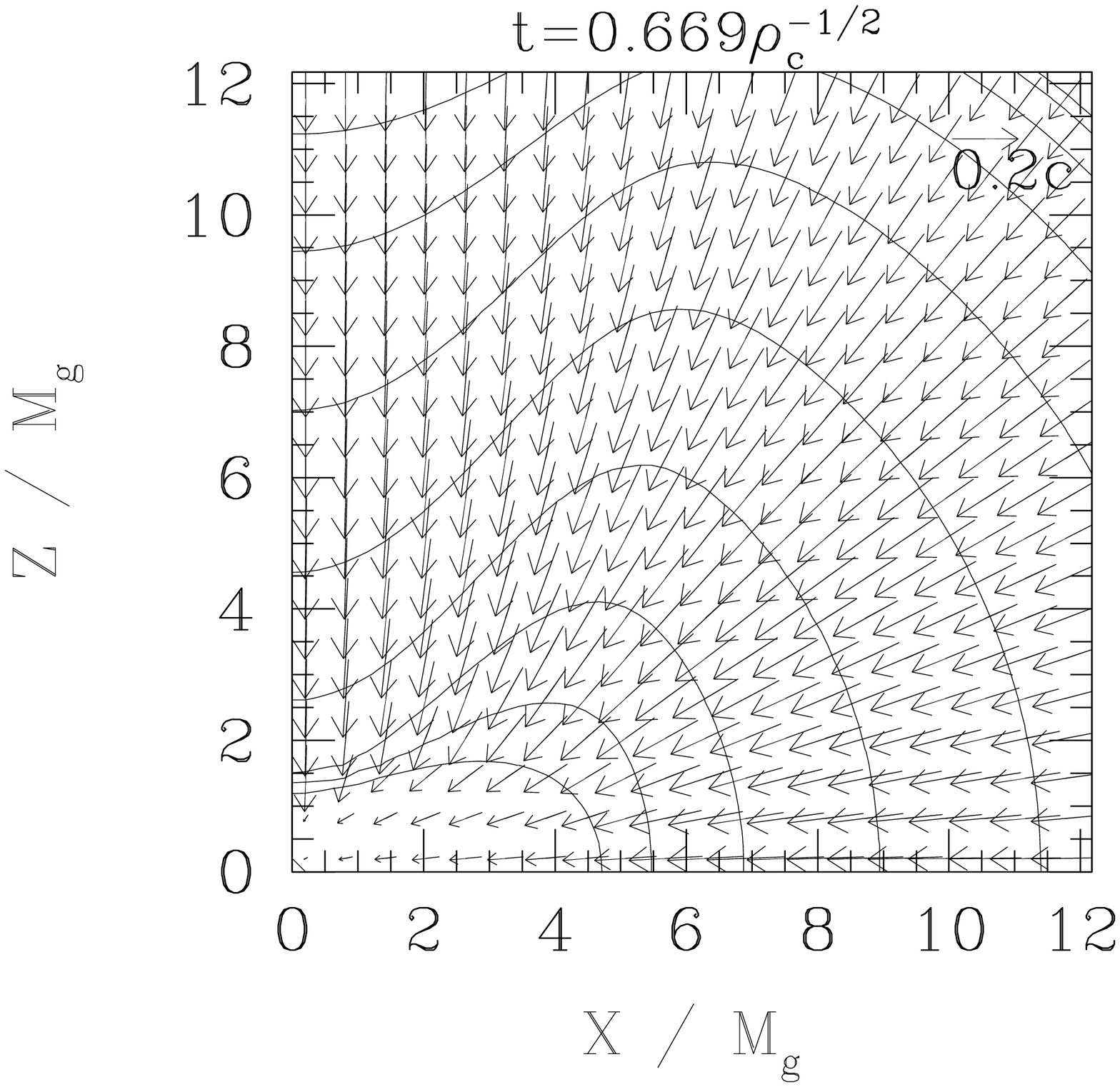}
\epsfxsize=1.8in
\leavevmode
\epsffile{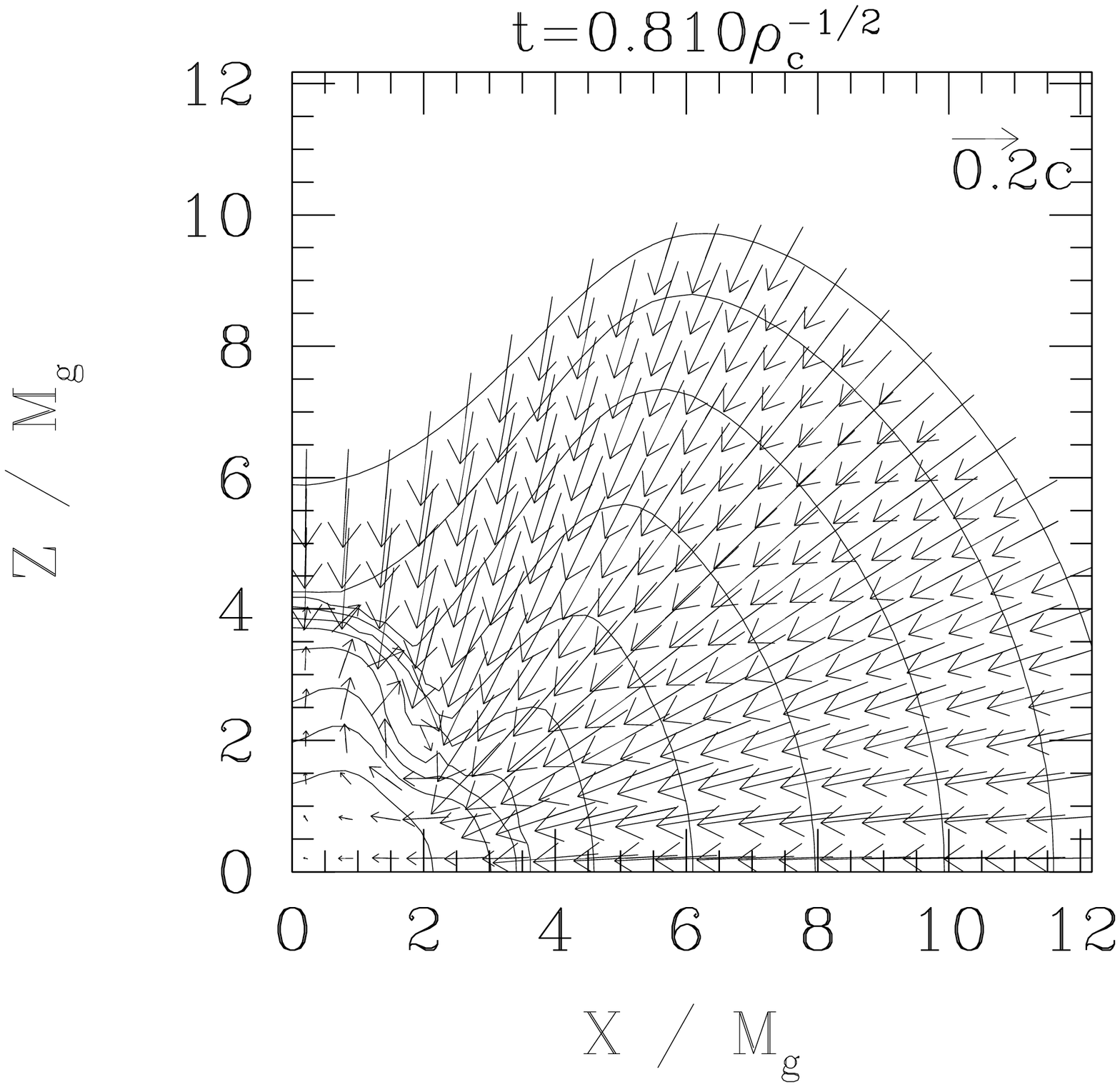}
\epsfxsize=1.8in
\leavevmode
\epsffile{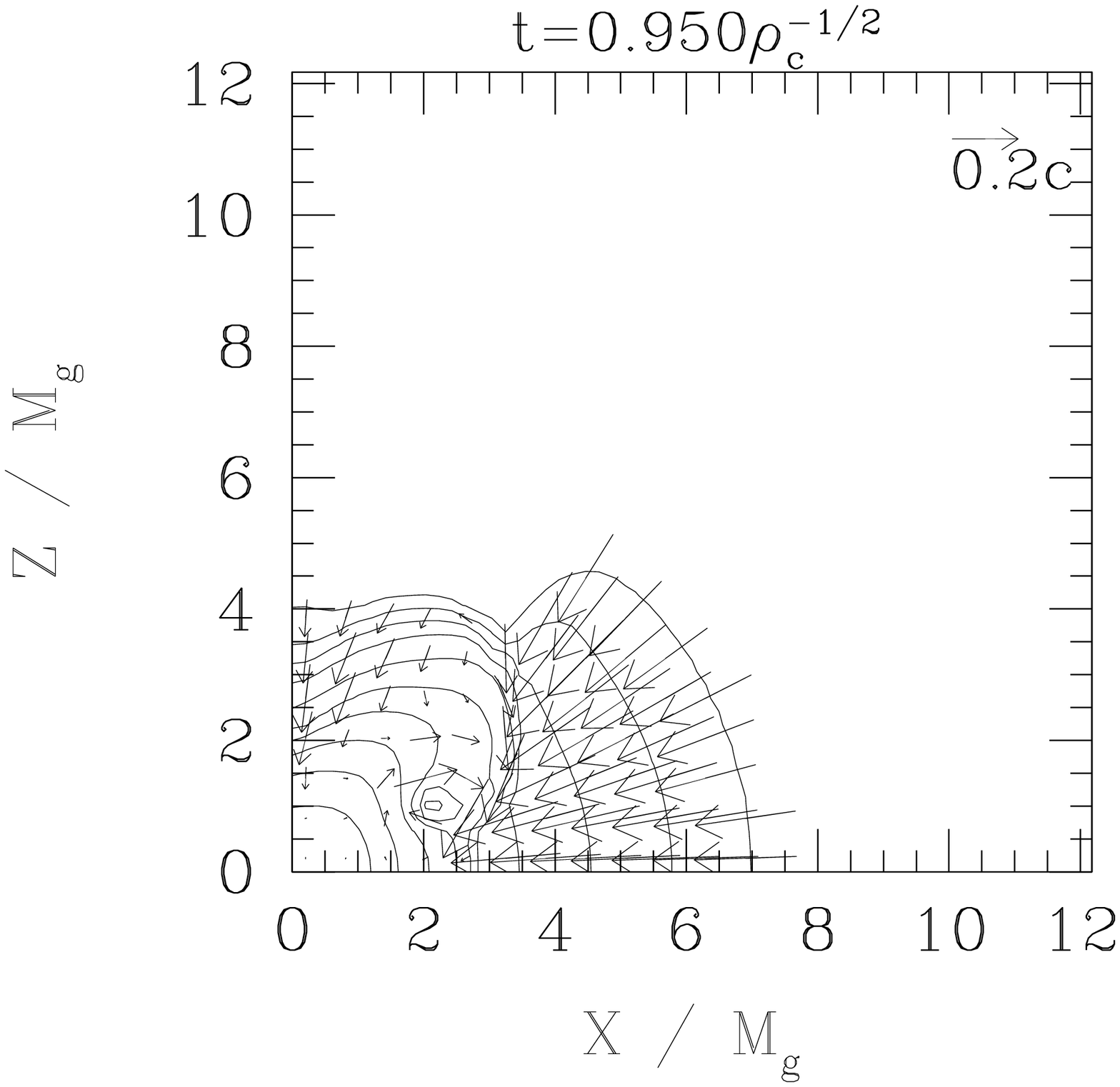}\\
\epsfxsize=1.8in
\leavevmode
\epsffile{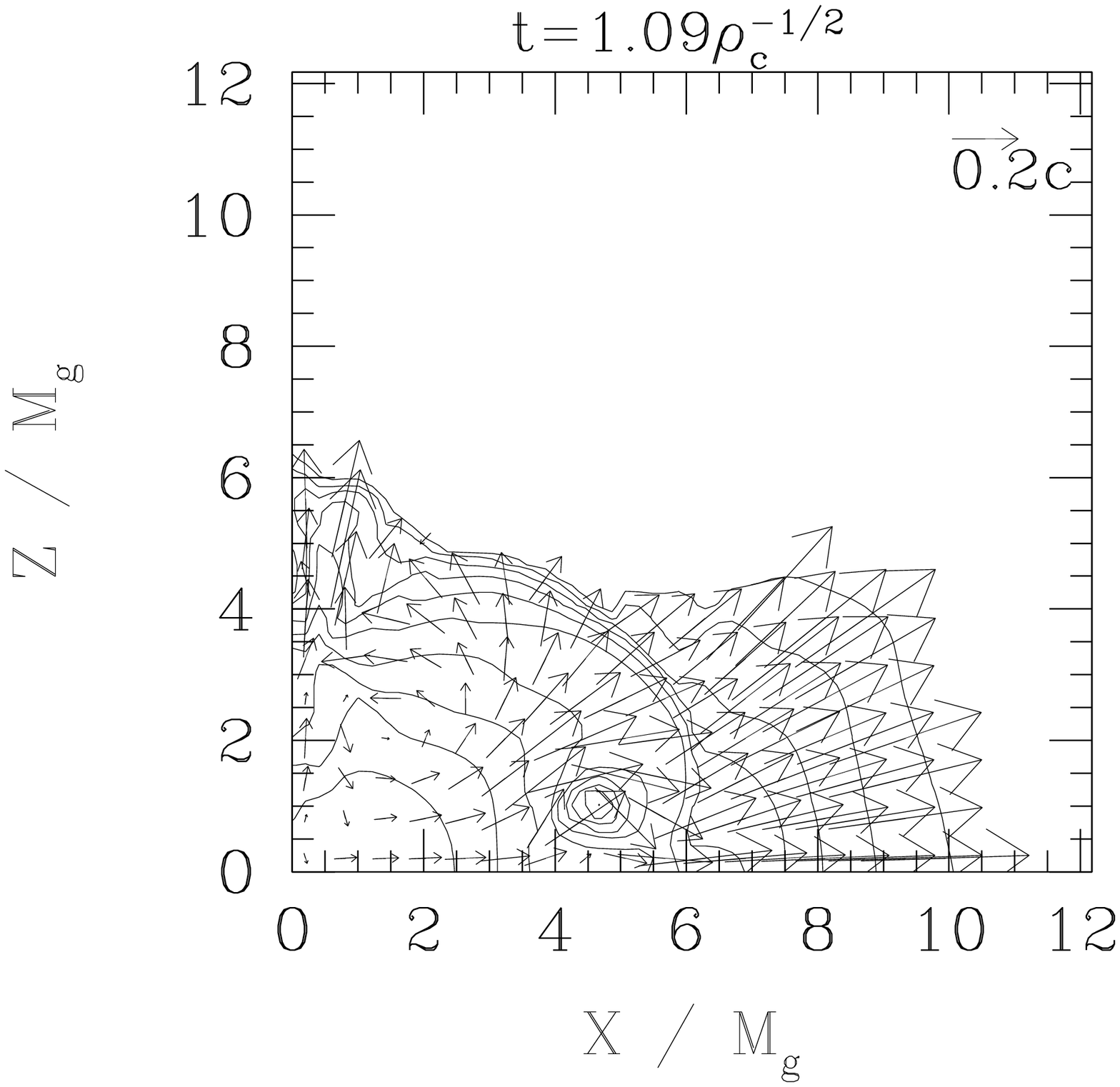}
\epsfxsize=1.8in
\leavevmode
\epsffile{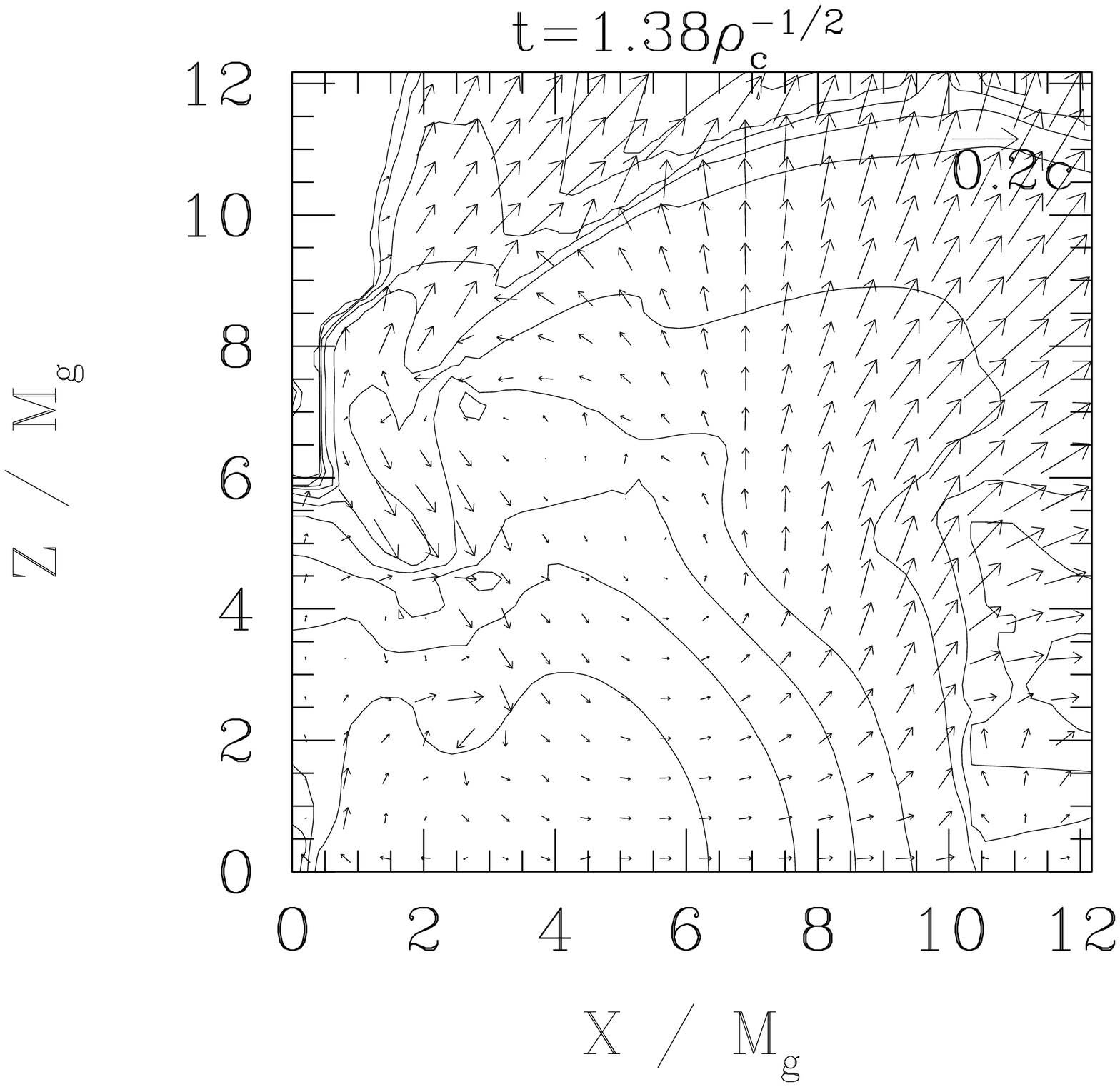}
\epsfxsize=1.8in
\leavevmode
\epsffile{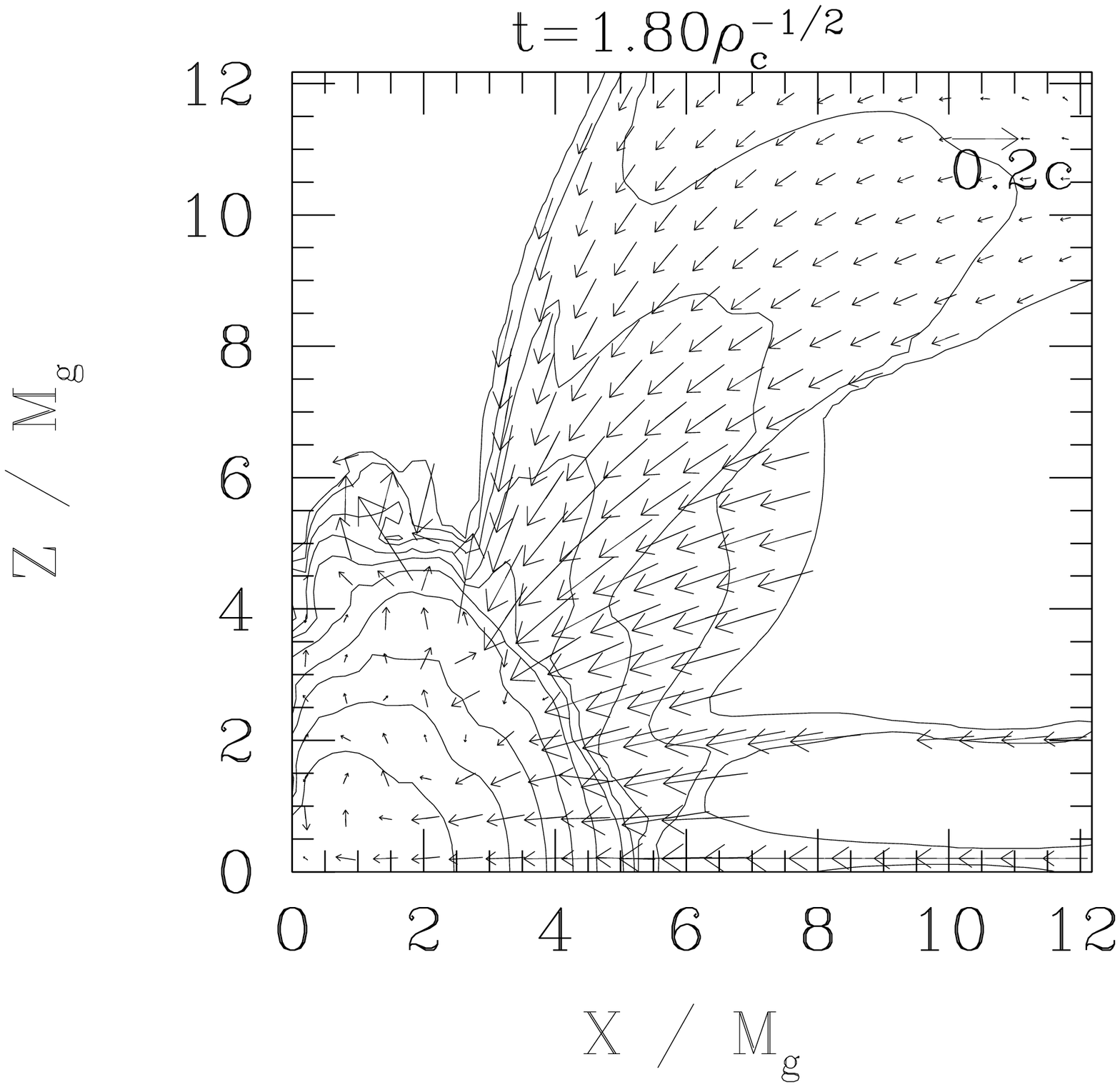}
\end{center}
\caption{The same as Fig.~4, but for non-adiabatic collapse 
for model (K) with $K^{-1/2}=4.00$. The contour lines 
are drawn for $\rho_*/\rho_{*~{\rm max}}=10^{-0.4j}$ 
for $j=0,1,2,\cdots,10$, where $\rho_{*~{\rm max}}$ 
is 45.9, 379, 1366, 256, 23.1, and 
219 times larger than $\rho_{*~{\rm max}}$ at $t=0$.
Note that $\rho_c=\rho_{\rm max}$ at $t=0$ in this case. 
} 
\end{figure}


\begin{thebibliography}{99}
\bibitem{foot00} But, for a special class of gauge conditions 
with appropriate prescriptions, it is still possible 
to preserve the accuracy and stability without 
artificial viscosity ({e.g.}, D. Garfinkle, personal communication).
\bibitem{foot} Even if we solve Eq.~(\ref{energy}) 
using the $\Gamma$-law equation of state, 
the equation of state is preserved in polytropic form 
with invariant $K$ in the case that we do not add 
artificial viscosity in our present choice of the energy equation. 
This is because the relation 
$e_*/\rho_*=\{P/(\Gamma-1)\}^{1/\Gamma}/\rho=$constant 
holds during the evolution in the absence of the 
artificial viscosity.
\bibitem{foot2} We note that 
even if $K^{-1/2}$ is smaller than 3.6, a black hole 
could be formed for the case in which the gravitational binding energy is 
sufficiently large initially. Hence, the value $3.6$ is not an 
absolute threshold, even for $q=0$, although it could be a good guideline.
\bibitem{foot5} We reached this conclusion because 
convergence was not well achieved when $N$ was increased.
\bibitem{foot4} Throughout this paper, $\rho_c$ and $\rho(r=0)$ are 
used as the central density at $t=0$ and the central density during 
the time evolution, respectively. $\rho_{\rm max}$ denotes 
the maximum density at $t=0$.
\bibitem{foot3} Note that $q$ of a formed black hole is expected to be 
$\sim 0.9$ in this case. This implies that the 
coordinate radius of the innermost 
stable circular orbit around the black hole is $\sim 2M_g$ 
in our coordinate systems, and hence, matter located 
at $\agt 2 M_g $ could maintain a stable orbit.
\end{thebibliography}
\end{document}